\documentclass[twocolumn,showpacs,prl]{revtex4}
\usepackage{graphicx}
\usepackage{dcolumn}
\usepackage{bm}
\usepackage{amsmath}
\begin{document}

\newcommand{\lsim}   {\mathrel{\mathop{\kern 0pt \rlap
  {\raise.2ex\hbox{$<$}}}
  \lower.9ex\hbox{\kern-.190em $\sim$}}}
\newcommand{\gsim}   {\mathrel{\mathop{\kern 0pt \rlap
  {\raise.2ex\hbox{$>$}}}
  \lower.9ex\hbox{\kern-.190em $\sim$}}}

\newcommand{\be}{\begin{equation}}
\newcommand{\ee}{\end{equation}}
\newcommand{\ba}{\begin{eqnarray}}
\newcommand{\ea}{\end{eqnarray}}
\def\bone{$B^{(1)}$}
\def\bone{B^{(1)}}
\def\etal{{\it et al.~}}
\def\eg{{\it e.g.~}}
\def\ie{{\it i.e.~}}
\def\DM{dark matter~}
\def\DE{dark energy~} 
\def\GC{Galactic center~} 
\def\susy{SUSY~}

\title{\hfill {\small{CERN-PH-TH/2007-173}}\break $~$ \hfill\break 
GZK Photons  as  Ultra High Energy Cosmic Rays}

\author{Graciela B. Gelmini$^{a,b}$, Oleg E. Kalashev$^{c}$ and Dmitry
  V. Semikoz$^{d,b}$ }
\affiliation{ $^a$ Department of Physics and Astronomy, UCLA, Los Angeles,
CA 90095-1547, USA \\
 $^b$ CERN, PH-TH, CH-1211 Gen\`eve 23, Switzerland\\
 $^c$INR RAS, 60th October
Anniversary pr. 7a, 117312 Moscow, Russia  \\ 
$^d$ APC, College de France, 11 pl. 
Marcelin Berthelot, Paris 75005, France } 
\vspace{0.5truecm}
\begin{abstract}
We calculate the flux of ``GZK-photons", namely the flux of Ultra
High Energy Cosmic Rays (UHECR) consisting of
photons produced by  extragalactic nucleons through the resonant
 photoproduction of pions,
 the so called GZK effect. We show that,
for primary nucleons, the GZK photon fraction of the total UHECR flux
is between   $10^{-4}$  and $10^{-2}$ above $10^{19}$ eV  and  up to the order of $0.1$  above $10^{20}$ eV. The GZK photon flux depends on the assumed
UHECR spectrum, slope of the nucleon flux at the source, distribution of
sources and  intervening backgrounds. Detection of this photon flux would
open the way for UHECR gamma-ray astronomy.   Detection of a larger photon
 flux would imply the emission of photons at the source or new physics.
We compare the photon fractions expected for GZK photons and the minimal predicted by Top-Down models. We find that
 the photon fraction above  $10^{19}$ eV is a crucial test for Top-Down models.

\end{abstract}

\pacs{ 
\hfill UCLA/04/TEP/17 }
\maketitle

\vspace{1truecm}

\section{I. Introduction}

The cosmic rays with energies beyond the Greisen-Zatsepin-Kuzmin (GZK)
cutoff~\cite{gzk} at $4\times 10^{19}$~eV present a challenging outstanding
puzzle in astroparticle physics and cosmology~\cite{agasa, hires}. Nucleons
cannot be confined to our galaxy for energies above the ``ankle", i.e.  above
10$^{18.5}$~eV. This and the absence of a correlation of arrival directions
with the galactic plane indicate that, if nucleons are the primary particles
of the ultra high energy cosmic rays (UHECR), these nucleons should be of
extragalactic origin. However, nucleons with energies above 
$5 \times 10^{19}$~eV could not reach Earth from a distance beyond 50 to 100
Mpc~\cite{50Mpc} because they scatter off the cosmic microwave background
(CMB) photons with a resonant photoproduction of pions: $p\gamma \rightarrow
\Delta^* \rightarrow N\pi$, where the pion carries away $\sim 20\%$ of the
original nucleon energy.  The mean free path for this reaction is only
$6$~Mpc. Photons with comparable energy pair-produce electrons and positrons
on the radio background and, likewise, cannot reach Earth from beyond 10 to 40
Mpc~\cite{40Mpc} (although the photon energy-attenuation length is uncertain,
due to the uncertainties in the spectrum of the absorbing radio
background). There only few known astrophysical sources within those distances 
that could produce such energetic particles, but they are not
 located along the arrival 
directions of observed cosmic rays.

Intervening sheets of large scale intense extra galactic magnetic fields
(EGMF), with intensities $B \sim 0.1 -1\times 10^{-6}$~G, could provide
sufficient angular deflection for protons to explain the lack of observed
sources in the directions of arrival of UHECR. However, recent realistic
simulations of the expected large scale EGMF, show that strong deflections
could only occur when particles cross galaxy clusters. Except in the regions
close to the Virgo, Perseus and Coma clusters the obtained magnetic fields are
not larger than 3$\times 10^{-11}$~G~\cite{dolag2004} and the deflections
expected are not important (however  see Ref.~\cite{Sigl:2004yk}).

Whether particles can be emitted with the necessary energies by astrophysical
accelerators, such as active galactic nuclei, jets or extended lobes of 
radio galaxies, 
or even extended object such as colliding galaxies and
clusters of galaxies, is still an open question. The size and possible
magnetic and electric fields of these astrophysical sites make it plausible
for them to produce UHECR at most up to energies of $10^{21}$~eV. Larger emission
energies would require a reconsideration of possible acceleration models or
sites.

Heavy nuclei are an interesting possibility for UHECR primaries, since they
could be produced at the sources with larger maximum energies (proportional to
their charges) and would more easily be deflected by intervening magnetic
fields. On the other hand, both AGASA and HiRes data favor 
a dominance of light hadrons, consisting with
being all protons, in the composition of UHECR above 10$^{19}$~eV. However, we
should keep in mind that the inferred composition is sensitive to the
interaction models used. Assuming a proton plus iron composition, HiRes Stereo
data show a constant or slowly changing composition of 80\% protons and 20\%
iron nuclei between 10$^{18.0}$~eV and 10$^{19.4}$~eV. This is consistent with
the change in composition from heavy to light in the 10$^{17}$~eV to
10$^{18}$~eV range found by HiRes Prototype~\cite{hires_composition}.  HiRes
monocular data show a 90\% proton composition between 10$^{17.6}$~eV and
10$^{20}$~eV~\cite{hires_composition_fit}. Similar results were found by
AGASA, which produced bounds on the iron fraction (again assuming an iron plus
proton composition) of 14~$(+16, -14)$\% and 30~$(+7, -6)$\% above
10$^{19.0}$~eV and 10$^{19.25}$~eV respectively, and 1 $\sigma$ upper bound
of 66\% above 10$^{19.5}$~eV~\cite{agasa_composition_2}.

 In fact, a galactic component of the UHECR flux, which could be important 
up to energies  10$^{19}$~eV, should consists of heavy nuclei, given the
 lack of
correlation with the galactic plane of events at this energy (outside the
galactic plane, galactic protons would be deflected by a maximum of 15-20$^o$
at this energies~\cite{galactic_magn_field}).   For nuclei the
dominant energy loss process is photodissociation through scattering with the
infra-red background below 10$^{20}$~eV~\cite{puget} and with the CMB 
above $10^{20}$~eV,
and pair creation on the CMB in a small energy interval around 10$^{20}$~eV
(at energies for which the typical CMB photon energy in the rest frame of the
nucleus is above threshold, i.e. above 1 MeV, but below the peak of the giant
resonance, 10-20 MeV)~\cite{epele}. The typical attenuation length in the
energy range 4$\times 10^{19}$ to $1 \times 10^{20}$~eV changes from several
10$^{3}$~Mpc for iron and silicon to be comparable to that of nucleons for
helium~\cite{epele, bertone}. At energies above $1\times10^{20}$~eV, the
 attenuation
length of heavy nuclei decreases and becomes less than 10 Mpc at about $3
\times 10^{20}$~eV for iron, $2 \times 10^{20}$~eV for silicon and $1 \times
10^{20}$~eV for carbon (see for example Fig.~1 of Ref.~\cite{bertone}). In the
realistically low EGMF of Ref.~\cite{dolag2004}, most of the heavy nuclei
with $E>10^{20}$ eV reaching us from more than 
 10 Mpc away with energies above those mentioned
would disintegrate into protons with energy ($1/A$) of the original nucleus
energy, where $A$ is the atomic number (this is 1/56 of the original energy
for iron nuclei).
 Note also that the same photodissociation  
processes  can destroy heavy nuclei near  their sources, if 
the intensity of the infrared
background near the sources is large enough.
 One should not forget 
 that all UHECR above $10^{18}$ eV could be due to extragalactic protons 
\cite{berezinsky2002}.

The GZK cutoff at $4\times 10^{19}$~eV seems not to be present in the data of
the AGASA ground array~\cite{agasa} but it appears in the data of the HiRes
air fluorescence detector~\cite{hires}. In any case, there are events above
the GZK cutoff, even in the HiRes data set, and these remain unexplained since
the local Universe ($\sim 100$~Mpc) is devoid of strong candidate sources in
the direction the events point to, and also of the large magnetic fields which
could deflect the incoming particles significantly.  Due to the limited
statistics and different systematic errors of both experiments the
discrepancy between them is not very significant. However,
the presence or absence of the GZK cutoff remains an open question.  This
controversy will be solved conclusively by the Pierre Auger
Observatory~\cite{Auger}, a hybrid combination of charged particles detectors
and fluorescence telescopes,  perhaps within the next one or two years.

The analysis of the muon content in air showers has been used by AGASA to
reject photon dominance in UHECR above 10$^{19}$~eV~\cite{agasa_composition_1,
agasa_composition_2}. Assuming a composition of protons plus photons, AGASA
quotes upper limits for the photon ratio of 34\%, 59\% and 63\% at
10$^{19}$~eV, 10$^{19.25}$~eV and 10$^{19.5}$~eV respectively at the 95\%
confidence level~\cite{agasa_composition_2}, and even above 10$^{20}$~eV they
find no indication that the events they observe are mostly
photons~\cite{agasa_composition_1}. Also a reanalysis of horizontal showers at
Haverah Park concluded that photons cannot constitute more that 50\% of the
UHECR above 4$\times 10^{19}$~eV~\cite{haverah}.

The GZK process produces pions. From the decay of $\pi^{\pm}$ one obtains
neutrinos.  These ``GZK neutrinos" have been extensively studied, from
1969~\cite{bere} onward 
(see for example~\cite{reviewGZKneutrinos,reviewGZKneutrinos2} and
references therein), and constitute one of the main high energy signals
expected in neutrino telescopes, such as ICECUBE~\cite{ICECUBE} 
ANITA~\cite{ANITA} and  SALSA~\cite{SALSA} 
or space based observatories such as EUSO~\cite{EUSO} and
OWL~\cite{OWL}.  From the decay of $\pi^0$ we obtain photons, ``GZK photons", with about 0.1 of
the original proton energy, which have been known to be a subdominant
component of the UHECR since the work of Wdowczyk {\it et al.} in the early
1970's~\cite{wdowczyk}.
In 1990  it was suggested that if the  extragalactic radio background
and magnetic fields are small ($B< 3 \times 10^{-11}$ G)
GZK photons could dominate over protons and explain the
super-GZK events~\cite{Aharonian1990}. The dependence
of the GZK photon flux on extragalactic magnetic fields was later
studied in Ref.~\cite{SiglOlinto95}. The argument of Ref.~\cite{Aharonian1990} and its dependence on extragalactic magnetic fields
was again discussed~\cite{astro_photons} in connection with the possible
correlation of UHECR arrival directions with BL Lacertae objects~\cite{Tinyakov:2001nr}.
However, to our knowledge, no complete study of the expected
fluxes of  GZK photons was done so far, including their dependence on
the initial proton fluxes, distribution of proton sources and UHECR spectrum,
besides intervening backgrounds.

With the
advent of the Pierre Auger Observatory, we expect to have in the near future
the high statistic data that may allow to study a subdominant component of
UHECR consisting of photons. The GZK photons provide a complementary handle to
GZK neutrinos and other signatures to try to determine the spectrum and
composition of the UHECR. The flux of GZK photons is necessarily correlated
with the flux of GZK neutrinos, although the former is affected by the
radio background and EGMF values which do not affect the latter.

In this paper we  show that if the UHECR are mostly protons,
  depending on the UHECR spectrum assumed,  the slope of
the proton flux, distribution of sources and intervening backgrounds,
 between   $10^{-4}$  and $10^{-2}$ of
 the UHECR  above $10^{19}$ eV  and between  $10^{-5}$  and $0.6$ of
 the UHECR  above $10^{20}$ eV are GZK photons, the range being much higher 
 for the AGASA spectrum than for the HiRes spectrum (see Fig.~\ref{F14}).
Detection of these photons would open the way for  UHECR photon astronomy.

Detection of a larger photon flux than expected for GZK photons 
would imply the emission of photons at the
source or new physics. New physics is involved in Top-Down models,
produced as an alternative to acceleration models to explain the
 origin of the highest energy cosmic rays. All of the Top-Down models
 predict photon dominance at the highest energies.
Here, we estimate  the minimum photon
fraction Top-Down models predict, not only assuming the AGASA spectrum
which these models were originally proposed to explain,  but
also assuming the HiRes spectrum.  We  show that at high energy, close to
10$^{20}$~eV,
the maximum expected  flux of GZK photons is comparable to (for the
 AGASA spectrum)
or much smaller than (for the HiRes spectrum)  the minimum flux of
 photons predicted
by Top-Down models which fit the AGASA or the HiRes data (see Fig.~\ref{F14}).
 We  try to minimize the photon
ratio  predicted by Top-Down models by assuming that these models explain only
the highest energy UHECR (if they do not explain even those events,
 the models are irrelevant
for UHECR). We show that the photon ratio
at energies close to $10^{20}$ eV is a crucial test for Top-Down models, since 
it is always higher than  about 0.5, independently of the UHECR spectrum assumed.

We also show that, surprisingly, in a limited energy range above
$10^{20}$~eV, GZK photons could become the dominant component of the UHECR
(assuming that protons could be accelerated at  the source to energies as
large as 10$^{22}$~eV). This result allows us to fit the AGASA data
with an original flux of only nucleons. This  seems to
 contradict previous
estimates of the GZK photon flux in which this flux is always subdominant,
  however one needs to take into account
the assumed initial spectrum and intervening radio background and 
magnetic fields (for example
in Ref.~\cite{reviewGZKneutrinos}  an average EGMF of
10$^{-9}$~G is assumed,  much larger than the fields found later in 
Ref.~\cite{dolag2004}).

In section II,  we explain our calculations and show the dependence of the 
GZK photon flux on the assumed 
initial proton flux and intervening background parameters. 
In section II we only normalize the fluxes 
we show to one point of the AGASA or HiRes spectrum, but we do
not fit these spectra (which we do in the following section).
In section III, we
estimate  the maximum and 
minimum GZK photon fractions expected either with the AGASA spectrum 
or with the HiRes spectrum. In section IV  we estimate the minimum
 photon fractions predicted by several
by Top-Down models and compare them with the maximum GZK photon fraction
 we find in section III.
We also include a comparison with experimental upper bounds on photon
 fractions.

\section{II. The GZK photon flux}

We use a numerical code developed in Ref.~\cite{kks1999} 
to compute the flux of GZK
photons produced by an homogeneous distribution of sources emitting originally
only protons.  It calculates the propagation of protons and photons using the
standard dominant processes, explained for example in Ref.~\cite{reviews1}).
For protons, it takes into account single and multiple pion production, and
$e^{\pm}$ pair creation. For photons, it includes $e^{\pm}$ pair production,
inverse Compton scattering and double $e^{\pm}$ pair production processes.
For electrons and positrons, it takes into account Compton scattering, triple
pair production and synchrotron energy loss on extra galactic magnetic fields
(EGMF).  The propagation of protons and photons is calculated
self-consistently. Namely, secondary (and higher generation) particles arising
in all reactions are propagated alongside with the primaries.  UHE protons and
photons lose their energy in interactions with the electro-magnetic
background, which consist of CMB, radio, infra-red and optical components, as
well as EGMF.  Protons are sensitive essentially to the CMB only, while for
photons all components of the electro-magnetic background are important.  
Notice that the radio background is not yet well known and that our
 conclusions 
depend strongly on the background assumed. We
include three models for the radio background: the  background based on
estimates by Clark {\it et al.}~\cite{clark} and the two models of
Protheroe and Biermann~\cite{PB}, both predicting larger background than the
first.  To calculate the infra-red/optical background we used the same
approach as in Ref.~\cite{Primack:2000xp}.
 In any  event,  the infra-red/optical background 
is not important  for the production and absorption of GZK photons at
 high energies.
This background is important to  transport the energy of secondary photons
 in the cascade process 
from the  0.1 - 100 TeV energy range  to the  0.1-100 GeV energy range 
observed by EGRET. 
The resulting flux in the EGRET energy range is not sensitive to details
of the infra-red/optical background models.

For the EGMF only the upper bound is established observationally, 
$B \lsim 10^{-9}
\,({\rm Mpc}/l_c)^{1/2}$~G~\cite{FR} (where $l_c$ is the
reversal scale of the magnetic field in comoving coordinates).  
It is believed that the magnetic fields in  clusters
 can be generated from a primordial ``seed'' if the later has
comoving magnitude $B \sim 10^{-12}$~G \cite{Dolag:2002,dolag2004}.  The
evolution of EGMF together with the large scale structure of the Universe has
been simulated recently by two groups using independent numerical procedures
\cite{Sigl:2004yk,dolag2004}. Magnetic field strengths significantly larger
than 10$^{-10}$~G were found only within large clusters of galaxies.  In our
simulations we vary the magnetic field strength in the range $B = 10^{-12} -
10^{- 9}$~G, assuming an 
unstructured field along the propagation path.

Notice that we assume that protons are produced at the source 
but the results at
 high energies would be  identical if we had taken neutrons instead. 
 The interactions of neutrons and protons with the intervening
backgrounds are identical and when  a neutron decays practically all
 of its energy goes to the final  proton (while the electron and neutrino are
produced with energies 10$^{17}$~eV or lower).

The resulting GZK photon flux depends on several
astrophysical parameters.  These parametrize the initial proton flux, the
distribution of sources, the radio background and the EGMF. In this section, to explore the
flux dependence on a given parameter, we fix all the other unknown parameters
to the  following values. For the radio background we take the lower
estimate of Protheroe and Biermann~\cite{PB}, which is intermediate between the
other two we consider. For the EGMF we take $B=10^{-11}$G which is the average
value found in Ref.~\cite{dolag2004}.  For the source distribution, we take a
uniform continuous distribution of sources with zero minimum distance to us
(i.e. a minimum distance comparable to the interaction length).
For the maximum energy of the injected protons we use $E_{\rm max}=10^{22}$~eV,
which is considered already a generous upper limit for acceleration in
astrophysical models \cite{hillas}.

With respect to cosmological parameters, we take  the Hubble constant 
$H=70$~km~s$^{-1}$~Mpc$^{-1}$, a dark energy density (in units of the critical
density) $\Omega_{\Lambda}= 0.7$ and a dark matter density  $\Omega_{\rm
m}=0.3$. We assume the sources extend to a maximum redshift $z_{\rm max}=2$
(although any $z_{\rm max} > 1$ gives the same results at the high energies we
consider) and disregard a possible evolution of the sources with redshift.

\subsection{A. Dependence of the GZK photon flux on the 
initial proton spectrum}

We parametrize the initial proton flux for any source with the following power
law function,
\begin{equation}
F(E) = f~ \frac{1}{E^\alpha}~ \theta(E_{\rm max} -E)~.
\label{proton_flux}
\end{equation}
The power law index $\alpha$ and maximum energy $E_{\rm max}$ are considered free
parameters. The amplitude $f$ is fixed by normalizing the final proton
flux from all sources to the observed flux of UHECR, which we take to be 
either the AGASA flux or the HiRes flux. 

We are
 implicitly assuming that the sources are astrophysical, since these are the
 only ones which could produce solely protons (or neutrons) as UHECR
 primaries. Astrophysical acceleration mechanisms often result in $\alpha
 \gsim 2$~\cite{AS2}, however, harder spectra, $\alpha \lsim 1.5$ are also
 possible, see e.g. Ref.~ \cite{AS1.5}. The 
resulting spectrum may differ from a power-law,
 it may even have  a peak at high energies~\cite{peaks}. 
 AGN cores could  accelerate protons with induced
electric fields, similarly to what happens in a
linear accelerator. This mechanism would produce  an 
almost monoenergetic proton flux, with energies as high as $10^{20}$~eV or
higher~\cite{mono}. Here, we will consider the
 power law index to be in the  range  $1 \le \alpha \le 2.7$.
\begin{figure}[ht]
\includegraphics[height=0.48\textwidth,clip=true,angle=270]{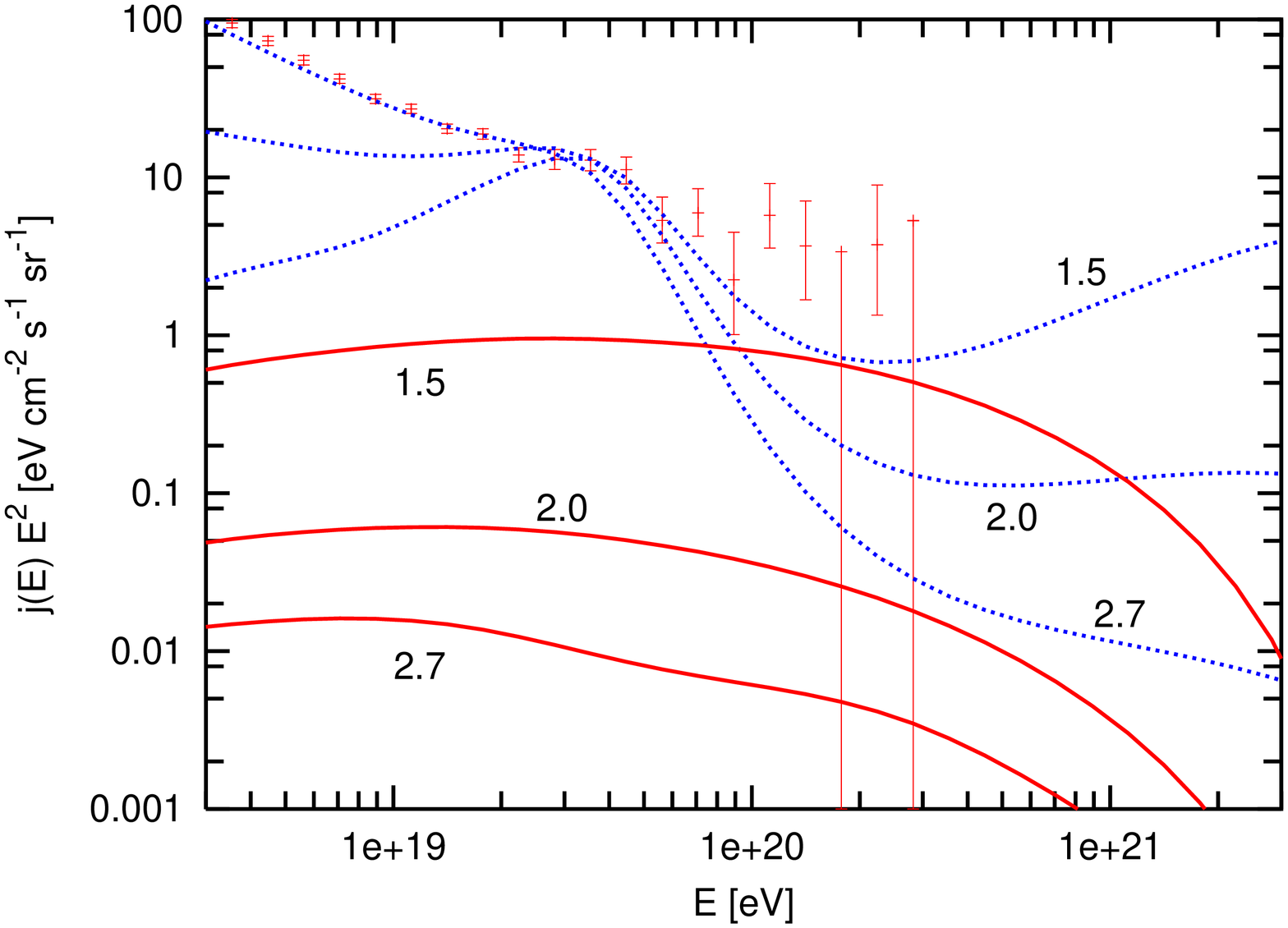}
\includegraphics[height=0.48\textwidth,clip=true,angle=270]{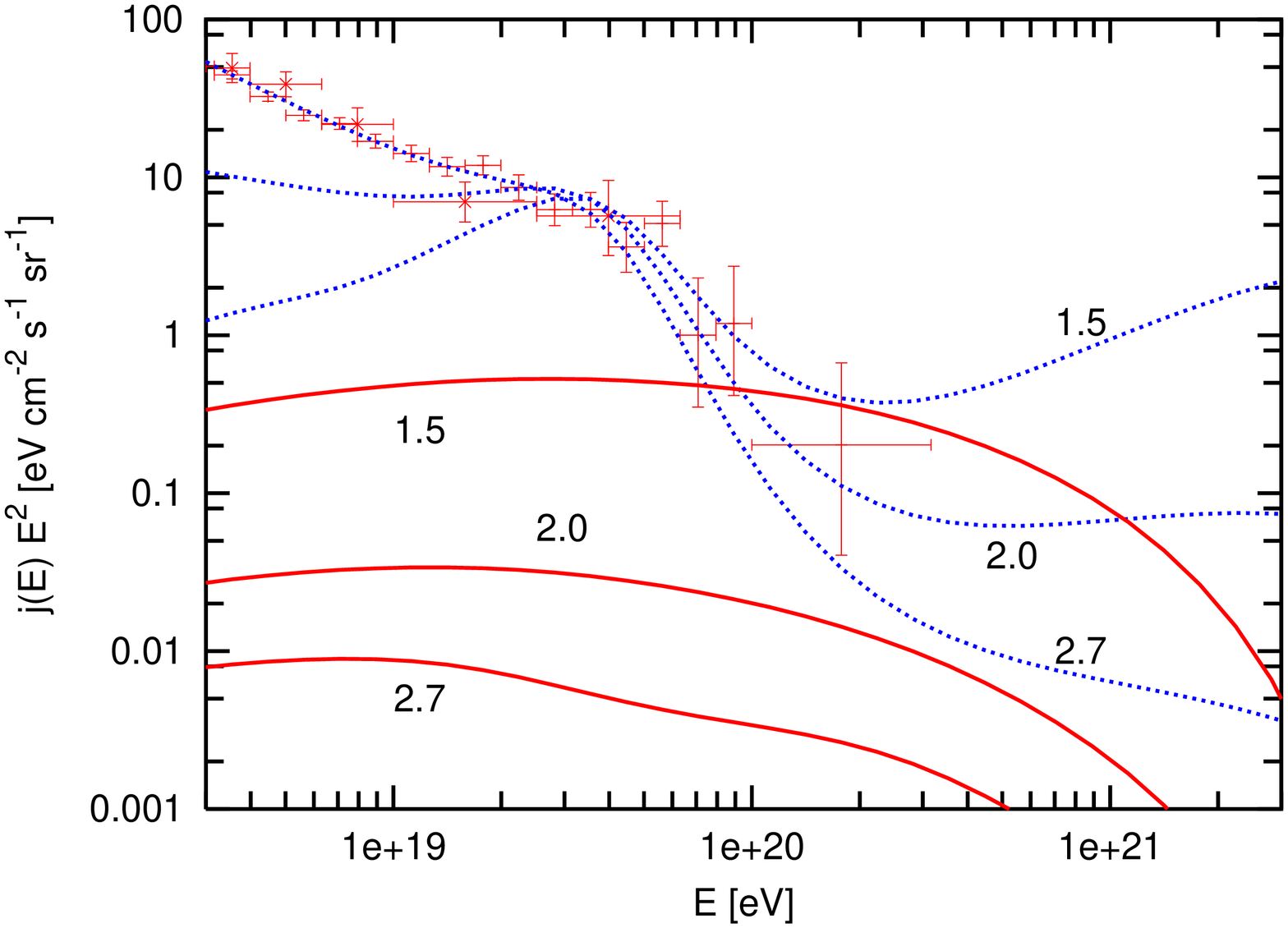}
\caption[...]{UHECR proton flux (blue dotted lines) normalized to the AGASA
data (upper panel) and HiRes data (lower panel) at $3\times~10^{19}$~eV and
GZK photon flux (red solid lines) for three values of the power law index
$\alpha$ of the initial proton flux at the source: $\alpha=$ 1.5, 2.0 and 2.7
(from highest to lowest fluxes at high energy).}
\label{F1}
\end{figure}

Fig.~\ref{F1} shows the GZK photon flux for three values of the power law
index in Eq.~(\ref{proton_flux}), $\alpha=$1.5, 2 and 2.7. Dotted (solid) lines
correspond to the resulting flux of protons (GZK photons) from all sources.  A
proton spectrum $\sim 1/E^{2.7}$ does not require an extra contribution
to fit the UHECR data, except at very low energies 
$E<10^{18}$ eV outside the range we study~\cite{Berezinsky:2002vt}.
 For $\alpha \le 2$ an extra low energy component
(LEC) is required to fit the UHECR data at $E<10^{19}$~eV.
 The LEC may be  a galactic contribution
(for example of iron nuclei, to explain the lack of correlation of arrival
directions with the galactic plane), which can be parametrized as 
 power law with an exponential cutoff as in Eq.(2) below.  In this
case, the ``ankle" is the energy where the extragalactic protons start to
dominate over the LEC.  The LEC could also be  due to
 a population of extragalactic
lower energy proton sources.
This latter contribution can be parametrized  again as in
Eq.~(\ref{proton_flux}), but with parameters different than those of the
extragalactic proton population which dominates above the GZK energy. 

Notice that in this section we just normalize the total flux to a
point of the AGASA or HiRes spectrum, but we do not fit 
these spectra, so we do not
add a LEC, even if it would be needed. We do fit the UHECR 
spectrum in the next section.

As seen in Fig.~\ref{F1}, the flux of super-GZK protons and, consequently,
 the flux of  the GZK photons they generate, depend strongly on the power law
index of the initial proton flux: they are lower for large values of $\alpha$.
In the most conservative case of a proton flux $\sim 1/E^{2.7}$ the GZK photon
flux at $E=10^{19}$~eV is as small as $0.03\%$ and it increases  to a few $\%$
at $E=2\times10^{20}$~eV. This means that even with the final
statistics of Auger it might be difficult to detect the GZK photons in this
case.  On the other hand, in the optimistic case of an injection spectrum
$\sim1/E^{1.5}$, the GZK photons can contribute as much as 1-3\% at
$E=10^{19}$ eV and  50\% or more at $E=10^{20}$ eV.
\begin{figure}[ht]
\includegraphics[height=0.48\textwidth,clip=true,angle=270]{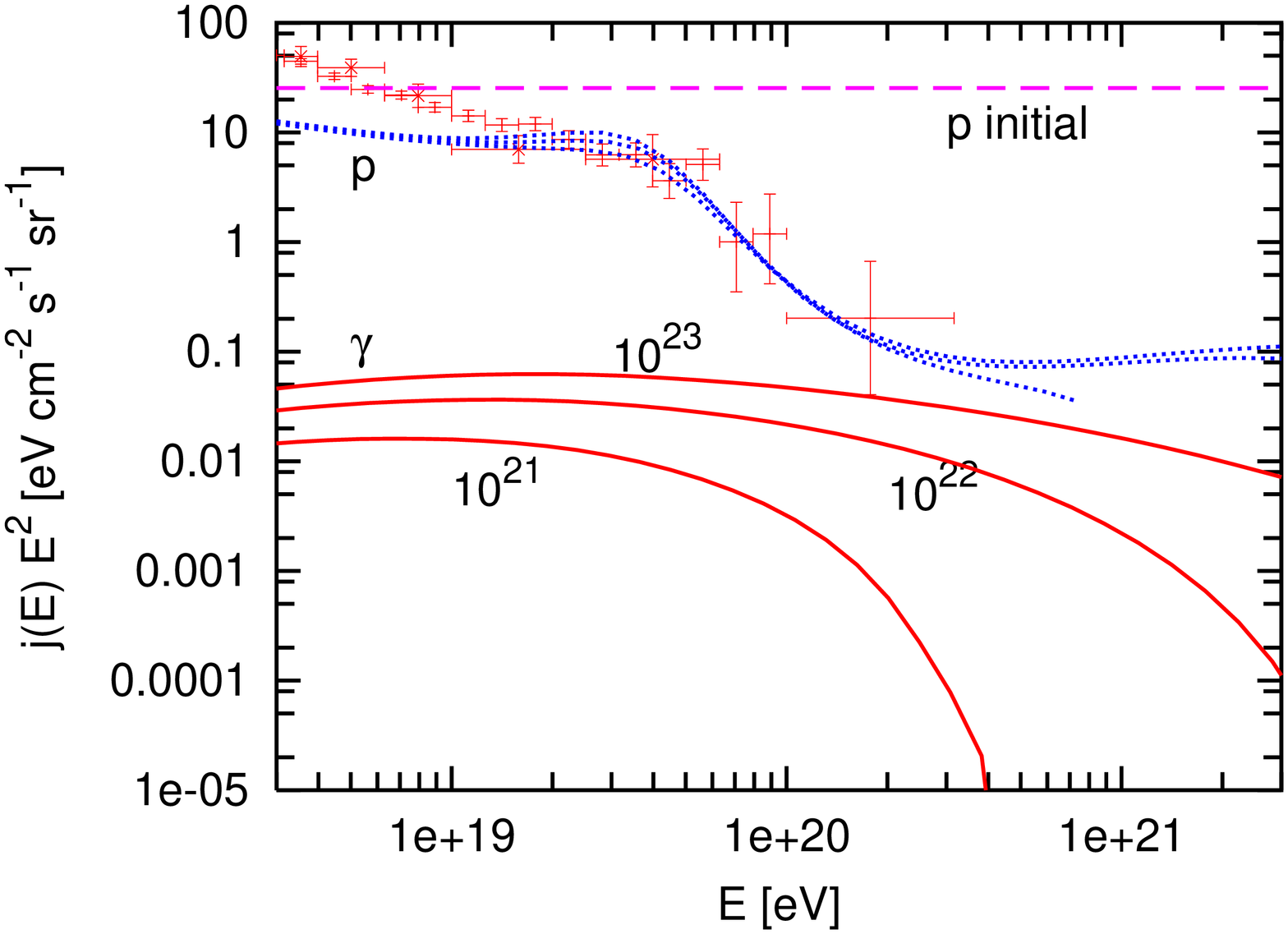}
\includegraphics[height=0.48\textwidth,clip=true,angle=270]{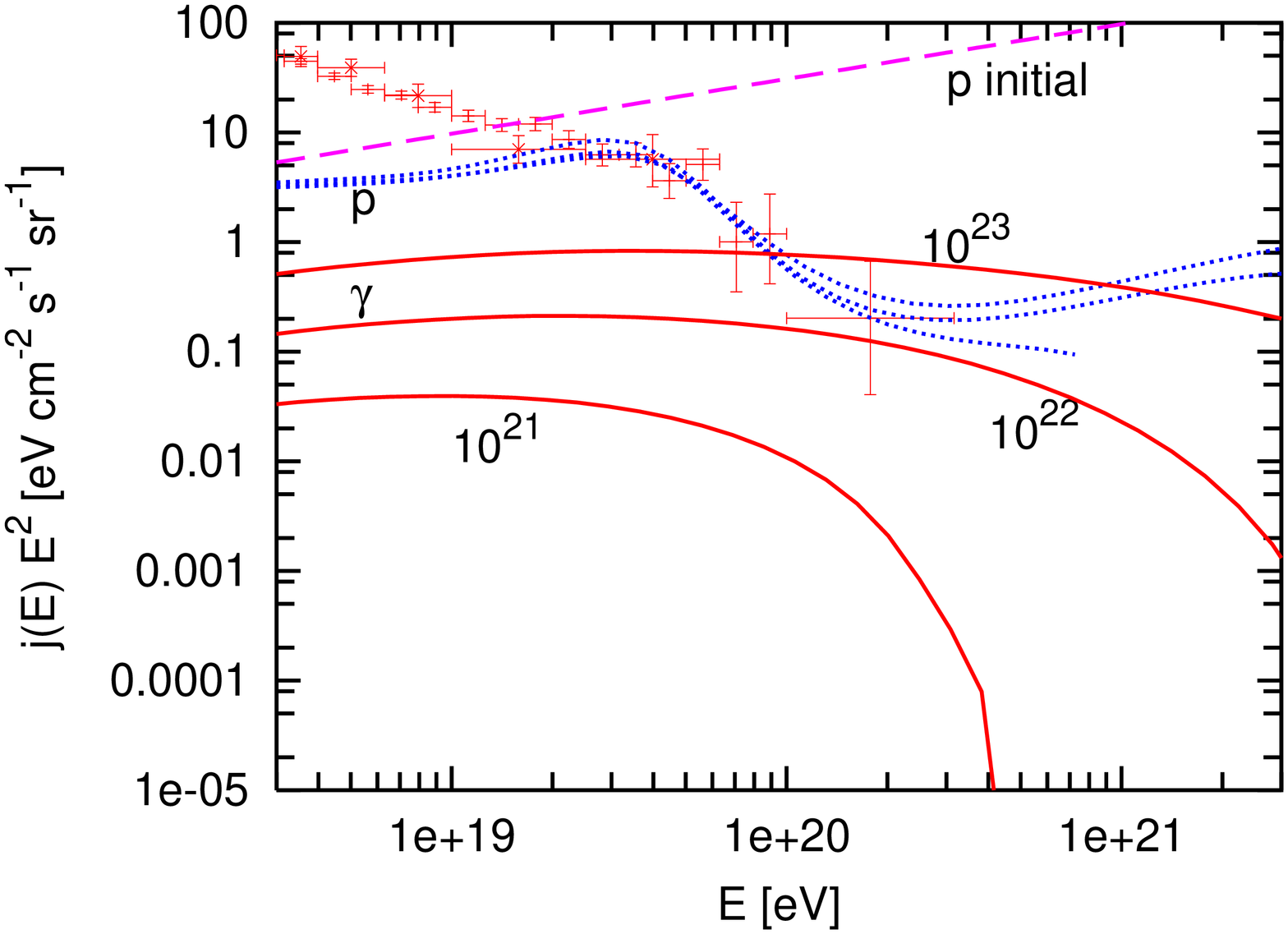}
\caption[...]{UHECR proton flux (dotted lines) normalized to the HiRes
 data at about $3\times~10^{19}$~eV and GZK photon flux (solid lines) for
three values of the maximal energy of the initial proton spectrum:
$E_{\rm max}=$ $10^{23}$~eV, $10^{22}$~eV and $10^{21}$~eV
 (from highest to lowest fluxes at high energy). The initial proton flux is
(a) $\sim 1/E^2$ (upper panel) and (b) $\sim 1/E^{1.5}$ (lower panel)}.
\label{F2}
\end{figure}

Let us note here, that most of the energy produced in the form of GZK
photons cascades down in energy to below the pair production 
threshold for photons on
the CMB. For $\alpha<2$ the diffuse extragalactic gamma-ray flux measured by
EGRET~\cite{EGRET} at GeV energies imposes a constraint on the GZK photon
flux at high energies, which we have taken into account.

The dependence of the GZK photon flux on the maximum energy $E_{\rm max}$ of
the initial proton flux (see Eq.~(\ref{proton_flux})) is shown in
Fig.~\ref{F2}, for $E_{\rm max}=$ $10^{21}$~eV,
 $10^{22}$~eV and $10^{23}$~eV.  We do not show here the case 
of $\alpha ={2.7}$ because for
such a steeply falling proton flux the GZK photon flux practically does not
depend on $E_{\rm max}$. Fig.~\ref{F2}a shows the case of $\alpha = 2$ and
Fig.~\ref{F2}b that of $\alpha ={1.5}$. These figures clearly show that the
dependence on $E_{\rm max}$ is more significant for smaller values of the
power law index $\alpha$.  Note that not only the photon flux,
 but also the final UHECR proton flux above the GZK
 cutoff depends strongly on $E_{\rm max}$.

For relatively small values of the maximal energy,
 such as $E_{\rm max}=10^{21}$~eV, the
GZK photon flux is very small
for any power law index
$\alpha$ (see the lowest curves in Fig.~\ref{F2}a and Fig.~\ref{F2}b).  For
larger values of the maximal energy, such as $E_{\rm max}=10^{22}$ eV and
$E_{\rm max}=10^{23}$~eV, the GZK photon flux increases considerably
for $\alpha \leq 2$.

\subsection{B. Dependence of the GZK photon flux on the minimal 
distance to the sources}

Quite often in the literature the minimal distance to the sources is taken to
be negligible (i.e. comparable to the interaction length). This is one of the
cases we consider as well. However we take also 50 Mpc, as inferred from the
small-scale clustering of events seen in the AGASA data~\cite{AGASA_clusters},
and 100 Mpc, to show how the fluxes
diminish with this assumption (what proves that most photons
 come from smaller distances).
Contrary to AGASA, HiRes does not see a clustering component in its own
data~\cite{HiRes_clusters}. The combined dataset shows that clustering still
exists, but it is not as significant as in the data of AGASA
alone~\cite{agasa_hires}.  Note, that the non-observation of clustering in the
HiRes stereo data does not contradict the result of AGASA, because of the
small number of events in the sample~\cite{agasa_hires_ok}.

\begin{figure}[ht]
\includegraphics[height=0.48\textwidth,clip=true,angle=270]{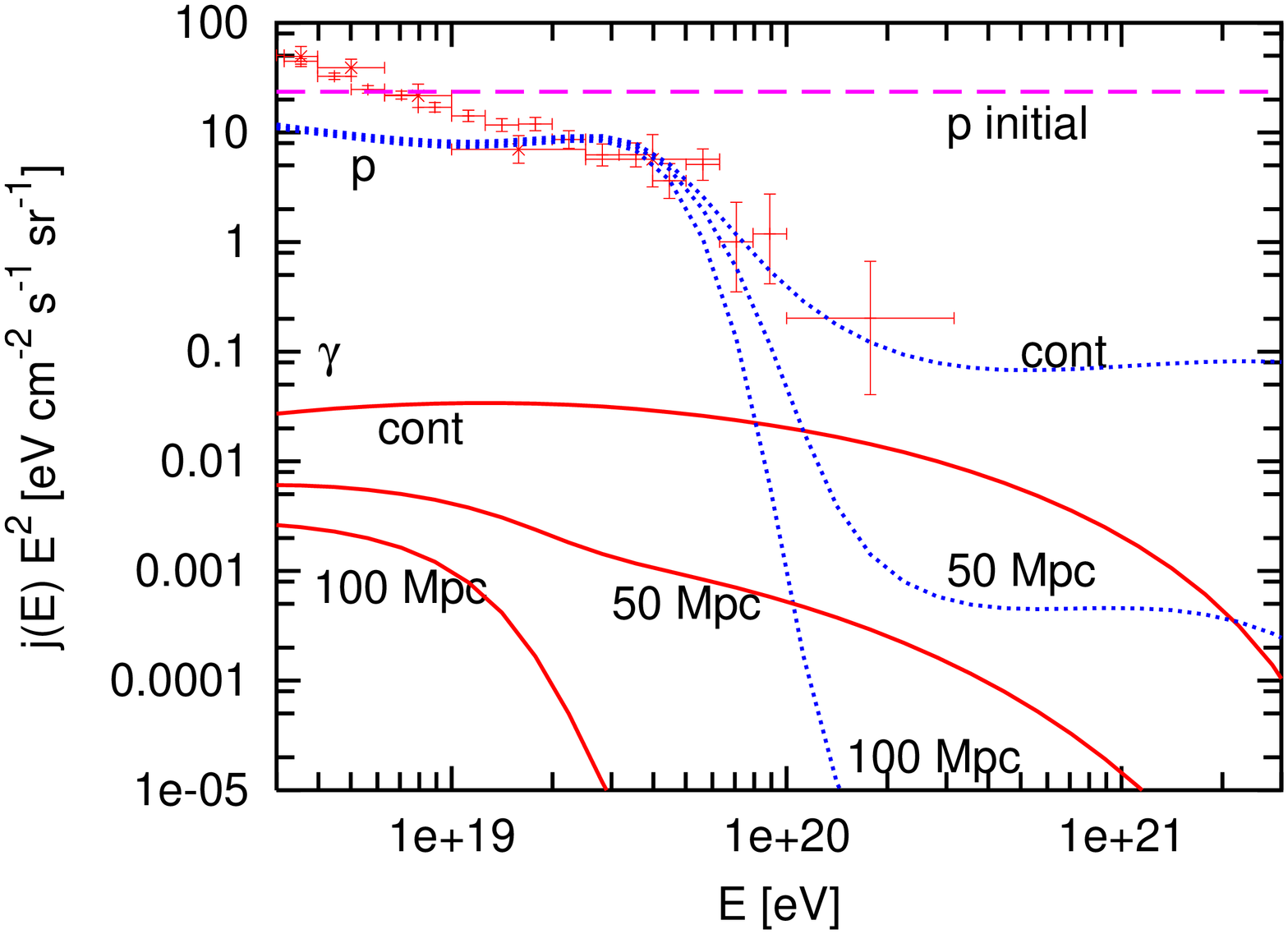}
\includegraphics[height=0.48\textwidth,clip=true,angle=270]{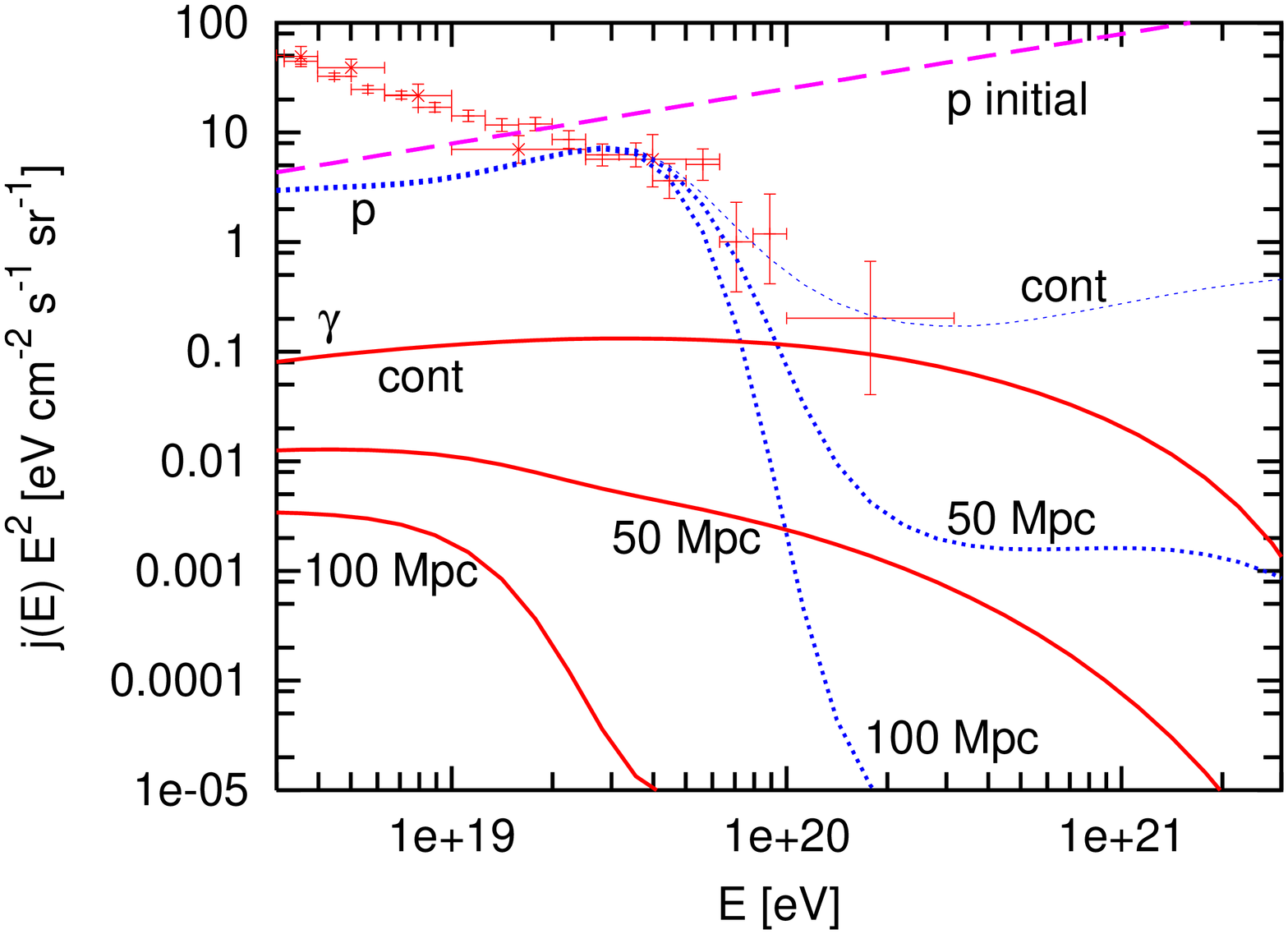}
\caption[...]{UHECR proton flux (dotted lines) normalized to
 the HiRes data at 4$\times 10^{19}$~eV and GZK photon flux (solid lines) for
three values of the minimal distance to the sources: 0, 50 Mpc and 100 Mpc,
 (from highest to lowest fluxes at high energy) for an initial 
proton flux (a)  $1/E^{2}$ (upper panel) and (b) $1/E^{1.5}$ (lower panel).}
\label{F3}
\end{figure}

Assuming proton primaries and a small EGMF (following Ref.~\cite{dolag2004}),
it is possible to infer the density of the 
sources~\cite{sources, agasa_hires_ok} 
from the clustering component of UHECR. AGASA data alone
suggest a source density of $2\times 10^{-5}$~Mpc$^{-3}$, which makes
plausible the existence of one source within 50 Mpc of us. However, the HiRes
negative result on clustering requires a larger density of sources and, as a
result, a smaller distance to the nearest one of them. Larger 
values of the EGMF
(as found in Ref.~\cite{Sigl:2004yk}) and/or some fraction of iron
in the UHECR have the effect of reducing the required number of sources and,
consequently, increasing the expected distance to the nearest one.

Fig.~\ref{F3} shows the dependence of the UHECR proton and GZK photon fluxes on
the assumed minimal distance to sources for an initial
 proton flux $\sim 1/E^{2}$ in Fig.~\ref{F3}a and $\sim 1/E^{1.5}$ 
 in Fig.~\ref{F3}b.  The highest, intermediate and lowest fluxes
correspond to a minimal distance of 0 
(labeled cont. for continuous), 50 and 100 Mpc, respectively. 
Notice that in all the examples presented in Fig.~\ref{F3} 
the protons dominate the flux (i.e. the total flux is 
practically the proton flux).
Only the highest proton fluxes shown in Fig.~\ref{F3} (with negligible minimal distance)
fit well the HiRes data.  The intermediate and lowest proton fluxes have a
sharp cutoff and do not fit the HiRes data any longer.
 We clearly see in the figures that most of the GZK photons with
energies $E>10^{19}$ eV should come from nearby sources within 100 Mpc (see
the impressive reduction in flux if we only take sources more than 100 Mpc
away).

\subsection{C. Dependence of the GZK photon flux on the radio background}

\begin{figure}[ht]
\includegraphics[height=0.48\textwidth,clip=true,angle=270]
{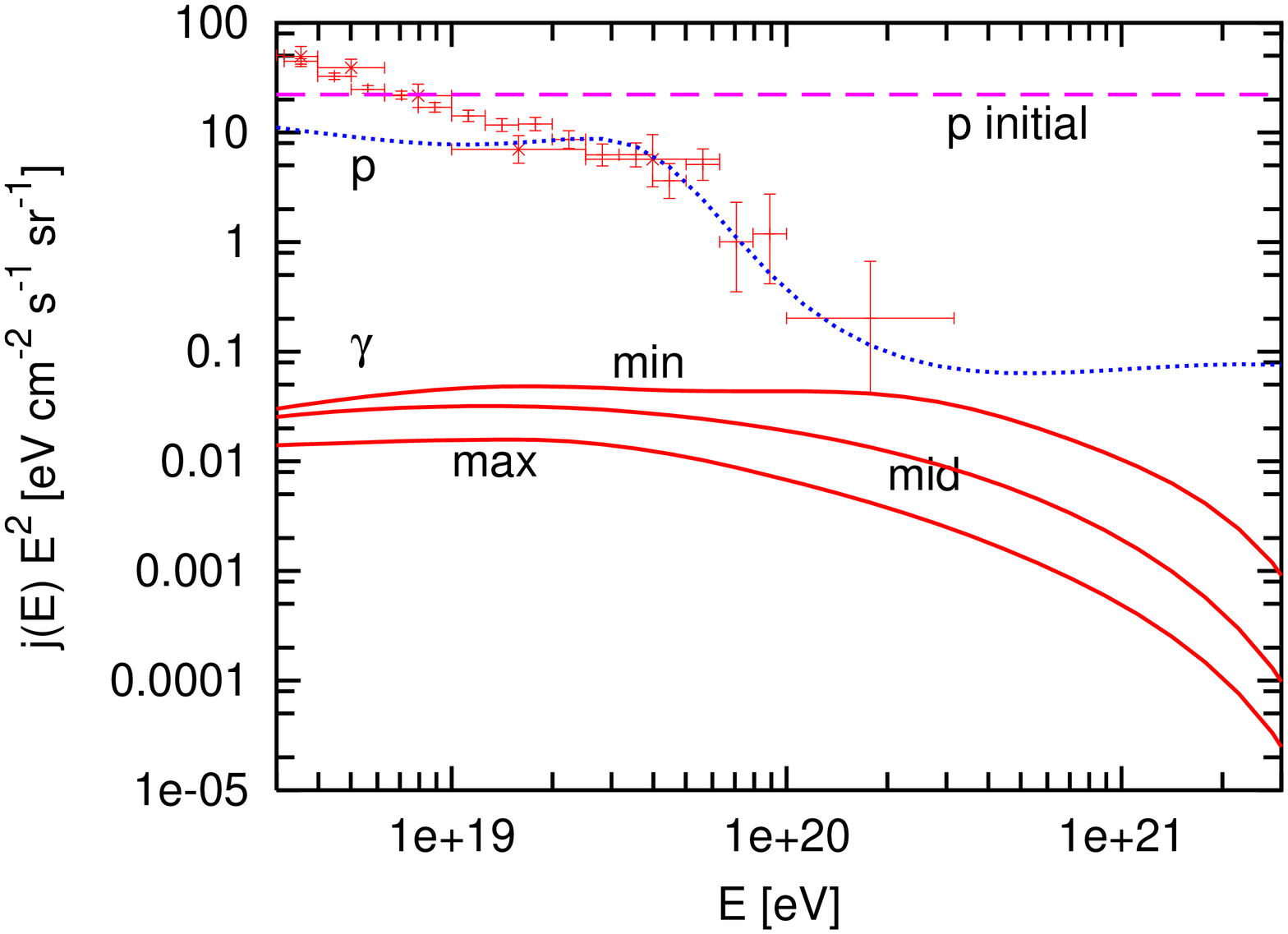}
\includegraphics[height=0.48\textwidth,clip=true,angle=270]
{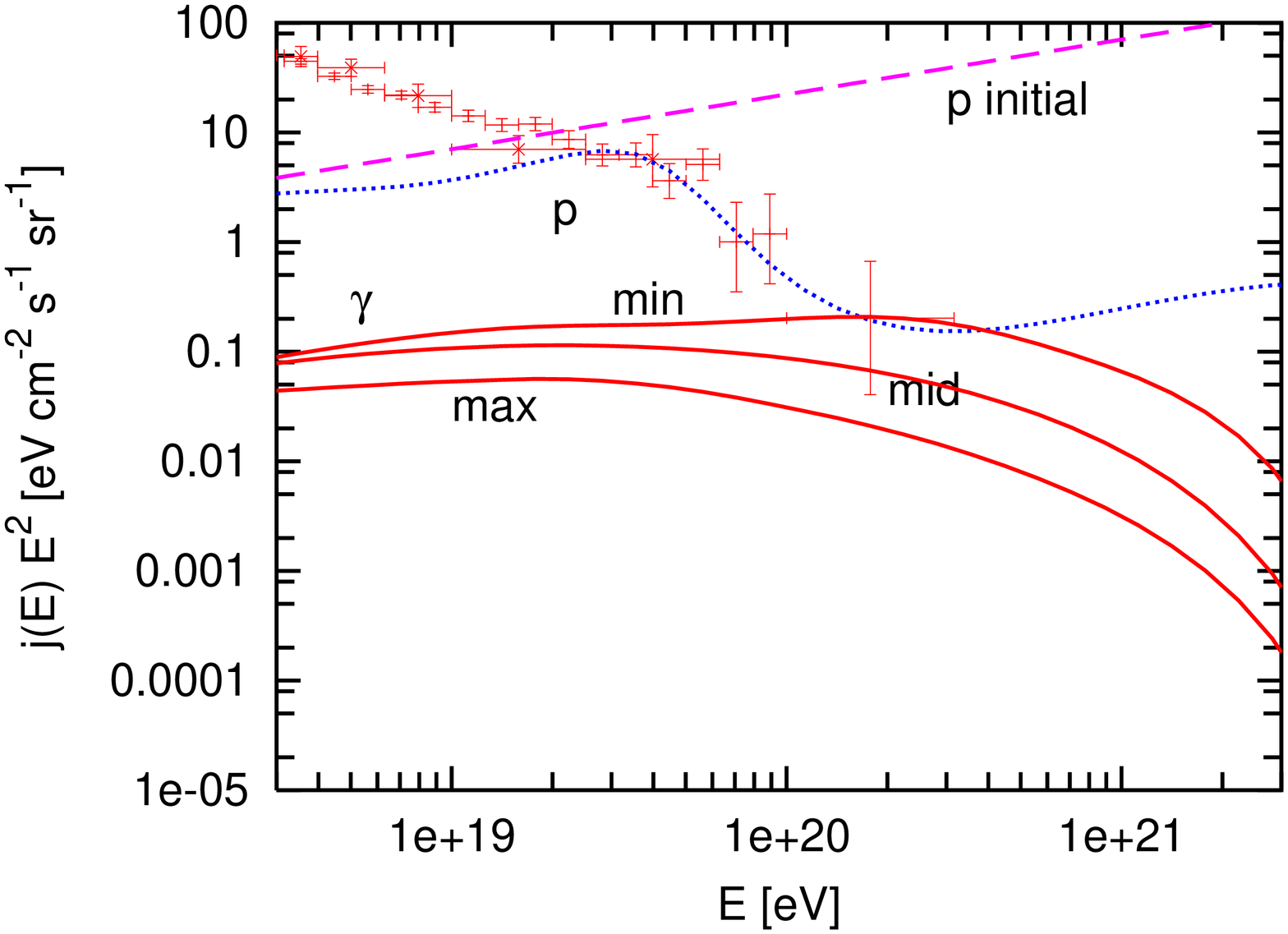}
\caption[...]{UHECR proton flux (dotted lines) normalized
 to the HiRes data at  4$\times10^{19}$~eV and GZK photon
 flux (solid lines) for the
three estimates of radio background considered in this paper. The initial proton 
spectrum in (a) is $\sim 1/E^2$ (upper panel) and in (b) is
$ \sim 1/E^{1.5}$ (lower panel).}
\label{F4}
\end{figure}
The main source of  energy loss of photons with $E>10^{19}$~eV is pair
production on the radio background (while at lower energies pair production
on the CMB is more important).  Fig.~\ref{F4} shows  GZK photon fluxes
for the three different estimates of the radio background we consider:
 the minimal background,
of Clark et al.~\cite{clark}, and the two estimates  of Protheroe
 and Biermann~\cite{PB}, both
larger than the first one. In Fig.~\ref{F4}a the injected proton spectrum $\sim
1/E^2$ and in Fig.~\ref{F4}b it is $\sim 1/E^{1.5}$.
These figures show that (for the EGMF assumed,  $B=10^{-11}$~G 
 as mentioned above)
 the GZK photon flux depends only mildly on the radio
background  at energies below $E<10^{20}$ eV, where we find a factor 2-3 of
difference between the highest flux (with the lowest radio background from
Ref.~\cite{clark}) and the lowest flux (with the highest background 
of Ref.~\cite{PB}).
However,  at energies above $E>10^{20}$ eV, the differences increase, reaching
one order of magnitude or more.  This behavior is due to the different shapes
of the assumed radio spectra.  As we see next, larger EGMF, $B >10^{-10}$~G,
increase the GZK photon absorption considerably at  
$E<10^{20}$ eV, but not close to 
 $E \simeq 10^{20}$ eV and above.
\begin{figure}[ht]
\includegraphics[height=0.48\textwidth,clip=true,angle=270]{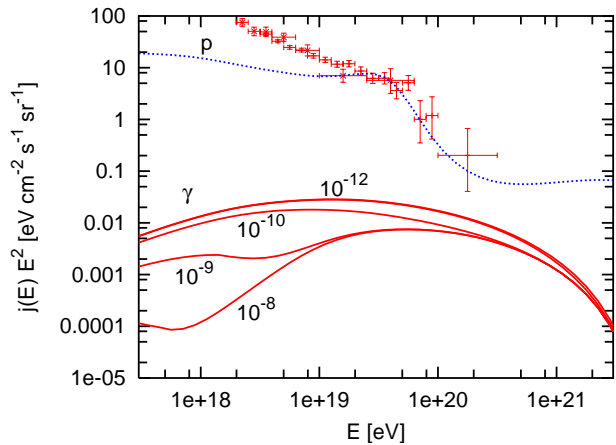}
\caption[...]{UHECR proton flux (dotted lines) normalized 
to the HiRes data at  $3\times10^{19}$~eV and GZK photon flux (solid lines) for
four values  of the average EGMF, $10^{-12}$~G, $10^{-10}$~G  $10^{-9}$~G
and $10^{-8}$~G (from highest to lowest fluxes), for a 
 proton flux  $\sim 1/E^2$.}
\label{F5}
\end{figure}

\subsection{D. Dependence of the GZK photon flux  on EGMF}

The spacial structure, amplitude and correlation length of the EGMF outside
clusters of galaxies are unknown. 
The existing models of the EGMF attempt to
evolve these fields together with the large scale structure of the Universe,
starting from certain (primordial) seed values. In these models, the EGMF in
the voids are close to the comoving value of the primordial field, while the
EGMF in clusters of galaxies and filaments are amplified.  Constrained
simulations of the ``local'' Universe 
(within 100 Mpc from Earth)~\cite{dolag2004},
 in which the magnetic field is normalized to the
values observed within clusters, yield an average
 $B_{\rm EGMF}=(10^{-11} - 10^{-12})$~G in
voids. Fig.~\ref{F5} shows
that for $B_{\rm EGMF} < 10^{-10}$~G, the resulting GZK photon flux changes
very little with $B$, but it decreases considerably at low energies for
 $B_{\rm EGMF} \gsim 10^{-9}$~G.
In Fig.~\ref{F5} an initial proton flux $\sim 1/E^2$ was
assumed and sources were integrated from zero distance.  Assuming a minimum
distance of 50~Mpc to the nearest sources (case not shown in the figures),
 the GZK photon fluxes differ at
most by a factor of 3 when  the EGMF magnitude is varied in 
the range $B < 10^{-10}$~G.

Fig.~\ref{F5} is the only place in this paper where we used 
$B_{\rm EGMF} = 10^{-8}$~G,
and this is
just to show how the photon  flux is affected by large $B$ fields.
 For EGMF $\sim 10^{-8}$~G or larger, the photon energy
is lost into synchrotron radiation as soon as the
UHE photon pair produces, even for energies $E < 10^{19}$~eV.
 Thus the shape of the
spectrum follows the energy dependence of the 
photon pair production interaction length (which is
dominated by the interaction with  the CMB below $10^{19}$~eV 
and with the radio background above this energy).  For
smaller magnetic field strengths, the length of synchrotron energy loss
increases and, at low energies, several steps of pair production
and inverse Compton decay happen.
 For large enough energies, the
synchrotron radiation length is smaller than the interaction length
for all the EGMF values considered (i.e. even as small as 
$B \geq 10^{-12}$~G) , so the
the photon energy
is lost into synchrotron radiation as soon the photon pair produces.
Thus, only the photons which do not interact with the radio background
can reach us and  the spectra
for all values of the EGMF converge.

Our results  in Fig.~\ref{F5}  for  $B_{\rm EGMF} \leq 10^{-9}$~G
are similar to  those in
Fig. 3 of Ref.~\cite{SiglOlinto95}. In particular,  both figures show that the
GZK flux does not depend strongly on the magnetic field for 
 $B_{\rm EGMF} < 10^{-10}$~G, 
and that for larger fields there is a suppression 
of the photon flux at energies $E<10^{19}$ eV (due to pair production
on the CMB followed by  synchrotron energy loss).

\subsection{E. Summary of  the GZK photon flux dependence on
 different parameters }

Figs.~\ref{F4} and \ref{F5} show that given a particular
UHECR proton flux the uncertainty in the resulting  GZK photon
 flux due to  our ignorance of the intervening backgrounds 
(minimum  to  maximum estimates of the radio background and 
EGMF from 10$^{-11}$~G, 
which is equivalent to zero, to $10^{-9}$~G) is within about
 one order of magnitude.

Figs.~\ref{F1} to \ref{F3}  show much larger changes in the  GZK photon flux
when  the parameters
 defining the
UHECR proton flux, i.e. the power law index $\alpha$, 
 maximum energy $E_{\rm max}$,
and minimal distance to the sources, are varied. However, once the
particular UHECR spectrum is fixed, these uncertainties due to the 
extragalactic proton model decrease and become comparable with
those due to our ignorance of the intervening background. In the next section,
 Figs.~\ref{F7} and \ref{F8}  show that
a particular proton dominated observed flux, the
 HiRes spectrum in this case, can be fitted with very different extragalactic
 proton fluxes, whose corresponding
GZK photon fluxes differ by about one order of magnitude, for a given fixed
background. In fact,
 the difference between the two photon lines in Fig.~\ref{F7} shows the 
uncertainty in the GZK photon flux due to intervening background (about one
 order of magnitude), given a particular extragalactic proton flux, 
while the difference between the lower photon line of
 Fig.~\ref{F7} and the lower photon line of  Fig.~\ref{F8} (both computed
 with the same background, i.e. maximum radio background
 and EGMF $B=10^{-9}$~G) shows the uncertainly due to the UHECR proton
 flux (which is one order of magnitude too).

This means that placing an upper limit  on the GZK photon flux,
 or measuring it, provides complementary information to that
 contained in the  UHECR proton flux itself. However, 
 extracting information on the extragalactic nucleon flux from the GZK photons
would require to have independent information on 
the extragalactic magnetic fields and
  radio background,  vice versa.

\section{III. Results: possible scenarios with GZK photons}

We show in Sect. II  that
 if the UHECR above 10$^{19}$~eV are mostly protons (or neutrons),
depending
 on the slope of the proton flux, the  distribution of sources and the
intervening backgrounds, between  $10^{-5}$  and  $10^{-2}$ of
 the UHECR  above $10^{19}$ eV are photons. Much
 larger photon factions are predicted
 at   $10^{20}$~eV in some cases.

  The largest GZK photon fractions in UHECR
 happen for small values of $\alpha$, large values of $E_{\rm max}$,
  small minimal distance to the sources (which is compatible with a
small frequency of clustering of the events) and small intervening backgrounds. 
In the most favorable
cases for a large photon flux, GZK photons could dominate the UHECR flux in an
energy range above 
10$^{20}$~eV. As we show below, this allows us 
to fit the AGASA data, at the expense of
assuming that the initial protons could have  a hard spectrum $\sim 1/E$
and be accelerated to energies as high
 as $10^{22}$~eV. In this extreme case, the AGASA data (as shown in subsection
 III-A below) can be explained
 without any new physics, except in what the mechanism of acceleration of
 the initial protons  is concerned. We also fit the HiRes spectrum 
(in III-B below).
With the HiRes spectrum the GZK photons are always subdominant
and  can be neglected for the fit. 
In both cases, AGASA or HiRes data,
 we evaluate the minimum and maximum GZK photon fractions
expected with each spectrum of UHECR.

We make a one-parameter $\chi^2$ fit to the assumed total spectrum,
obtained by summing up the contributions of protons, GZK photons and  a
low energy component (LEC) when needed. 

In this section we parametrize the
LEC with
\begin{equation}
F_{\rm LEC} \sim E^{- \beta} \exp (-E/E_{\rm cut})~.
\label{LEC_flux}
\end{equation}
 and we fit the amplitude  to the lowest 
energy bin in the figures. 
 We choose the parameter $\beta=2.7-2.8$ to fit the low energy spectral points,
 and the parameter $E_{\rm cut}$ so that the minimum $\chi^2$
value per degree of freedom of the fit is smaller than one. 

 We use the 18
highest energy data  bins of  AGASA  and the 16 highest 
energy  data bins of HiRes-1 monocular data.
We also separately check  the $\chi^2$ for the AGASA events above the  GZK 
cutoff, i.e. for the 3 highest energy AGASA data bins, with $E>10^{20}$ eV. 
We do this  to exclude models which do not fit well the highest energy
events but whose minimum $\chi^2$ considering all the 18 bins could be good
due to the LEC assumed. Additionally, we check  that the
number of events predicted above the end point of the AGASA spectrum (the energy 
above which  AGASA has observed no events), i.e. at $E> 2.5 \times 10^{20}$ eV,
  is not larger than 4 (predicting 4 events 
and observing none   has a very small Poisson probability of 1.8\%).
 The  number of events we predict above the end point of the 
HiRes spectrum, at $E> 3.2\times 10^{20}$~eV,
is always much smaller than 4.

\subsection{A. GZK photons with the AGASA spectrum}
\label{AGASA}

In this subsection, we will discuss fits to the AGASA data with
extragalactic protons, their secondary GZK photons and a LEC as in
Eq.~(2) when needed.
Unless we mention otherwise, here we take a zero
(i.e. comparable with the interaction length) minimum distance to the sources.

The fits to the AGASA spectrum at high energy with  a proton dominated flux
are very poor.  As shown in Fig.~\ref{F1},
for  $\alpha<2.7$  a low energy component (LEC) which we parametrize as in
Eq.~(\ref{LEC_flux}),
possibly consisting of galactic or extragalactic Fe and protons,  is
necessary to fit the data.
It is well known that with extragalactic protons plus a  LEC
one can fit  the AGASA data  below the GZK cutoff, at energies 
$3 \times 10^{18}~{\rm eV}~ < E < 10^{20}~{\rm eV}~$. In fact,
 we tried power law indexes $\alpha=2.7, 2, 1.5, 1$ and we obtained fits with
 minimum $\chi^2 = 36, 17.7, 14, 14$ for 15 degrees of freedom,
 respectively. The first fit (with $\alpha=2.7$, 
which does not require a LEC) is bad, but
 the others (which do require a LEC) are good. Even the first fit could be
 improved to a minimum $\chi^2 =18$ by changing
 the power index slightly to $\alpha=2.6$
and increasing the number of sources in  the early universe as $(1+z)^3$.
However, the same proton fluxes fit the AGASA data at $E>10^{20}$ eV
 very poorly. We found
 minimum $\chi^2 = 12, 12, 9.8, 7.8$ for 3 degrees of freedom,
respectively. The reason for these bad fits is that
for  $\alpha \ge 2$ the proton flux at super-GZK energies 
is very small, and even for
 $\alpha < 2$ it is still not enough.

These fits can be improved by adding  a large component of GZK photons.
We try to maximize  the GZK photon flux by reducing the
radio background and EGMF, and increasing the  maximum proton energy 
 in Eq.~(\ref{proton_flux}) up to
 $E_{\rm max}=10^{22}$~eV.

\begin{figure}[ht]
\includegraphics[height=0.48\textwidth,clip=true,angle=270]{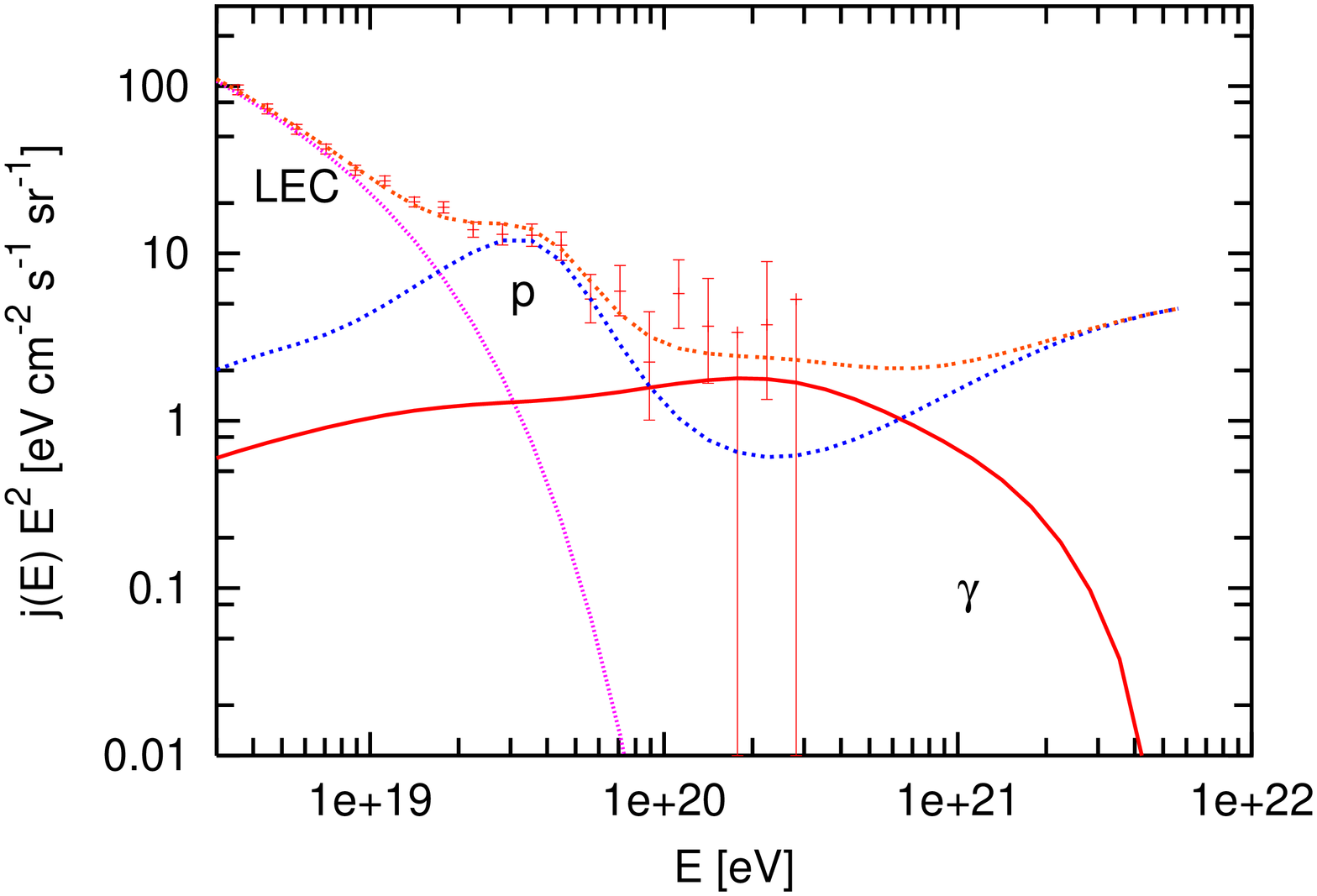}
\includegraphics[height=0.48\textwidth,clip=true,angle=270]{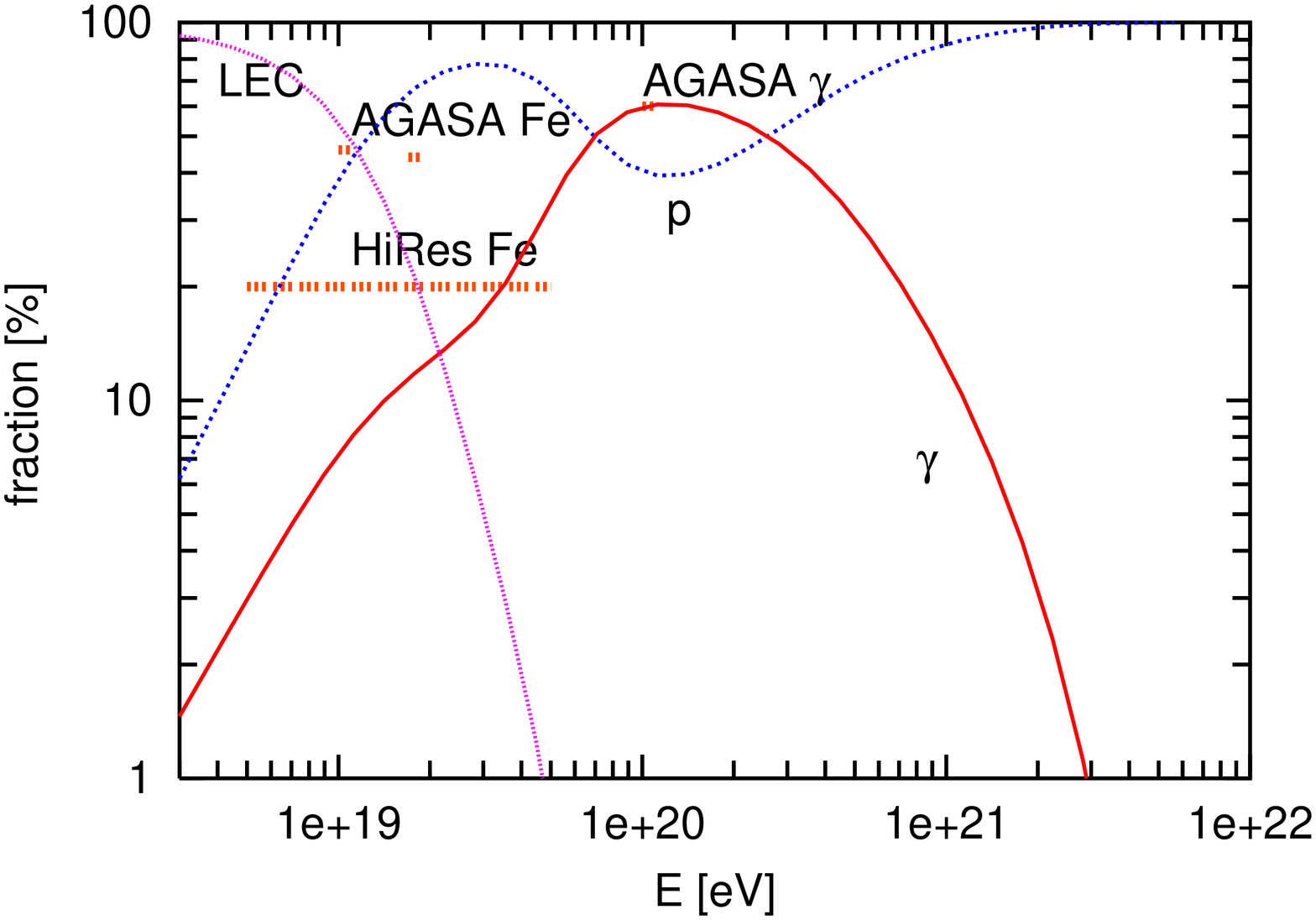}
\caption[...]{ Example of a fit to the  AGASA data with
 extragalactic  protons, the GZK photons they produce and a  low energy  
component (LEC) at $E<10^{19}$~eV.
 (a) differential spectra (upper panel) and (b) fraction  in percentage
 of the integrated flux above the energy $E$ 
 of every component (lower panel). Here we try to {\bf maximize} the photon
 component  thus we take an
extragalactic proton spectrum 
$\sim 1/E$ with maximum energy 
$E_{\rm max}=10^{22}$~eV,  $B_{\rm EGMF}=10^{-11}$~G and the
 minimum radio background.  
 Also shown in (b) are  the 2-$\sigma$ AGASA upper bounds on the Fe fraction
above 10$^{19.0}$~eV and 10$^{19.25}$~eV~\cite{agasa_composition_2},  the
 HiRes limits on Fe component~\cite{hires_composition_fit}
  and  the bound on  the photon fraction
obtained with AGASA data at 10$^{20}$~eV~\cite{agasa_photon}.}
\label{F6}
\end{figure}

\begin{figure}[ht]
\includegraphics[height=0.48\textwidth,clip=true,angle=270]{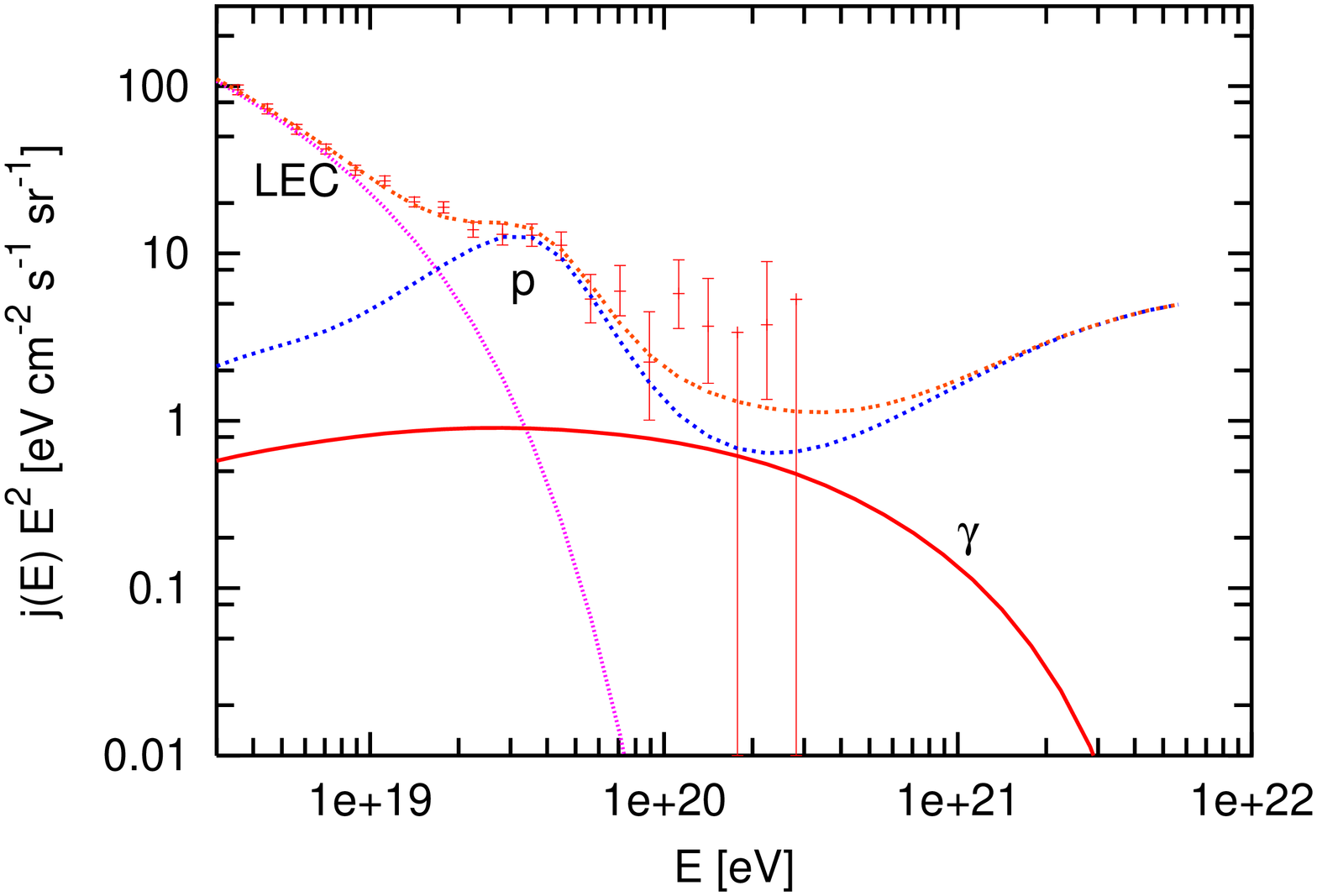}
\includegraphics[height=0.48\textwidth,clip=true,angle=270]{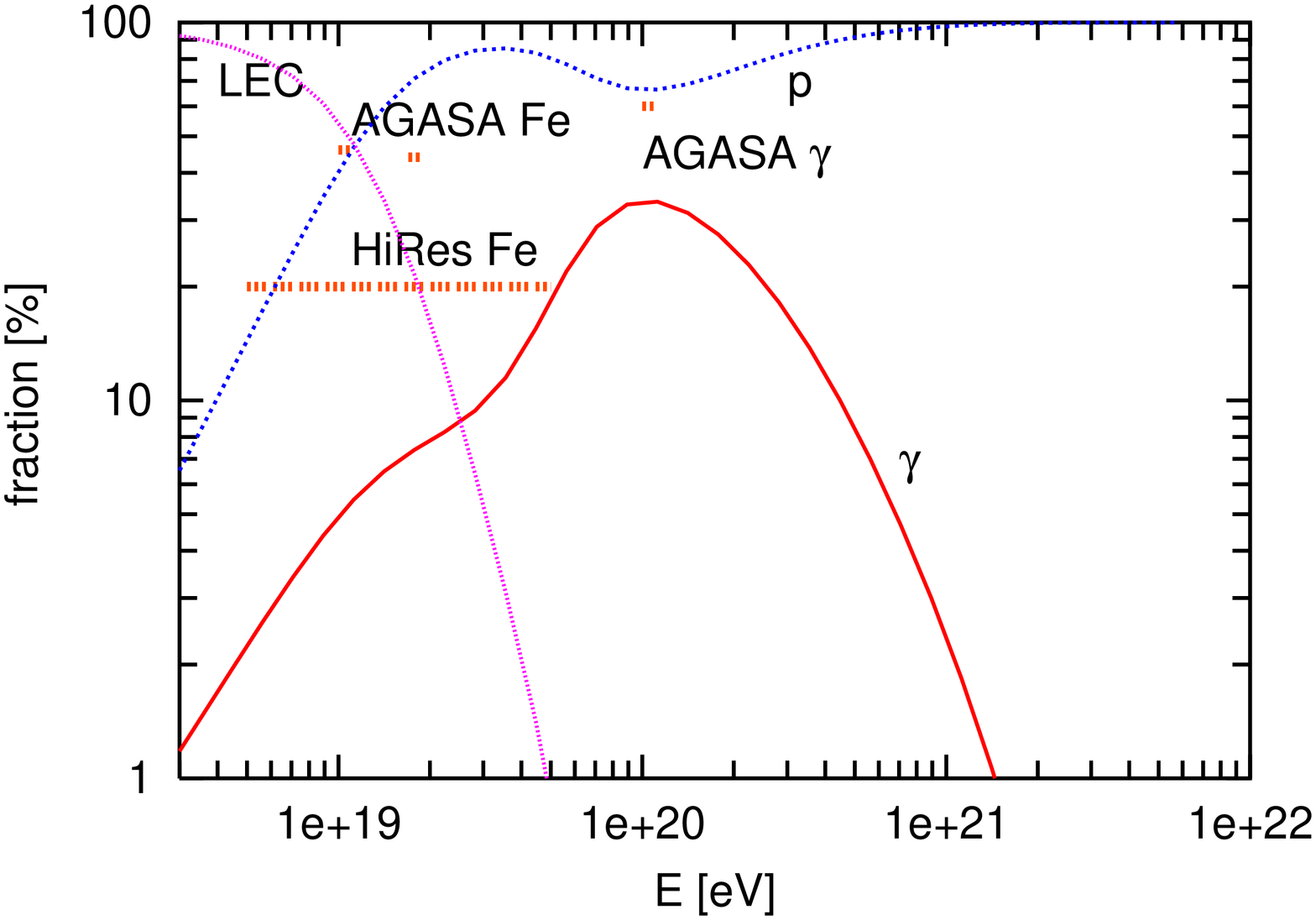}
\caption[...]{Same as Fig.~\ref{F6} but with reduced GZK photon 
flux due to assuming the
intermediate extragalactic radio background (instead of the lowest).
 Here we try to 
{\bf minimize} the photon component  while still providing a good fit
 to the AGASA data.
}
\label{F6BIS}
\end{figure}

In  Figs.~\ref{F6} and \ref{F6BIS}  we show
(a) the differential spectra, 
 of each component (i.e. extragalactic $p$,  LEC and GZK $\gamma$)
and total (upper panels), and  (b) the integrated flux fractions  
of different components in percentage of the  
total  predicted flux above the energy $E$ (lower panels). 
The extragalactic
  protons have here an initial 
spectrum $\sim 1/E$, with maximum energy 
$E_{\rm max}=10^{22}$~eV (see Eq.~(1)).
The particular LEC shown has parameters $\beta=2.7$ and
 cutoff energy $E_{\rm cut}=10^{19}$ eV
(see Eq.~(\ref{LEC_flux})). In both Figs. ~\ref{F6} and \ref{F6BIS} 
the EGMF is B=10$^{-11}$~G.
The only difference between both figures is the radio background:
 we took the lowest one for 
Fig.~\ref{F6} and the intermediate one for Fig.~\ref{F6BIS}. 
This is the only change we can impose
between the maximum and the minimum GZK photon flux while 
not reducing the goodness of fit to the AGASA
data to unacceptable levels.

The fit to the super-GZK AGASA events in Fig.~\ref{F6}a  is now perfect,
 due to the GZK photons: it has
a minimum $\chi^2 = 2.6$ for 3 degrees of freedom and at  $E>10^{20}$ eV 
there are 11.5 events (6.8 photons and 4.5 protons) where AGASA has observed 11.
The   spectrum predicts 4 events (2 photons and 2 protons) at
 energies above 2.5$\times 10^{20}$~eV, 
where AGASA has seen none, which we take as acceptable (the probability is
small, 1.8\%). Larger 
$E_{\rm max}$ or lower $\alpha$ values 
would lead to predict even more events where AGASA has seen none and
 would therefore 
not fit well the AGASA spectrum any longer.

The fit to the super-GZK AGASA events 
 in Fig.~\ref{F6BIS}a, where we try to lower
the GZK flux,  is not as
 good as that in Fig.~\ref{F6}a: 
it has  a minimum $\chi^2 = 5.5$ for 3 degrees of freedom and at 
 $E>10^{20}$ eV 
there are 7 events (2.5 photons and 4.5 protons). But, this fit is better 
 than that is  Fig.~\ref{F6}a
above the end-point of the AGASA spectrum: it predict only 2.7 events
 above the highest energy
AGASA point, which has a 6.7\% Poisson probability.

As we see, a good fit to the  AGASA data at $E>10^{20}$ eV
 with GZK photons is strongly restricted
by the total number of events on one side and by the number
 of events above the  end-point of the AGASA 
spectrum on the other. Thus,   Figs.~\ref{F6}-\ref{F6BIS}
 provide an estimate of the maximum and minimum
 GZK photon flux which fit the AGASA data.

Notice in Fig.~\ref{F6}b that with  the maximum
 GZK photon flux prediction, the 
photon ratio increases from about 7~\% at $10^{19}$~eV 
 to more than 50 \% above $10^{20}$~eV, and that
the total differential flux is dominated by GZK photons at   
energies between 1 and 7 $\times 10^{20}$~eV. This large
 GZK photon flux is possible only 
under the extreme conditions chosen here. A larger radio background,
 or a smaller maximum proton energy quickly diminish the
 GZK photon flux, as Fig.~\ref{F6BIS} demonstrates.

The EGRET bound on the photon energy which cascades down to
 the GeV energies
has been taken into account. We found that the flux predicted 
is about one order of magnitude
below the level measured by EGRET.

The 2-$\sigma$ AGASA upper 
bounds on the Fe fraction in the integrated fluxes, of  
46\% and 44\% above 10$^{19.0}$~eV and
 10$^{19.25}$~eV respectively~\cite{agasa_composition_2} 
are  shown in   Fig.~\ref{F6}b and Fig.~\ref{F6BIS}b.
 The LEC could  respect these bounds 
(so that the LEC could 
consist entirely of galactic Fe), if we assumed a somewhat
 softer  proton spectrum
than we choose for  Figs.~\ref{F6} and \ref{F6BIS}, possibly
 with  $\alpha \gsim 1.5$. With our choice, the extragalactic 
 proton spectrum is  a bit too low at energies 
below the GZK energy and, consequently, the LEC is too large. 
 The lower HiRes limit on a possible Fe low energy
component~\cite{hires_composition_fit}, rejects entirely 
a  LEC consisting mostly of iron.
 In this case  the  LEC should consist mostly of extragalactic 
protons with a soft
 spectrum  $\sim 1/E^{2.7}$ and a small maximum energy 
$E_{\rm max}  \ll 10^{20}$~eV which should
 come from a different class of UHECR sources (than those which produce
the super-GZK UHECR). 

Also shown in   Fig.~\ref{F6}b and  Fig.~\ref{F6BIS}b   is the 
bound on the photon fraction  obtained with AGASA 
data at 10$^{20}$~eV~\cite{agasa_photon},
which is saturated by the photon flux in
 Fig.~\ref{F6}.

\subsection{B. GZK photons with the HiRes spectrum}
\label{HiRes}
\begin{figure}[ht]
\includegraphics[height=0.48\textwidth,clip=true,angle=270]{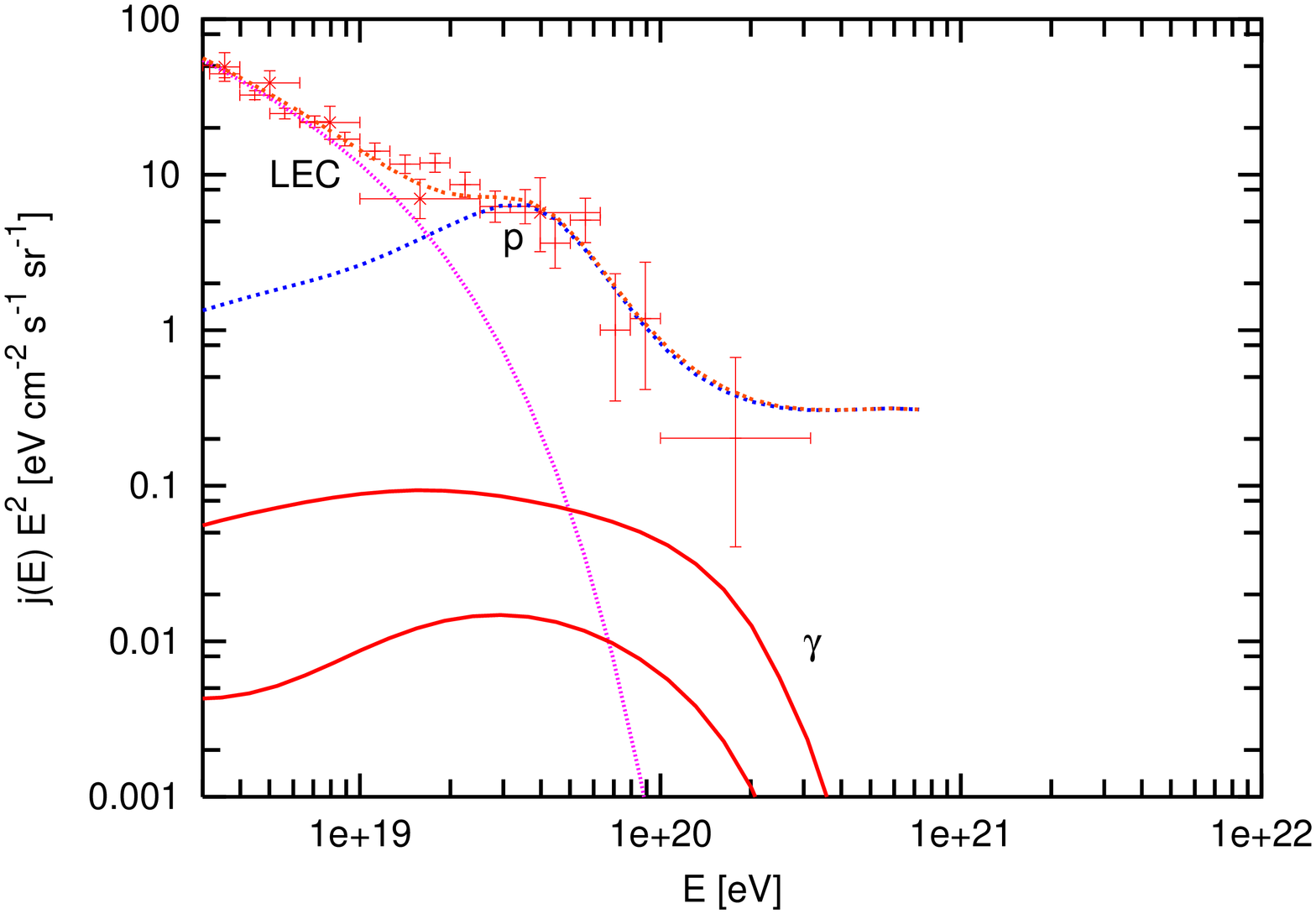}
\includegraphics[height=0.48\textwidth,clip=true,angle=270]{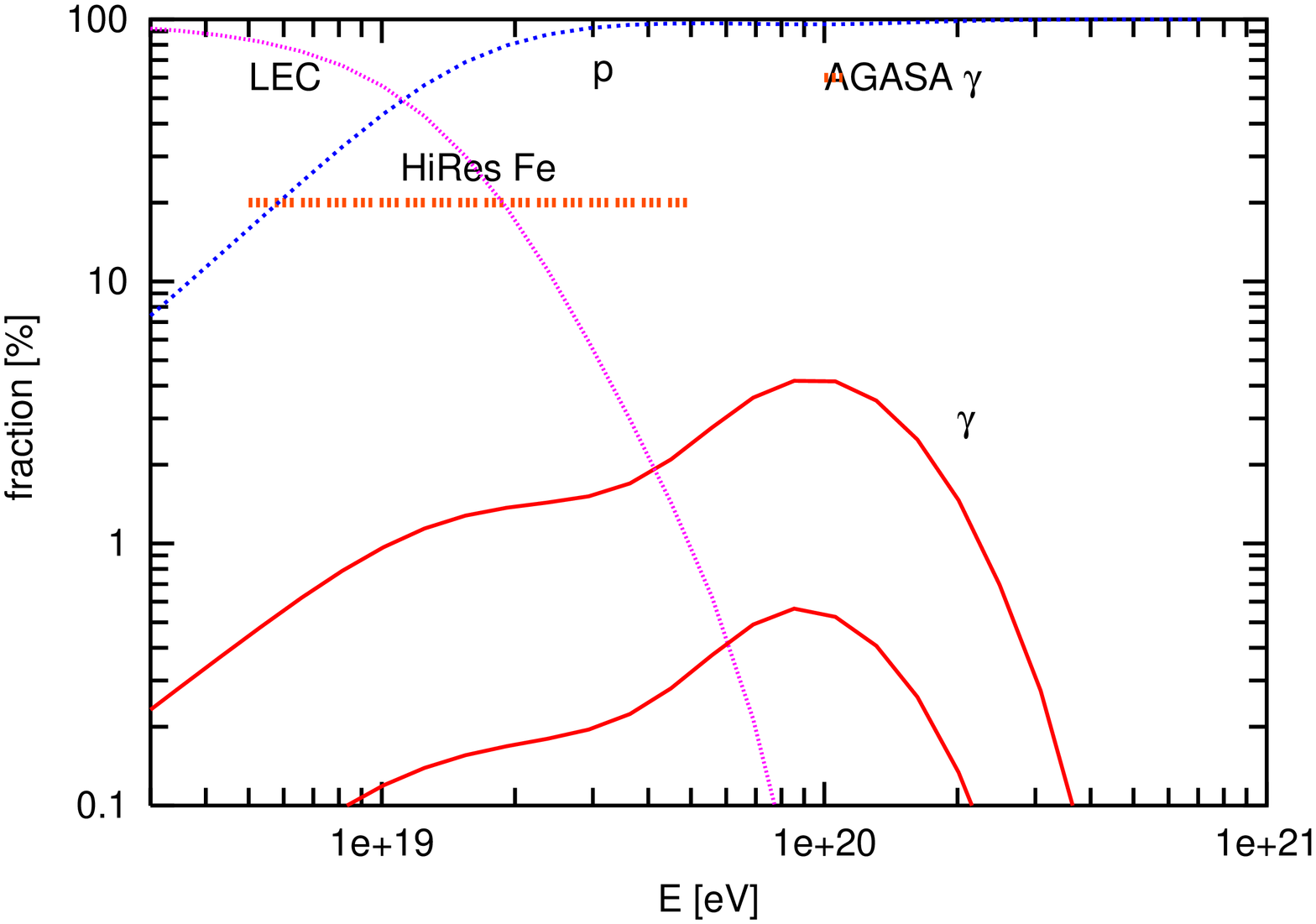}
\caption[...]{ 
Example of a fit to the  HiRes data with
 extragalactic  protons, the GZK photons they produce and a  low energy  
component (LEC) at $E<10^{19}$~eV.
 (a) Differential spectra (upper panel) and (b) fraction  in percentage
 of the total integrated predicted flux above the energy $E$ 
 of every component (lower panel). Here we try to {\bf maximize} the photon
 component  thus we take
an extragalactic proton spectrum $\sim 1/E$ with maximum energy 
$E_{\rm max}=10^{21}$~eV, minimum radio background and
 $B_{\rm EGMF}=10^{-11}$~G  for
the higher photon curve (maximum radio background and
 $B_{\rm EGMF}=10^{-9}$~G for the lower photon curve).
Also shown  in (b) are the HiRes limits on a possible Fe low energy
component~\cite{hires_composition_fit}
 and the bound on  the photon fraction
obtained with AGASA data at 10$^{20}$~eV~\cite{agasa_photon}.}
\label{F7}
\end{figure}
\begin{figure}[ht]
\includegraphics[height=0.48\textwidth,clip=true,angle=270]{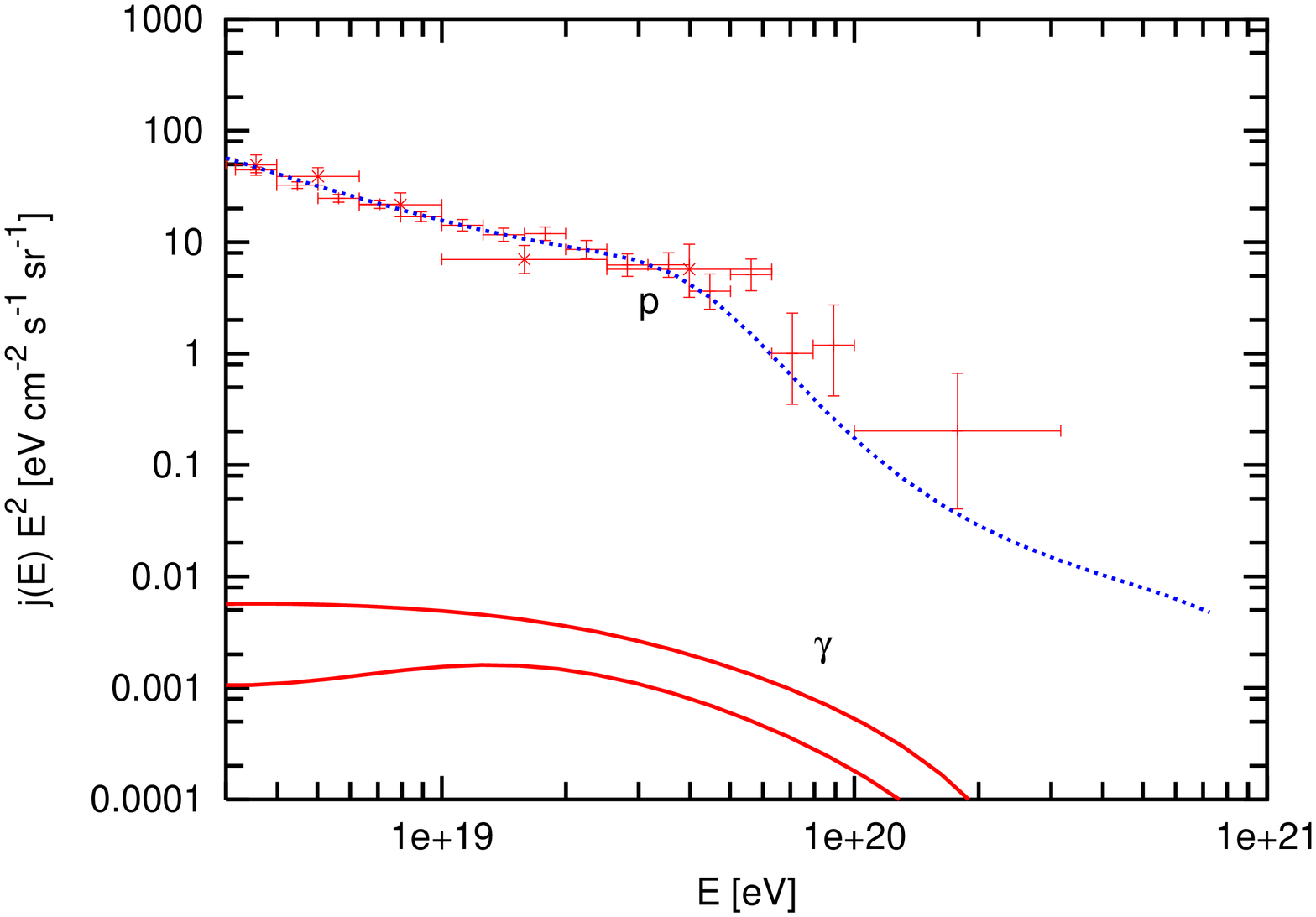}
\includegraphics[height=0.48\textwidth,clip=true,angle=270]{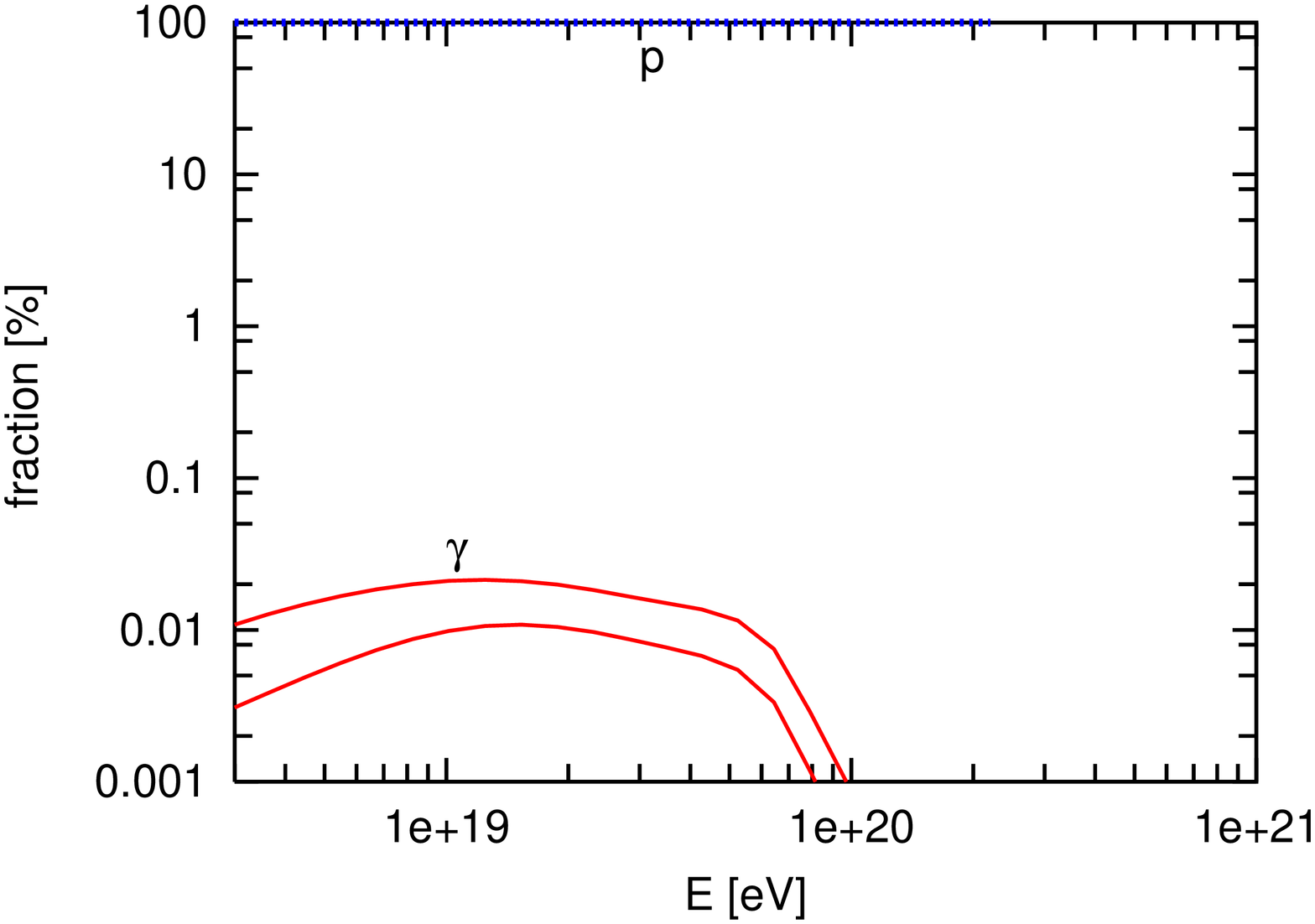}
\caption[...]{
Example of a fit to the  HiRes data with
 extragalactic  protons, the GZK photons they produce and a  low energy  
component (LEC) at $E<10^{19}$~eV.
 (a) Differential spectra (upper panel) and (b) 
 fraction  in percentage
 of the  integrated total predicted flux above the energy $E$ 
 of every component (lower panel). Here we try to {\bf minimize} the photon
 component  thus we take
an extragalactic proton spectrum $\sim 1/E^{2.7}$ with maximum energy 
$E_{\rm max}=3 \times 10^{20}$~eV, maximum radio background and
 $B_{\rm EGMF}=10^{-9}$~G  for the lower photon curve
 ($B_{\rm EGMF}=10^{-11}$~G and intermediate radio flux for the
 higher photon curve). The total flux is dominated by
 nucleons at all energies and is lower than the HiRes data at high
 energies. This is about the best fit that can be done to the HiRes
spectrum with  a
 proton dominated flux.
Also shown in (b) are the  HiRes limits on a possible LEC Fe
component~\cite{hires_composition_fit}
 and the bound on  the photon fraction
obtained with AGASA data at 10$^{20}$~eV~\cite{agasa_photon}.}
\label{F8}
\end{figure}

To estimate the possible range
of photon fractions compatible with the HiRes spectrum
 we will here present two  fits to the HiRes data, one maximizing 
and one minimizing the GZK photon flux. These fits  are presented in
 Figs.~\ref{F7} and \ref{F8} respectively.
 
 Figs.~\ref{F7} show (a) the differential spectra
 of each component (i.e. extragalactic protons,  LEC and GZK photons)
 and total
and  (b) the integrated flux fractions  of different components with
 respect to the  
total predicted flux shown in Fig.\ref{F7}a.
In order to maximize the flux of GZK photons we
need a relatively hard proton spectrum, thus a LEC is needed to fit 
the data at energies $E<10^{19}$~eV.
The particular LEC shown has parameters $\beta=2.7$ and
 cutoff energy $E_{\rm cut}=2 \times 10^{19}$ eV
(see Eq.~(\ref{LEC_flux}).
Here we assume an
extragalactic proton spectrum $\sim 1/E$ with maximum energy 
$E_{\rm max}=10^{21}$~eV, to maximize the number of super-GZK protons,
and to minimize  the photon absorption by the intervening medium, we
 assume the minimum radio background and  $B_{\rm EGMF}=10^{-11}$~G. 
This results in the higher photon curve in the figures. 
 The lower photon curve shows how much the photon flux decreases
 if we keep the same proton flux and change the intervening background
 from minimum to maximum, i.e. if we use 
 $B_{\rm EGMF}=10^{-9}$~G and
maximum radio background. The change is about an order of magnitude.

The total flux shown in Fig.~\ref{F7}a is dominated by protons and is
 insensitive to the GZK photon contribution.
 With this flux only one event (a proton event)
 is predicted above  1$\times 10^{20}$~eV.

 Also shown in Fig.~\ref{F7}b are the HiRes limits on a possible LEC
 Fe component~\cite{hires_composition_fit} and the bound on  the photon fraction
obtained with AGASA data at 10$^{20}$~eV~\cite{agasa_photon}.

In Fig.~\ref{F8} we fit the HiRes data with a conservative model with a
soft extragalactic proton spectrum, which does not require 
a low energy component.  Thus, the power law index of the 
required proton spectrum
is fixed by the observed UHECR  at energies below $10^{19}$ eV, where the 
spectrum  is $\sim1/E^{2.7}$. This model has
practically no freedom in the choice of  the proton 
flux power law index $\alpha$,
although this could be slightly varied
in the range $\alpha=2.4-2.7$ by changing the  redshift dependence of the 
distribution of sources.  For Fig.~\ref{F8} we conservatively
 choose $\alpha =2.7$ and the
smallest cutoff energy which provides  a good fit, which is 
  $E_{\rm max}=3 \times 10^{20}$~eV.
We  assume
zero minimal distance to the sources (larger values do no 
provide a good fit at high energies),
and, to maximize the absorption of photons, the maximum radio background and
 $B_{\rm EGMF}=10^{-9}$~G for the lower photon curve.
  We also give the result for 
 $B_{\rm EGMF}=10^{-11}$~G and intermediate radio background 
(higher photon curve) to show how
the photon flux increases with a less absorbing intervening background.
 The total flux is insensitive to the GZK photon contribution.

The difference between the lower photon line of
 Fig.~\ref{F7} and the lower photon line of  Fig.~\ref{F8} (both computed
 with the same background) shows the uncertainly due to the UHECR proton
 flux (which is one order of magnitude too) for models that fit 
the HiRes spectrum.

Also shown in Fig.~\ref{F8}b are the  HiRes limits on a possible LEC Fe
component~\cite{hires_composition_fit}
 and the bound on the photon fraction obtained with AGASA data 
at 10$^{20}$~eV~\cite{agasa_photon}.

We see in Fig.~\ref{F8}b  that in this case, in which we try to minimize the GZK photons, 
 these could contribute only $1-2 \times 10^{-4}$ at $10^{19}$ eV,  and
$1-2 \times 10^{-5}$ at $10^{20}$ eV,  of the total integrated flux.
 These levels of photon fraction are out of reach for the
present generation of experiments. At best Auger would detect a few GZK photons
in several years of observations, while HiRes would only  obtain upper limits
on the number of photons at all energies.

\section{IV. Discussion: Comparison of GZK photons, minimum
Top-Down photon predictions and experimental bounds}

In this section we discuss  the present  experimental bounds on and 
theoretical predictions for
UHECR photons, and  discuss the implications of a
 possible future photon detection
or future experimental upper limits on the photon fraction.

We start by comparing the minimal amount of photons predicted by
 Top-Down models
of UHECR with the expected range of GZK photons  discussed in Sect.III. 
We  show that, at high energies, close to 10$^{20}$~eV,
the maximum expected  flux of GZK photons is comparable to 
(for the AGASA spectrum)
or much smaller than (for the HiRes spectrum)  
the minimum flux of photons predicted
by Top-Down models which fit the AGASA or the HiRes data.
Thus, detection of a larger photon flux than expected for GZK photons, 
at those energies, would point to a Top-Down model 
(or to the emission of a large flux of photons at the sources).
The estimate of  the minimum photon ratio predicted by Top-Down models is 
also essential when applying to these models  already existing and 
possible future upper bounds on the fraction of photons in UHECR

Let us recall that Top-Down models were introduced as
an alternative to acceleration models to explain the
 highest energy cosmic rays, which the latter models have
 difficulty explaining. The spectra of  the UHECR produced in Top-Down models
are determined by the elementary particle physics of Z-boson
 decays  and of QCD fragmentation, which predict photon
 domination of the spectrum at high energies.

In  order to {\bf minimize} the photon fraction predicted by
 Top-Down models while fitting the UHECR spectrum, we ask 
 Top-Down models  to explain only the highest energy
events, those close to 10$^{20}$~eV while invoking a more
 conventional Bottom-Up 
extragalactic component (which we assume consists of nucleons)
 to dominate the flux at energies just below. This is an unnatural
 possibility which would require two completely independent mechanisms
to provide UHECR at comparable levels. We consider it only because
it provides the minimum amount of Top-Down photons.  We will present
 here fits to the AGASA and HiRes data following this strategy to
minimize the predicted photons for three Top-Down models:  Z-bursts,
 topological defects (necklaces) and super
heavy dark matter particles (SHDM).

 \subsection{A. Z-bursts}

\begin{figure}[ht]
\includegraphics[width=0.3\textwidth,clip=true,angle=270]{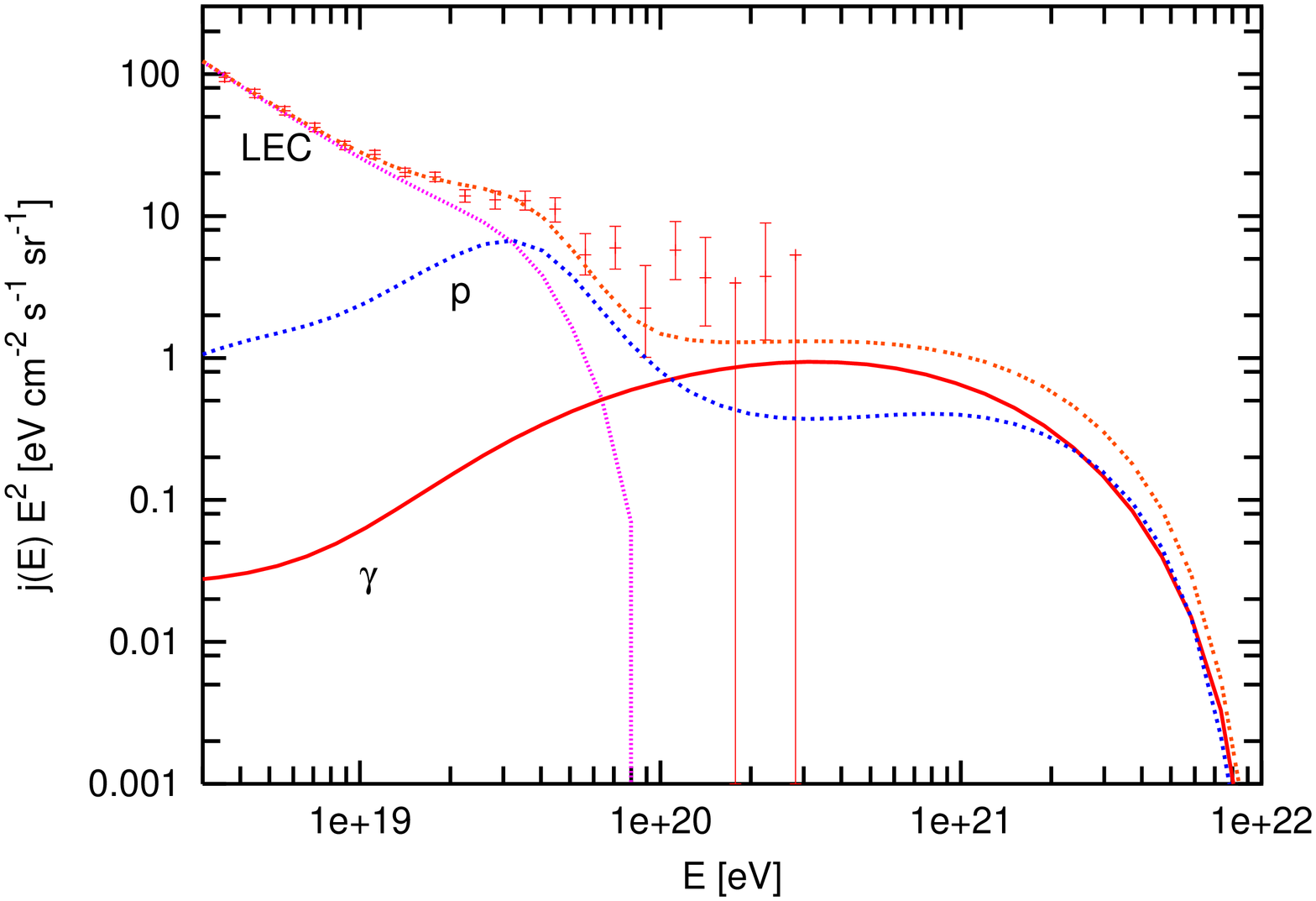}
\includegraphics[width=0.3\textwidth,clip=true,angle=270]{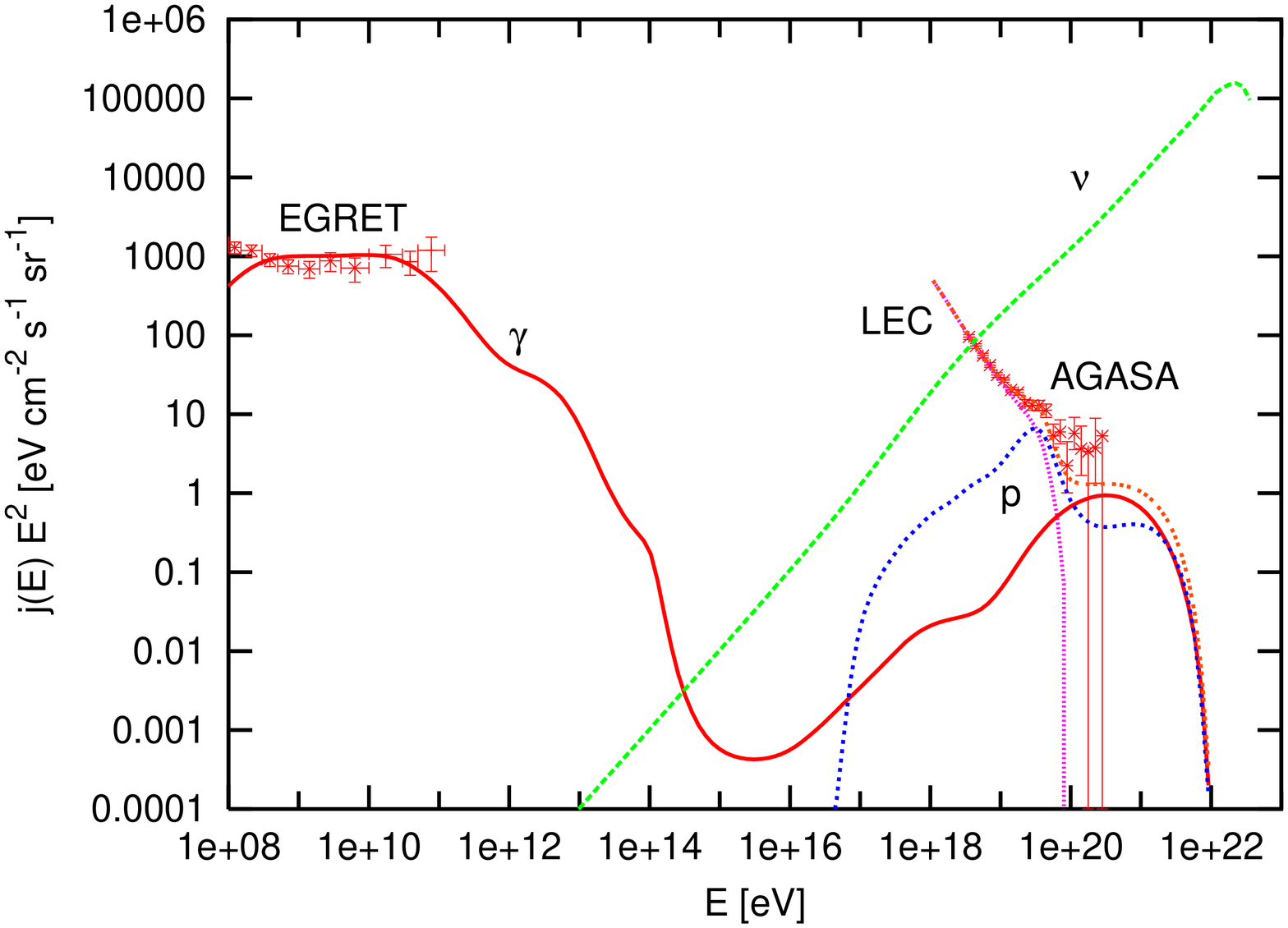}
\caption[...]{Example of a fit to the AGASA data with  a LEC plus a
 flux of protons and photons produced by Z-bursts   (a) showing
 the highest energies and (b) showing also the EGRET energy region.
  LEC due to protons from astrophysical sources. Also shown is the
 assumed initial neutrino spectrum (green curve); only its 
value at the resonance energy is important.}
\label{F9}
\end{figure}

\begin{figure}[ht]
\includegraphics[width=0.3\textwidth,clip=true,angle=270]{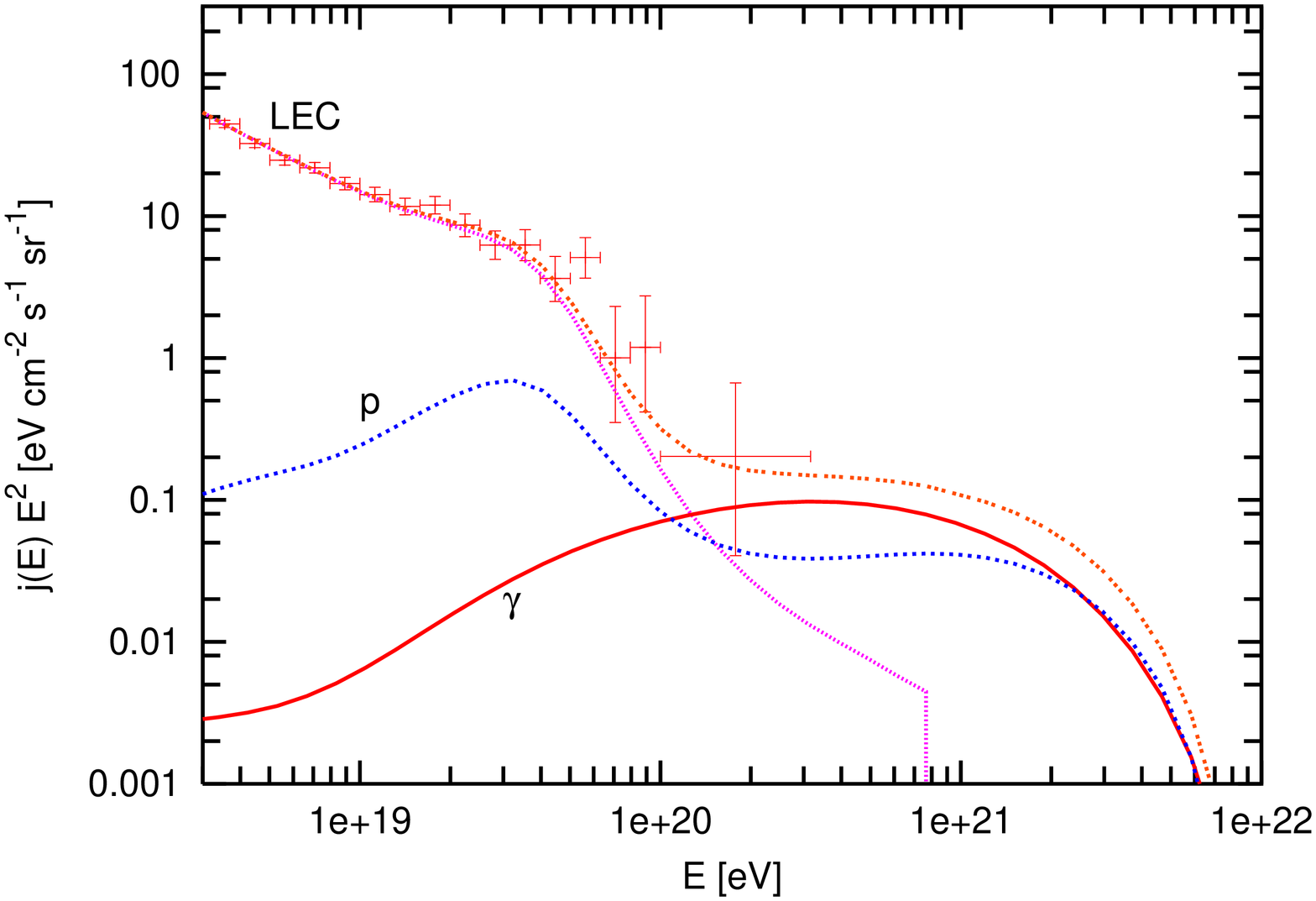}
\includegraphics[width=0.3\textwidth,clip=true,angle=270]{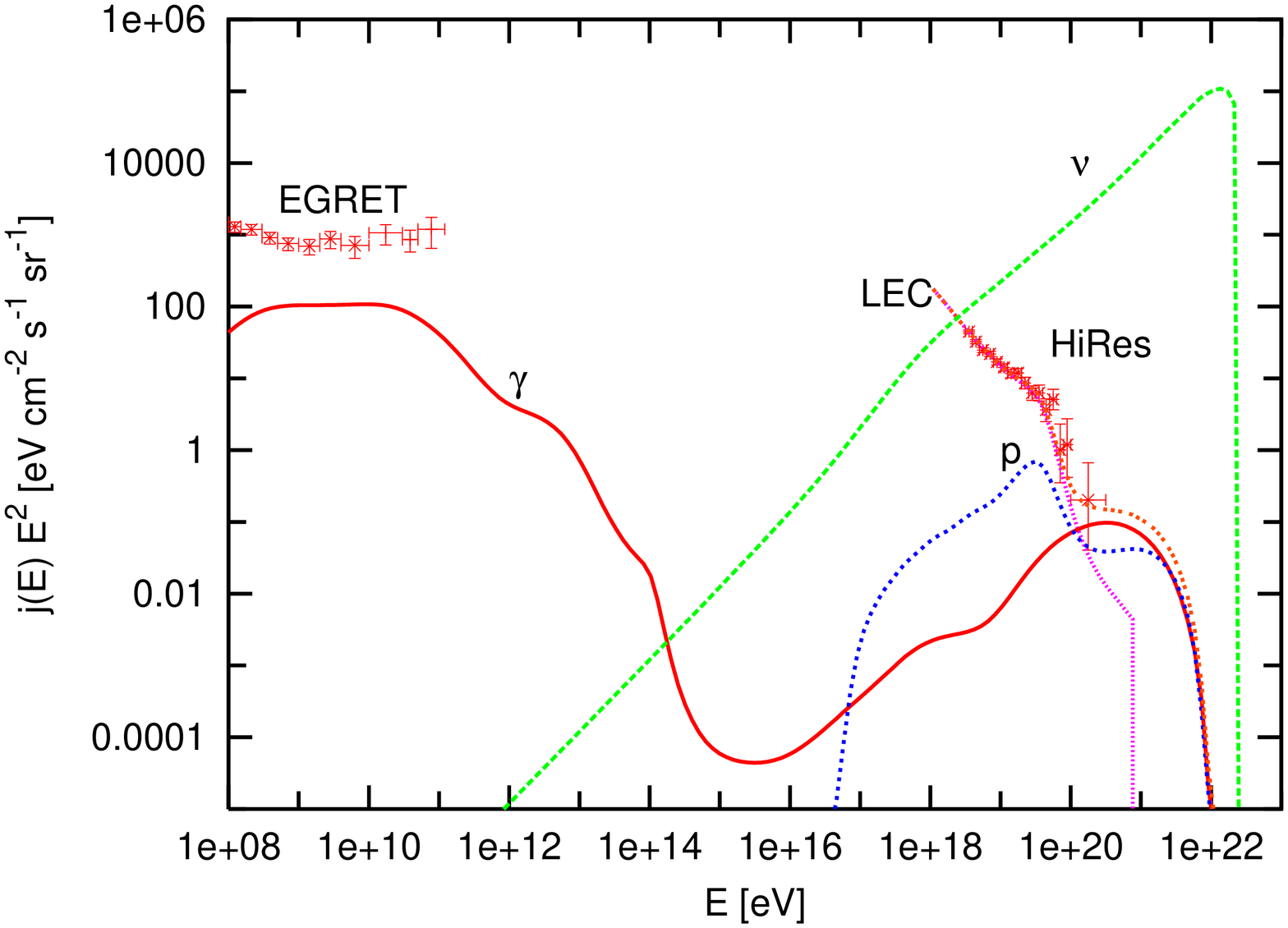}
\caption[...]{As in Fig.~\ref{F9} but for the HiRes data.}
\label{F10}
\end{figure}

In the Z-burst model~\cite{zburst} ultra-high energy
 (UHE) neutrinos coming from 
remote sources annihilate at the Z-resonance with
 relic background neutrinos.
The  Z bosons then decay,  producing 
secondary protons, neutrinos and photons.
The Z-resonance, which acts as a new cutoff,
occurs when the energy of the incoming $\nu$ is
$E_{res} = M_Z^2/ 2~m_{\nu}
	= 4\times 10^{21}{\rm eV}({\rm eV}/{m_\nu})$

So far Z-burst models have been studied mostly  to explain the AGASA spectrum
(however, see Ref.~\cite{Fodor-K-R}).
Many problems have been found, which are alleviated if one assumes the HiRes
spectrum. One of them is that practically no photons should be
 produced at the source together with the  UHECR neutrinos,
 otherwise too many low 
 energy photons in the EGRET region are predicted.  For example, with
sources emitting equal power in neutrinos and photons, the
EGRET bound~\cite{EGRET} on the diffused  
GeV- $\gamma$ ray background is violated by 
two orders of magnitude (see Fig.~3  of Ref.~\cite{zburst_problem}), when
the AGASA spectrum is considered.
Also bounds by the GLUE~\cite{GLUE} and FORTE~\cite{FORTE} 
experiments on the primary neutrino flux,
as well as the non-observation of UHECR events at energies above 
2.5 $\times 10^{20}$ eV  by the AGASA Collaboration
imply a lower bound $\sim 0.3$~eV on the relic neutrino mass
~\cite{reviewGZKneutrinos2, Fodor-K-R, GVW}.
Since this mass exceeds the square root of
 mass-squared differences inferred from
oscillation physics, the bound in fact applies to all three neutrino masses.
Together with the upper bound provided by  CMB anisotropy and 
large-scale structure observations, this bound
leaves only a small interval for neutrino masses around 0.3 eV,
if Z-bursts are to explain the existing UHECR AGASA spectrum.
These problems are somewhat alleviated if Z-bursts are to explain
 the ultra-GZK events in the HiRes spectrum instead of the AGASA spectrum,
as can be seen in Fig.~\ref{F10}. 

The  $p$ and $\gamma$ curves in Fig.~\ref{F9} 
 and Fig.~\ref{F10} show the predictions of a
Z-burst model computed as in Ref.~\cite{reviewGZKneutrinos2} but 
with a relic neutrino mass $m_\nu =0.4$~eV. We assume
 a maximum redshift $z_{\rm max}=3$ for
the UHE neutrino sources (which emit only neutrinos and 
 have not evolved), maximum intervening radio
background and  $B_{\rm EGMF} = 10^{-9}$~G.
In our calculation we do not consider the effect of local
inhomogeneities, such as the Virgo cluster~\cite{RWW-new}.
 The assumed spectrum of UHE neutrinos is shown in the 
figures. Only the part of this spectrum close to
 the resonance energy is relevant. Here we try to minimize the photon
 fraction predicted by Z-bursts by incorporating a low 
energy component of extragalactic nucleons.

In Fig.~\ref{F9}, a low energy component (LEC curve) parametrized
as a power law (as in Eq. (1)) with index $\alpha=2.8$, cutoff energy 
$E_{\rm max}=10^{20}$~eV 
and a minimum distance to the sources of 50 Mpc, has been 
added to the contribution
of the Z-bursts to fit the AGASA data. The fit 
 has minimum $\chi^2=15$ for 15 bins
with $E<10^{20}$ eV.  At higher energies, $E>10^{20}$ eV, 
the fit is not good, it has a min. 
$\chi^2=6.4$ for 3 degrees of freedom. The reason is that the predicted
flux is too low at these energies. However,  the fit to the spectrum
 above the end-point of the AGASA spectrum, 
 $E>2.5\times 10^{20}$~eV,  is  good: only
 two (mostly photon) events are predicted (where none were seen). 

 If we try to increase the Z-burst flux by minimizing
 the absorption of photons by the background, the fit is
 worse at high energies. If we take
 the lowest radio background
and a small EGMF $B=10^{-12}$ G, the fit to the AGASA
 spectrum at $E>10^{20}$ eV
is better, with min. $\chi^2=4$ for 3 degrees of freedom. However,
  5.8 events (mostly photons) are predicted above the
 AGASA end point, which we consider unacceptable.  

As shown in Fig.~\ref{F9}b, the gamma ray flux at low energies 
saturates the EGRET data. Also, as shown Fig.~\ref{F14}a,
the predicted photon fraction saturates the
upper bound on the photon fraction obtained
with AGASA data at 10$^{20}$~eV~\cite{agasa_photon}.

In Fig.~\ref{F10}, a low energy component (LEC curve) parametrized
as a power law (see Eq.~(1)) with index $\alpha=2.7$, maximum energy 
$E_{\rm max}=10^{21}$~eV
and zero  minimum distance to the sources,  has been added to the contribution
of the Z-bursts to fit the HiRes data. The spectrum of this model fits 
perfectly that of  HiRes. Only 1.8 events (1 proton  and 0.8 photon)
 are predicted above the end point of HiRes, were none were seen. 

Because the super-GZK nucleon flux is here lower than with the AGASA 
spectrum, the predicted gamma ray flux at low energies is well under
the EGRET data (see Fig.~\ref{F10}b).
As can be seen in Fig.~\ref{F14}a,
the predicted photon fraction is just under the upper bound obtained
with  AGASA data at 10$^{20}$~eV~\cite{agasa_photon}.

\subsection{B. Topological defects (necklaces)}

\begin{figure}[ht]
\includegraphics[width=0.3\textwidth,clip=true,angle=270]{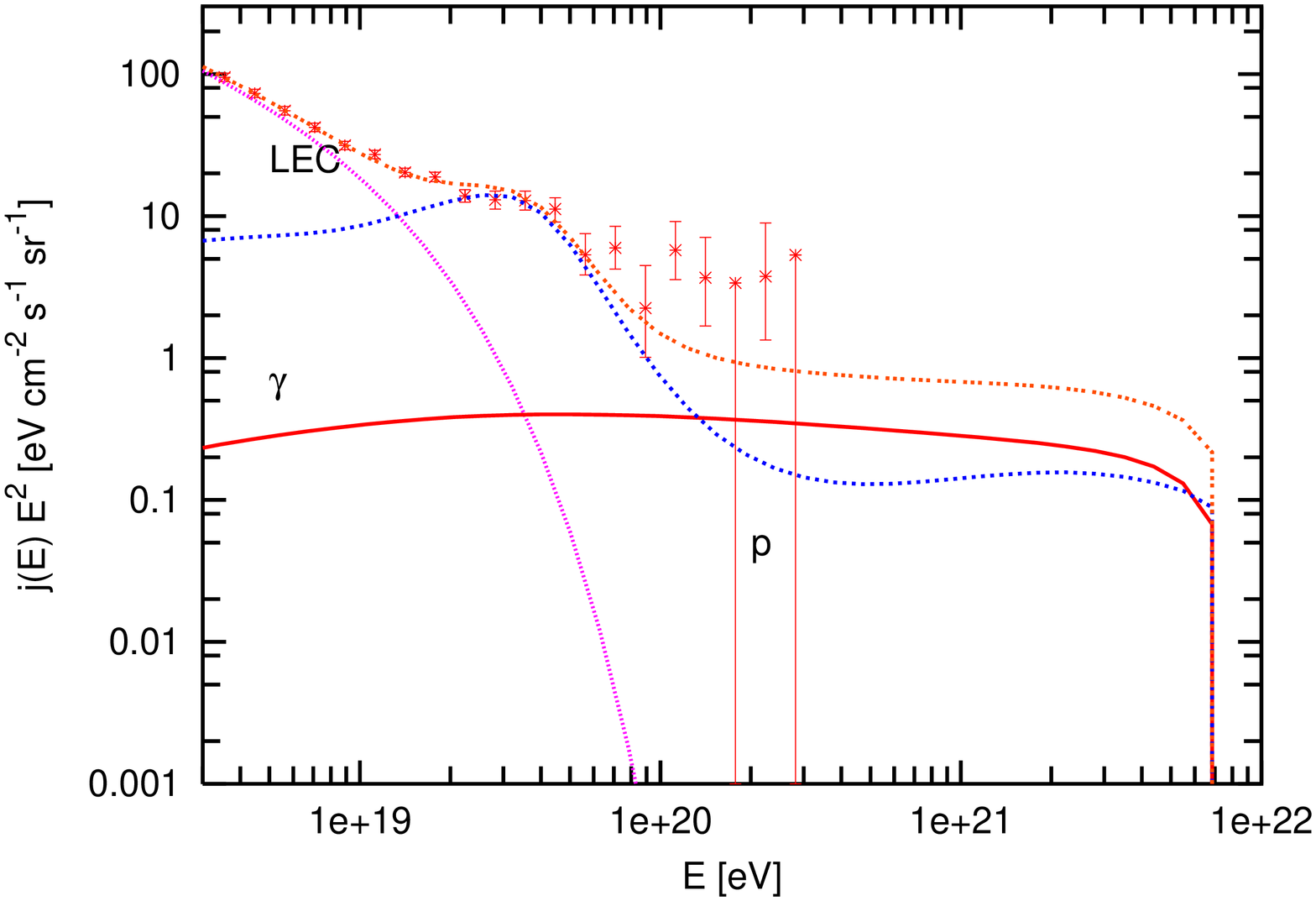}
\includegraphics[width=0.3\textwidth,clip=true,angle=270]{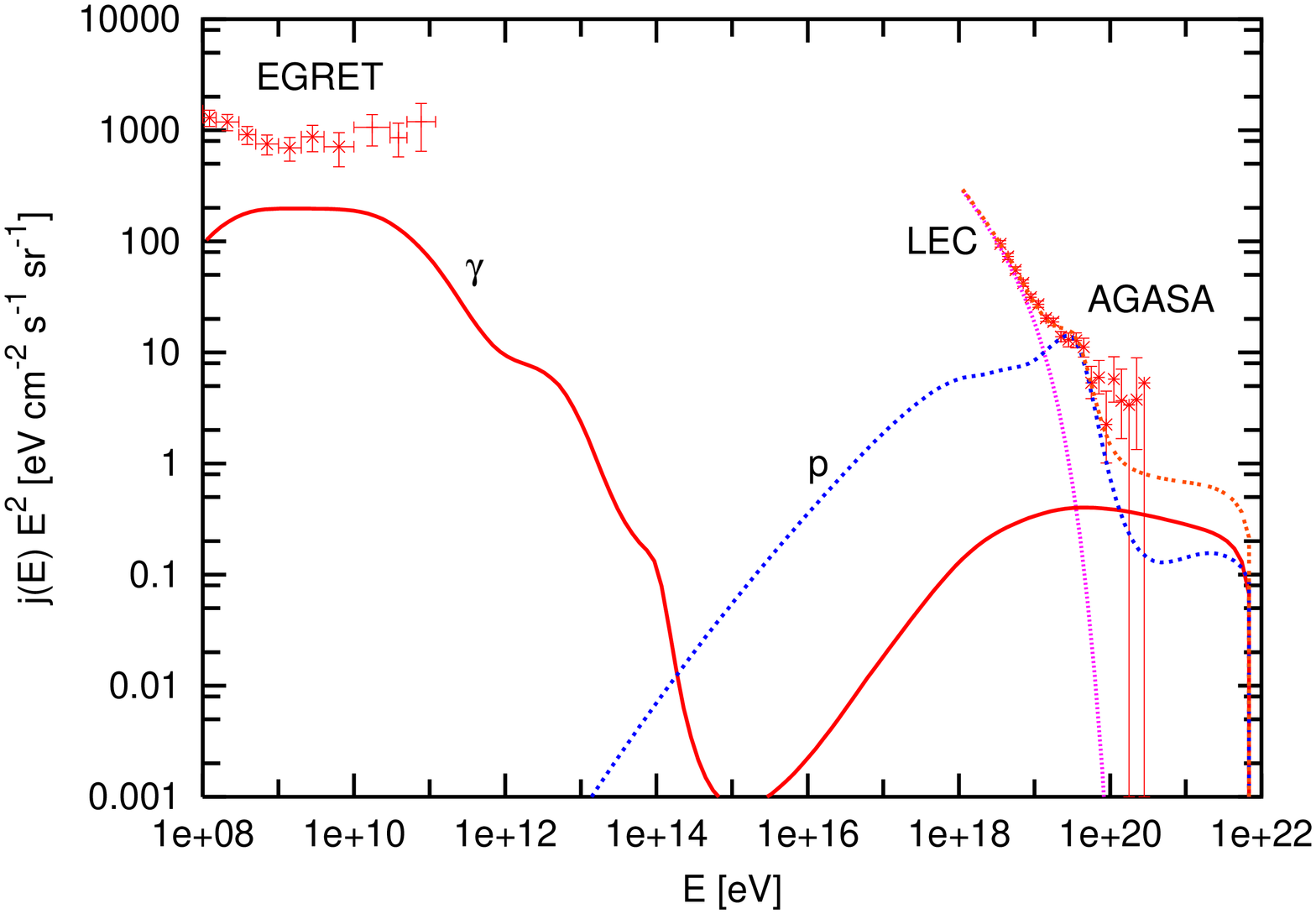}
\caption[...]
{Example of a fit to the AGASA spectrum with
 a LEC  plus secondary protons and photons 
in a topological defect (TD) model, showing (a) the highest
 energies  and (b) also the EGRET
energy range. The LEC, as in Eq.(2), is due to
 nucleons from astrophysical sources. The photon
 over nucleon ratio in  the decay products is about 3. }
\label{F11BIS}
\end{figure}

\begin{figure}[ht]
\includegraphics[width=0.3\textwidth,clip=true,angle=270]{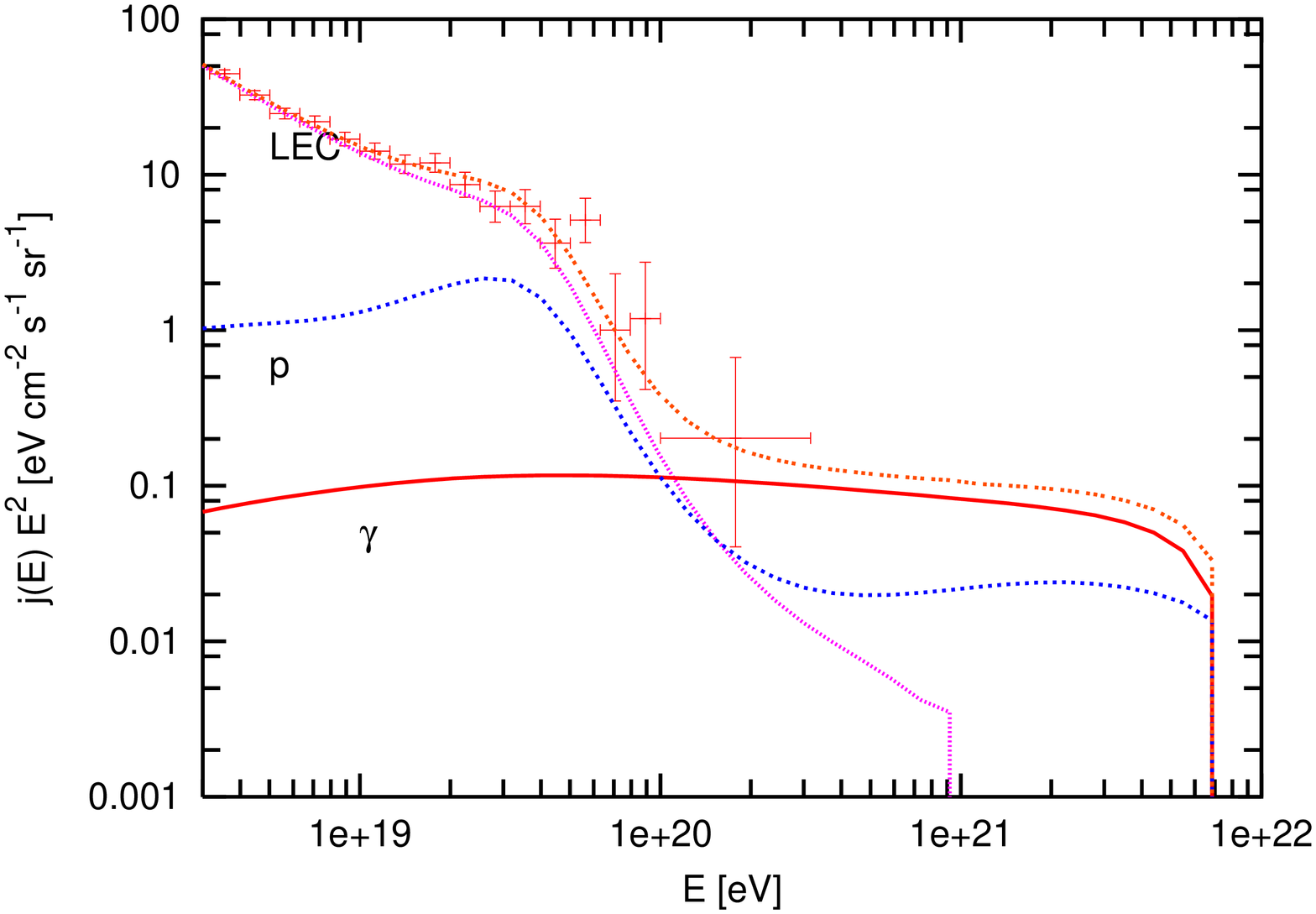}
\includegraphics[width=0.3\textwidth,clip=true,angle=270]{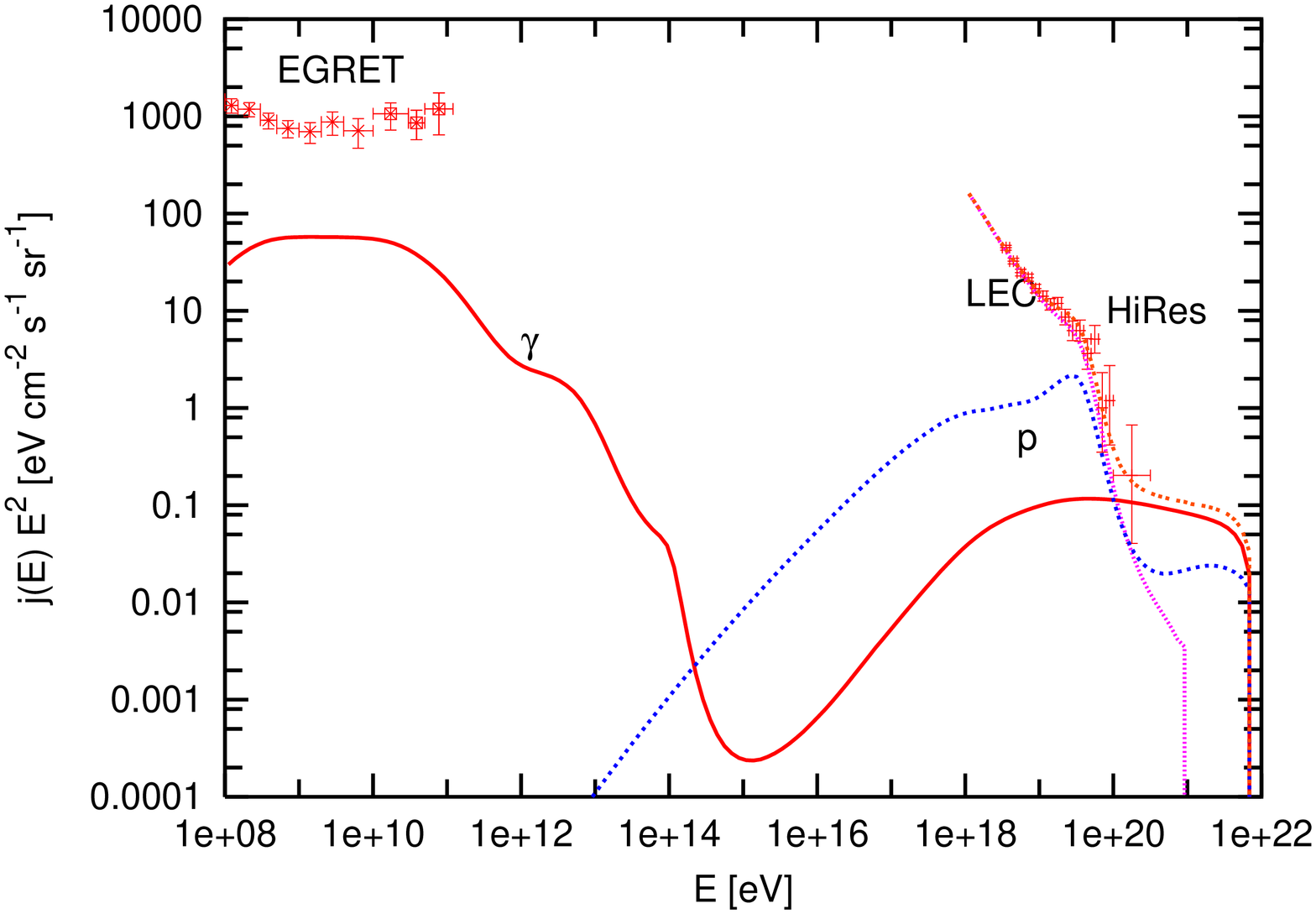}
\caption[...]{As in Fig.~\ref{F11BIS} but for the HiRes spectrum.}
\label{F12}
\end{figure}

The curves $p$ and $\gamma$  in Figs.~\ref{F11BIS}-\ref{F12} correspond to
secondary protons and photons 
in a particular top-down model, in which topological
 defects (TD), such as necklaces,
produce GUT-scale mass particles, which in turn decay into  
quarks, leptons etc (for a review see for example Ref.~\cite{td_review}).
 The mass scale of the parent 
particles provides the maximum energy of the UHECR, $E_{\rm max} = m_X$, thus these scenarios
 avoid the difficulty in astrophysical objects of accelerating  the  UHECR 
to the highest energies observed. As in  
Z-burst models, TD scenarios predict, therefore,  a new cutoff given by
the parent particle mass
  at energies above $10^{20}$ eV. The parent particles
typically decay into leptons and quarks. The quarks hadronize, and some
leptons decay resulting in a large cascade of photons, neutrinos, light
leptons and a smaller amount of  nucleons.

TD models may also have difficulties
 with the EGRET flux~\cite{EGRET, EGRET_NEW} on the diffused  
GeV- $\gamma$ ray background.  We have taken this possible bound into account.

The TD model of Figs.~\ref{F11BIS}-\ref{F12} assumes a parent particle mass 
$m_X= 2 \times 10^{13}$~GeV,
an EGMF of $10^{-12}$~G and the low radio background predicted by Protheroe
and Biermann, which is the intermediate radio background among the three we
consider in this paper. Even if we are trying to minimize
 the photon flux at high energies,
 the radio background and EGMF value are not the maximal
 we used in this paper. This is so
because, as we show here, a smaller amount of
 ultra-high energy photons yields a worse fit to 
the AGASA data. The heavy particle injection rate 
is assumed to be $\sim m_X t^{-3}$, 
where $t$ is the cosmic time.

The QCD spectrum used for Figs.~\ref{F11BIS} and \ref{F12} (shown in Fig.11 of
 Ref.~\cite{reviewGZKneutrinos}) corresponds to the decay of
the heavy particles into two quarks without supersymmetry~\cite{QCD-spectrum}.
 Originally, this decay model predicts a ratio of about 10  photons 
per nucleon in the decay products (as does Ref.~\cite{BK_2001}), while in more recent
 models  ~\cite{SHDM-other,  Barbot-Drees, SHDM_2004} this 
ratio is only  2 - 3. So, 
for Fig.~\ref{F11BIS} and  Fig.~\ref{F12}
the ratio was brought to be equal to 3.
 Here we fit the LEC with 
the function in Eq. (2) with $\beta=2.7$ and an exponential energy
 cut with $E_{\rm cut}=8 \times10^{19}$~eV,
 in order to increase the contribution of  the TD model to the AGASA flux, 
which is still too low
at high energies. Again, at energies $E<10^{20}$ eV the
fit is good, with minimum  $\chi^2=14$ for 15 degrees of freedom. 
However,  the fit of the AGASA 
spectrum above the GZK energy
is bad, with  minimum $\chi^2=7.4$ per 3 degrees of freedom. 
This is due to the strong
reduction of the TD  flux above the GZK energy (due to the
 GZK effect, because there are more protons than in
 Fig.~\ref{F11}), which means that in
 order to have a good fit at energies
below the GZK energy, the flux is too small at higher energies.
Now, there are only 3.7 events at  $E>10^{20}$ eV (of which 2.7 are photons), 
while AGASA observed 11 events. But, if we take the 
minimum radio background  (not shown in figures) instead of the intermediate
one we use for the figures,
the fit to the AGASA occupied bins above the GZK energy  
is  good (with  minimum $\chi^2 = 2.2$ per 3 
degrees of freedom), but the number of events predicted above the end-point  
of the AGASA spectrum (where no events were observed) 
becomes 10, which is again unacceptable.

From  Fig.~\ref{F11BIS}  we conclude
 that the representative TD models
we study are barely consistent 
with the AGASA  data. They either predict a  flux too
 low at super-GZK energies or too many events 
above the highest energy events observed by AGASA.
For the TD curve in Fig.~\ref{F14}a the model of 
 Fig.~\ref{F11BIS} was used. We see in Fig.~\ref{F14}a 
that the predicted photon ratio is somewhat 
above the upper bound on the photon fraction obtained
with  AGASA data at 10$^{20}$~eV~\cite{agasa_photon}.

In Fig.~\ref{F12},  a low energy component (LEC curve), parametrized
as a power law (see Eq.~(1)) with index $\alpha=2.7$ 
and cutoff energy $E_{\rm max}=10^{21}$~eV
and zero  minimum distance to the sources,  has been 
added to the contribution
of the TD model to fit the HiRes data. The spectrum 
of this model (with a $\gamma/ p$ ratio of 3) fits 
well the HiRes data. 
 This model predicts 0.4 events above the end point of the  HiRes spectrum.
It is clear that the fit would be good too with 
 a  larger $\gamma/ p$ ratio in the TD 
decay products, since one can  redistribute the protons 
between the LEC and the TD contribution without a significant 
change in the fit (but the photon fraction at the highest energies would be
somewhat larger).

\begin{figure}[ht]
\includegraphics[width=0.3\textwidth,clip=true,angle=270]{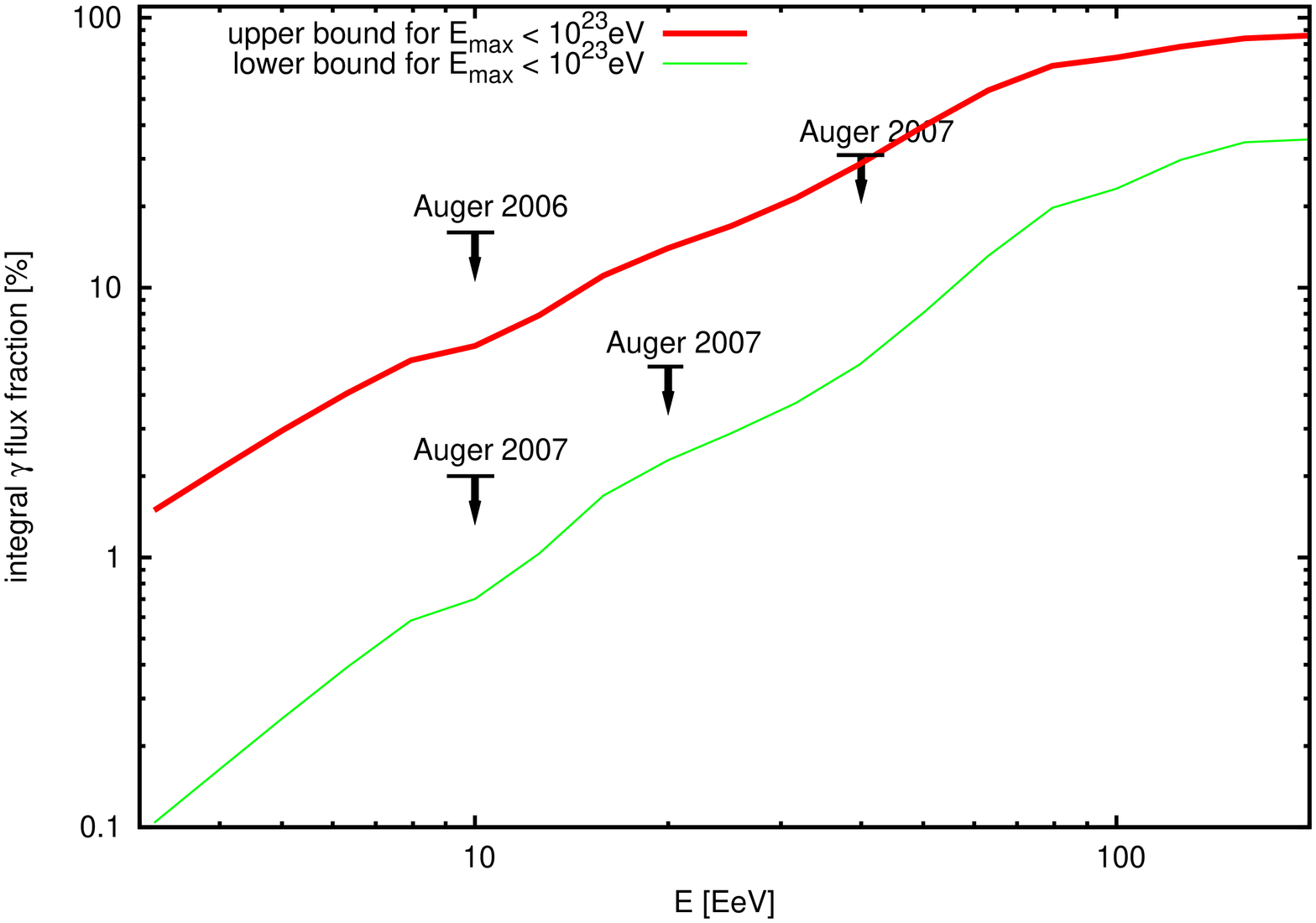}
\includegraphics[width=0.3\textwidth,clip=true,angle=270]{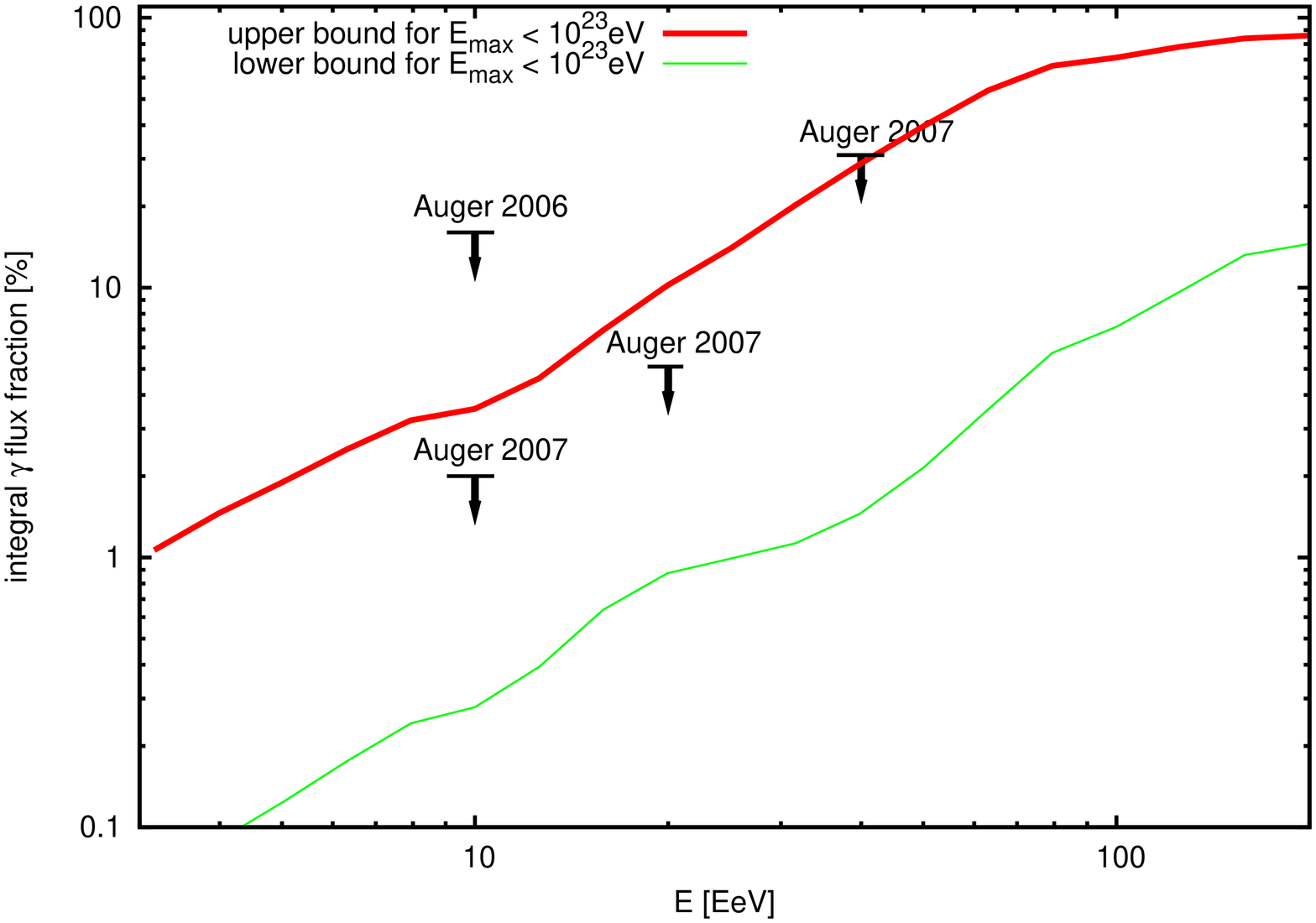}
\caption[...]{Maximum and minimum GZK photon fractions in the integral flux above the energy $E$ for
the topological defect (TD) model described in the text  and with  (a)  the AGASA spectrum (upper panel) and (b)  the HiRes spectrum (lower panel). The 2006~\cite{Auger-photon-06} and 2007~\cite{Auger-photon-07} Auger upper bounds on the photon fraction are also shown.}
\label{F12-BIS}
\end{figure}

 As mentioned above the QCD model used so far in this subsection predicts a ratio of  
 about 10  photons  per nucleon in the decay products~\cite{QCD-spectrum} (although we brought it artificially to 3)  while in more recent
 models  ~\cite{SHDM-other, Barbot-Drees, SHDM_2004} this 
ratio is considerably smaller. We include here also the results obtained with one of these more recent models. The heavy particle decay  spectrum used in Fig.~\ref{F12-BIS} corresponds to the decay of
the heavy particles into quark and antiquark  pairs with the ``gaugino set of supersymmetric parameters"  taken from Ref.~\cite{Barbot-Drees}.
We choose this particular decay mode because it is one in which the initial number of photons per nucleon produced is one of the lowest (since we want to estimate the minimum GZK photon flux
produced).
 This decay model predicts a ratio of about 2 or less photons 
per nucleon in the decay products.  At low energies the fragmentation functions were suppressed following Fig.~2.11 of Ref.~\cite{Barbot-thesis}.  For $(E/ E_{\rm max}) < R_o$ the suppression factor used is  $R^{-log_{10}(R/W^2)}$, where  $R=R_o/ (E/E_{\rm max})$ and $W$ is the width in decades at which the spectrum is suppressed by a factor 0.1 (for $(E/ E_{\rm max}) > R_o$ there is no suppression). From the figure just mentioned, one can find the values of the parameters $R_o$ and $W$. We used $Ro = 10^{-6}$ and $W=3.5$. 

Fig.~\ref{F12-BIS} shows the maximum and minimum photon fractions  found using the method of Ref.~\cite{Gelmini:2007jy} for $E_{\rm max} < 10^{23}$~eV.   In Ref..~\cite{Gelmini:2007jy} the maximum and minimum GZK photon fractions were found  assuming a power law spectrum  of protons is injected by astrophysical sources and fitting the AGASA and HiRes UHECR spectra for energies $E > 4 \times 10^{19}$~eV. It was also assumed that any possible low energy component is irrelevant at this energies. Notice that the LEC in Fig. 12 fulfills this latter condition but that in Fig. 13 does not.  To produce Fig.~\ref{F12-BIS}  we use the same procedure but replace the injected spectrum by that produced in the heavy particle decay.  We choose the value of the amplitude of the injected spectrum by maximizing the Poisson likelihood function
using the UHECR data from 4 $\times 10^{19}$~eV  up to the last published bin  of each spectrum  plus one  extra bin with zero observed events at  higher energies. This extra bin and the highest energy empty published bins,  take into account   the non-observation of events above the highest occupied energy bin in the data of each  collaboration, the end-point energy of each spectrum
  (i.e. at $E> 2.3 \times 10^{20}$ eV for AGASA~\cite{agasa_spec} 
and $E> 1.6 \times 10^{20}$~eV for HiRes~\cite{hires_mono_spec}),
although their aperture remains constant with increasing energy. 
 We then compute using a Monte Carlo technique the goodness of the fit, or $p$-value, of the distribution. Only the models with goodness of fit p-value larger than 0.05 are considered, as in Ref.~\cite{Gelmini:2007jy}.
  The maximum and minimum GZK photon fluxes depend on the intervening radio background and EGMF $B$ and  on the value of $E_{\rm max} = m_X/2$. 
The 2006~\cite{Auger-photon-06} and 2007~\cite{Auger-photon-07} Auger upper bounds on the photon fraction are also shown in Fig.~\ref{F12-BIS}.
The models with  the minimal photon fraction for the AGASA spectrum change with energy. For 
$E< 1.3 \times 10^{20}$~eV the minimum photon fraction  results from choosing   $E_{\rm max} =8 \times 10^{22}$,  intermediate radio background and $B=10^{-9}$G,
while for $E>1.3 \times10^{20}$~eV the model with minimum photon fraction has the same  $E_{\rm max}$ but maximal radio background, 
and $B=10^{-11}$G. The model with the minimal photon fraction for the HiRes spectrum
has also $E_{\rm max} =8 \times10^{22}$ and maximal radio background but $B=10^{-9}$G. 

\subsection{C. Super Heavy Dark Matter (SHDM)}

\begin{figure}[ht]
\includegraphics[width=0.3\textwidth,clip=true,angle=270]{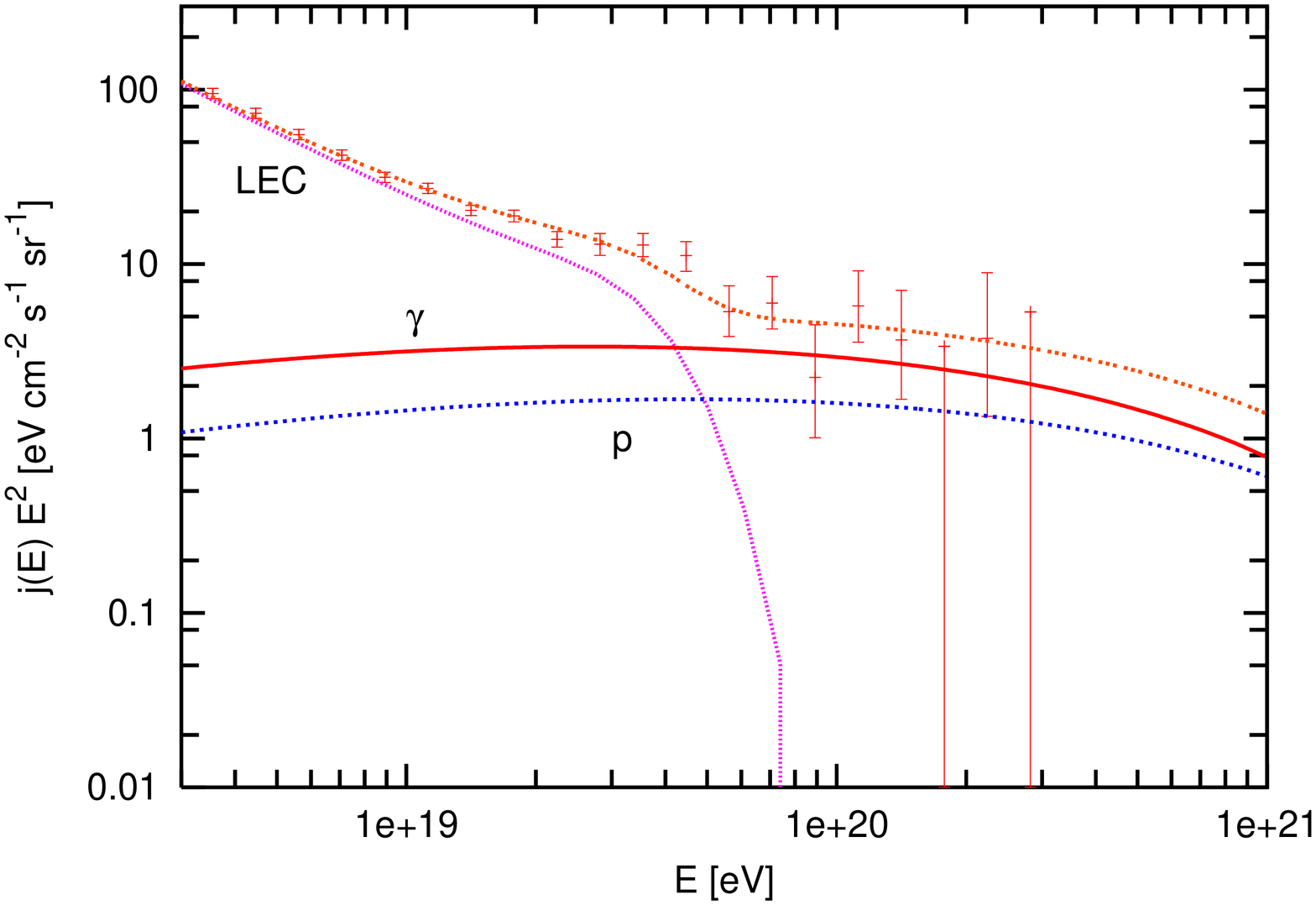}
\includegraphics[width=0.3\textwidth,clip=true,angle=270]{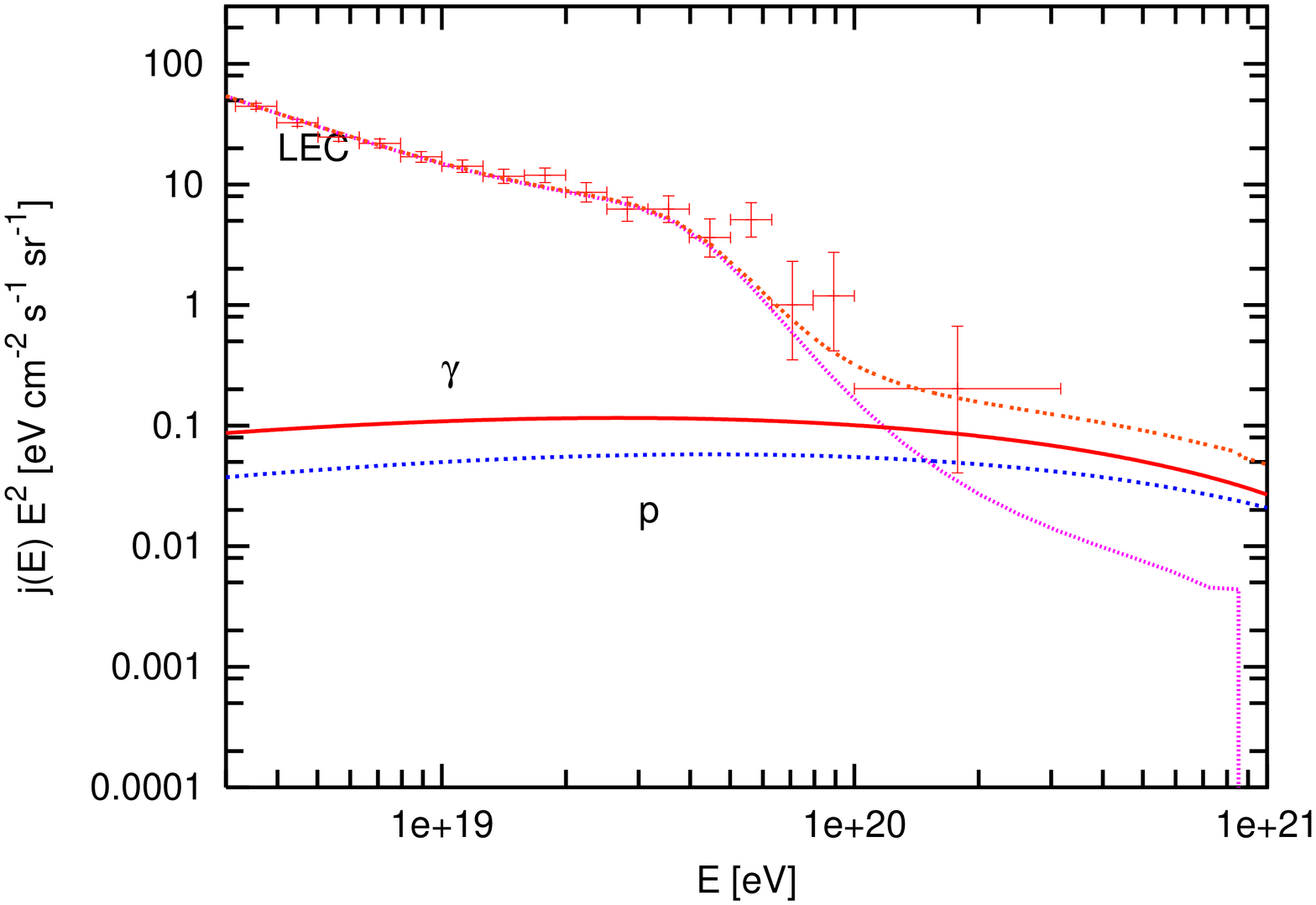}
\caption[...]{ Example of a fit (a) to AGASA 
(upper panel) and (b) HiRes (lower panel) data at high energies
with a LEC plus 
 protons and photons decay products in 
 a super heavy dark matter model. The parent particle
 mass is $ 2 \times10^{12}$~GeV. 
The low energy component (LEC) is due to nucleons from astrophysical sources.}
\label{F13}
\end{figure}

In this scenario super heavy metastable particles are produced 
in early Universe, and they remain 
 at present. They form part of  the  dark matter
 of the Universe and, in particular of
the dark halo of our galaxy. These particles (with colorful names such as
`cryptons' or `wimpzillas')
may decay~\cite{bkv97, birkel-sarkar, kr97} or annihilate~\cite{BDK} into the
observed UHECR.  The spectra of the decay  or annihilation products are
essentially determined by the physics of QCD fragmentation and  this
implies photon domination of the flux at the highest energies. 

The  UHECR in these models
are produced predominantly within the dark halo of our own galaxy. Thus
these  models predict an excess of UHECR events from 
the galactic center~\cite{dt1998}.
This anisotropy  is in conflict with  the data on arrival directions of the 
SUGAR experiment~\cite{SUGAR}, unless SHDM are responsible for the
majority of UHECR events only at energies above 
$6\times10^{19}$~eV~\cite{ks2003}. 
Even in this case, annihilating SHDM models are disfavored
 at least at the 99\% C.L. by the SUGAR
data, while decaying SHDM models  have a probability of
$\sim 10\%$ to be consistent with the SUGAR data~\cite{ks2003}.

As seen in Fig.~\ref{F14}a the model we present is barely consistent with the
upper bound on the photon fraction obtained
with  AGASA data at 10$^{20}$~eV~\cite{agasa_photon}.

 The $p$ and $\gamma$ curves in Fig.~\ref{F13} are   the predictions of a
 supersymmetric SHDM  model taken from  a recent
calculation in Ref.~\cite{SHDM_2004}, obtained by averaging over all possible decay channels,
including decays into  quarks, squarks, gluons and gluinos.
These predictions we use  here as  an example,
are similar  to those of previous
calculations~\cite{SHDM-other}  (see Fig. 17 of Ref. ~\cite{SHDM_2004}).
In particular, the ratio of  SHDM produced photons over nucleons  is about 2.

Here we reduced the mass of the parent particle  to $m_X= 2 \times
10^{12}$~GeV because, with the 10$^{14}$~GeV mass used in Ref.~\cite{SHDM_2004}
to fit the AGASA data, we find that
too many events are predicted above the end point of the AGASA  spectrum.
To be more precise, the model of Fig.~\ref{F13}, 
with $m_X= 2\times 10^{12}$~GeV, predicts 3.0 events above the
 end-point of the AGASA spectrum, i.e.   at
$E> 2.5 \times 10^{20}$ eV.  The
fit has a min. $\chi^2 = 2$ for the 3 occupied bins
 at energies $E> 10^{20}$ eV.

 For  $m_X= 10^{14}$ GeV, as used in Ref.~\cite{SHDM_2004}, the 
SHDM model predicts instead  8.5 events above the  AGASA end-point.
With the HiRes spectrum, there would not be any problem 
in using the higher $m_X$,
since  only 0.16 events are predicted with
 $m_X= 2\times 10^{12}$~GeV and 0.8 events  are predicted 
with  $m_X= 10^{14}$~GeV above the  HiRes end-point 
 (i.e. at  $E> 3.2\times 10^{20}$~eV).

We can turn this argument around and set a bound on the SHDM mass by requiring
that no more than, say,  3 events are predicted above 
 the end-point of the  AGASA spectrum.
  At the $95 \%$ C.L. this limit is $m_X < 2 \times 10^{21}$~eV.
This should be taken  as an order of magnitude
limit,  because AGASA assigned an energy to the 
events assuming proton primaries
and for photon primaries  the energy of some of the highest 
energy events can be higher~\cite{Teshima_privat}.
 A way to alleviate this bound, at the expense of 
reducing the goodness of the fit, is to
 reduce  the contribution of the SHDM model to the
total UHECR spectrum. For example, one could allow for
 $m_X= 10^{14}$~GeV by
 reducing by force the SHDM
contribution above the AGASA end-point   to 3 events.
In this case only  7 events  would be predicted 
 at $E> 10^{20}$ eV, where AGASA observed 11. The
fit has a min. $\chi^2 = 6.7$ for the 3 occupied bins
 at energies $E> 10^{20}$ eV.  Thus, reducing
the contribution of the SHDM flux to the AGASA flux to
 allow for larger $m_X$ values brings
SHDM models close to  just extragalactic  protons with a 
hard spectrum $\sim1/E$,
(with min.  $\chi^2 = 7.8$, see subsection III.A) in terms of goodness of fit.

The nucleon and photon spectra produced by the SHDM model we use is too  hard,
 thus an additional  low energy component (LEC), which we assume consists
of extragalactic nucleons, is needed to fit the data.
In Fig.~\ref{F13}a,   a LEC, parametrized
as a power law (see Eq.~(1)) with index $\alpha=2.8$,
 maximum energy $E_{\rm max}=10^{20}$~eV,
and with a zero minimum distance to the sources, has 
been added to the contribution
of the SHDM model to fit the AGASA data. In Fig.~\ref{F13}b,
 the LEC shown, added to fit the 
HiRes spectrum,   has 
$\alpha=2.7$, $E_{\rm max}=10^{21}$~eV and an assumed zero
 minimum distance to the sources.

 Note that the SHDM model studied so far, with the AGASA 
spectrum predicts a 
significant photon fraction, about 10-20 \%,
 at  energies $E>10^{19}$ eV (see Fig.~\ref{F14}a)
 which are too high for the recent Auger limits on the
the photon component of the UHECR. We discuss
 this issue in the following section.
 
 \begin{figure}[ht]
\includegraphics[width=0.3\textwidth,clip=true,angle=270]{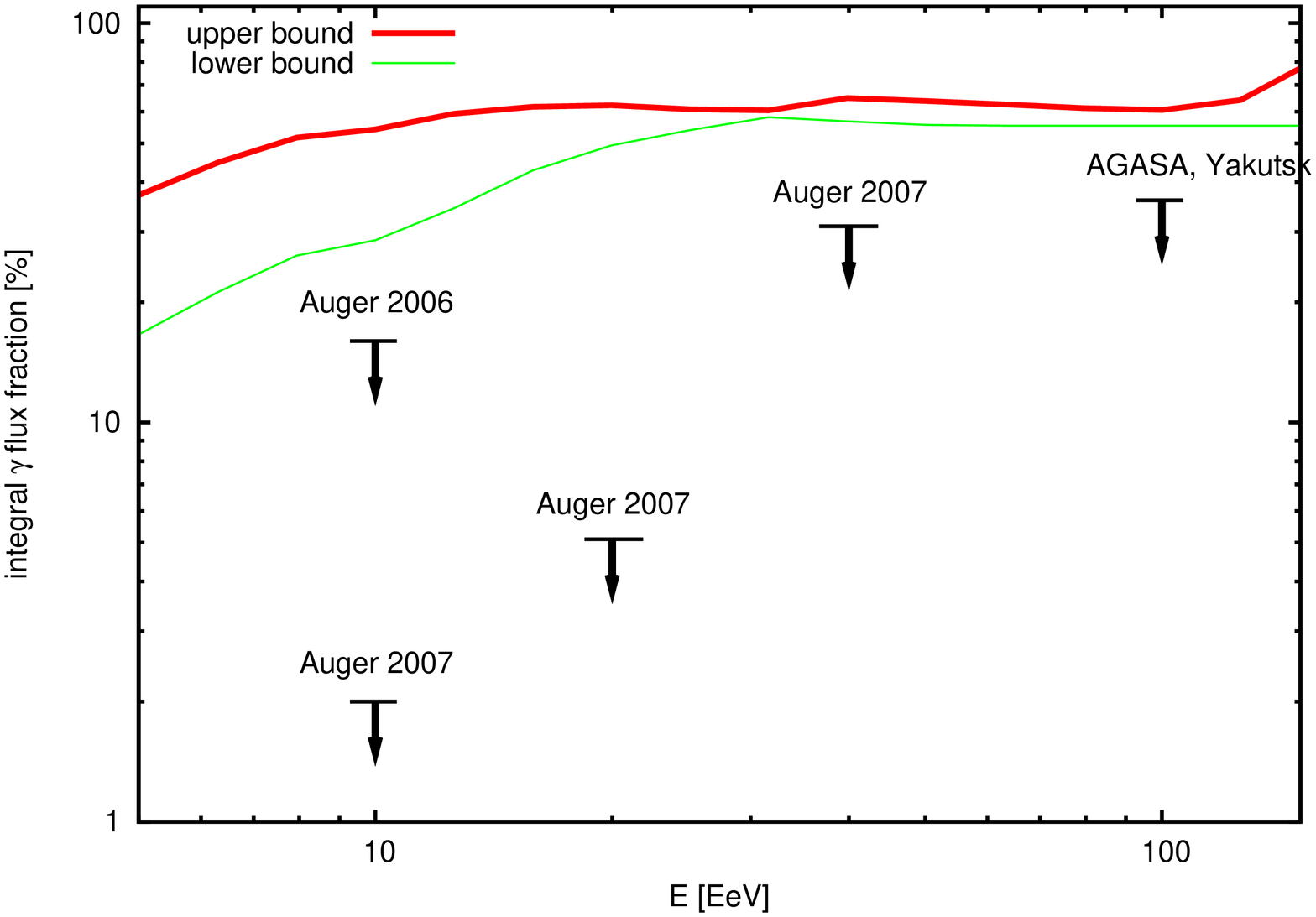}
\includegraphics[width=0.3\textwidth,clip=true,angle=270]{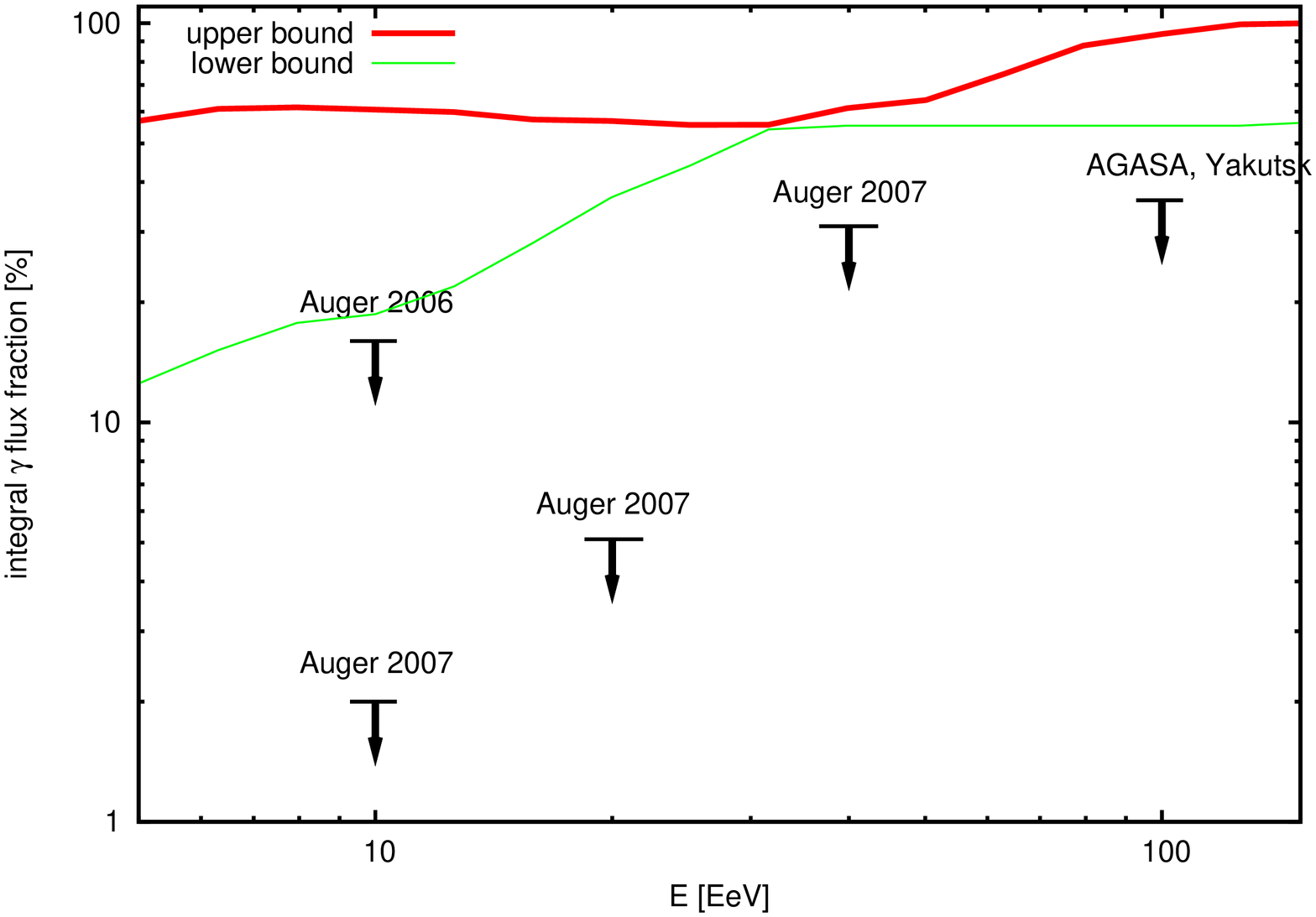}
\caption[...]{Maximum and minimum GZK photon fractions in the integral flux above the energy $E$ for
the SHDM model with the fragmentation function of Ref.~\cite{Barbot-Drees} mentioned in Sect. IV B using the statistical method of Ref.~\cite{Gelmini:2007jy}
and with  (a)  the AGASA spectrum (upper panel) and (b)  the HiRes spectrum (lower panel). 
The 2006~\cite{Auger-photon-06} and 2007~\cite{Auger-photon-07} Auger upper bounds on the photon fraction are also shown.
}
\label{F13-BIS}
\end{figure}

Using the statistical method of Ref.~\cite{Gelmini:2007jy} and the heavy particle decay  spectrum used in Fig.~\ref{F12-BIS} (taken from Refs.~\cite{Barbot-Drees,Barbot-thesis}- see the explanations in the last paragraph of the previous subsection) we fitted  the UHECR spectrum above 4 $\times 10^{19}$ eV  just with the spectrum resulting from the superheavy particle decay, with no absorption or redshift, and obtained the maximum and minimum photon fractions of the integrated flux  shown in  Fig.~\ref{F13-BIS}. We assumed that the LEC is negligible at energies 4 $\times 10^{19}$ eV and above.  Notice that the LEC  in Fig.15b, chosen above to fit the HiRes spectrum,  violates this assumption (what leads to lower predicted photon levels, since the SHDM model dominates only at higher energies).  In SHDM models the maximum and minimum photon fractions depend only on the value of $E_{\rm max}=m_X/2$ and for each energy $E$ the values of $E_{\rm max}$ giving the maximum of the minimum photon ratio are different.  We considered the range    
$1\times 10^{20}$ eV $<E_{\rm max}< $$1 \times 10^{23}$ eV, However the fitting procedure shows that only the ranges 3.5 $\times 10^{20}$ eV $<E_{\rm max}< 1.4 \times 10^{21}$ eV and 1.2$\times 10^{20}$ eV  $<E_{\rm max} < 7.1 \times 10^{20}$ eV provide acceptable models. 

Notice that when the spectrum of SHDM is assumed to dominate the UHECR spectrum only at the highest energies, i.e. close the 10$^{20}$ eV as is the case of the model in Fig.15b, the resulting minimum photon fractions are smaller (about 1\% at 1$\times 10^{19}$ eV - see Fig.17b) while if SHDM are assumed to reproduce the UHECR spectrum already at 4 $\times 10^{19}$ eV and above, the minimum expected photon fractions are larger  (above 10\% at 1$\times 10^{19}$ eV-see Fig.18b).

\subsection{D. Photon fractions}

\begin{figure}[ht]
\includegraphics[width=0.48\textwidth,clip=true,angle=0]{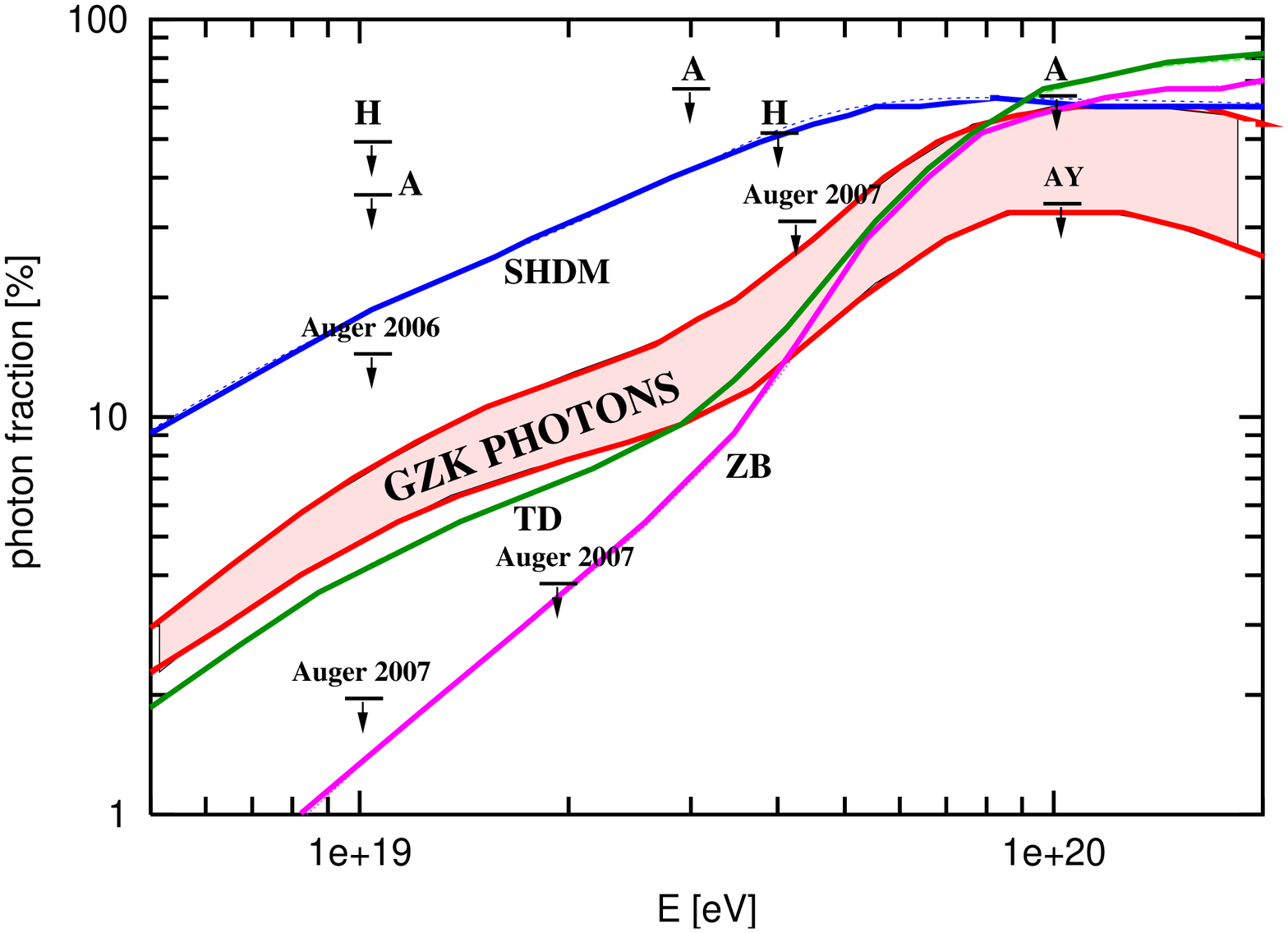}
\includegraphics[width=0.5\textwidth,clip=true,angle=0]{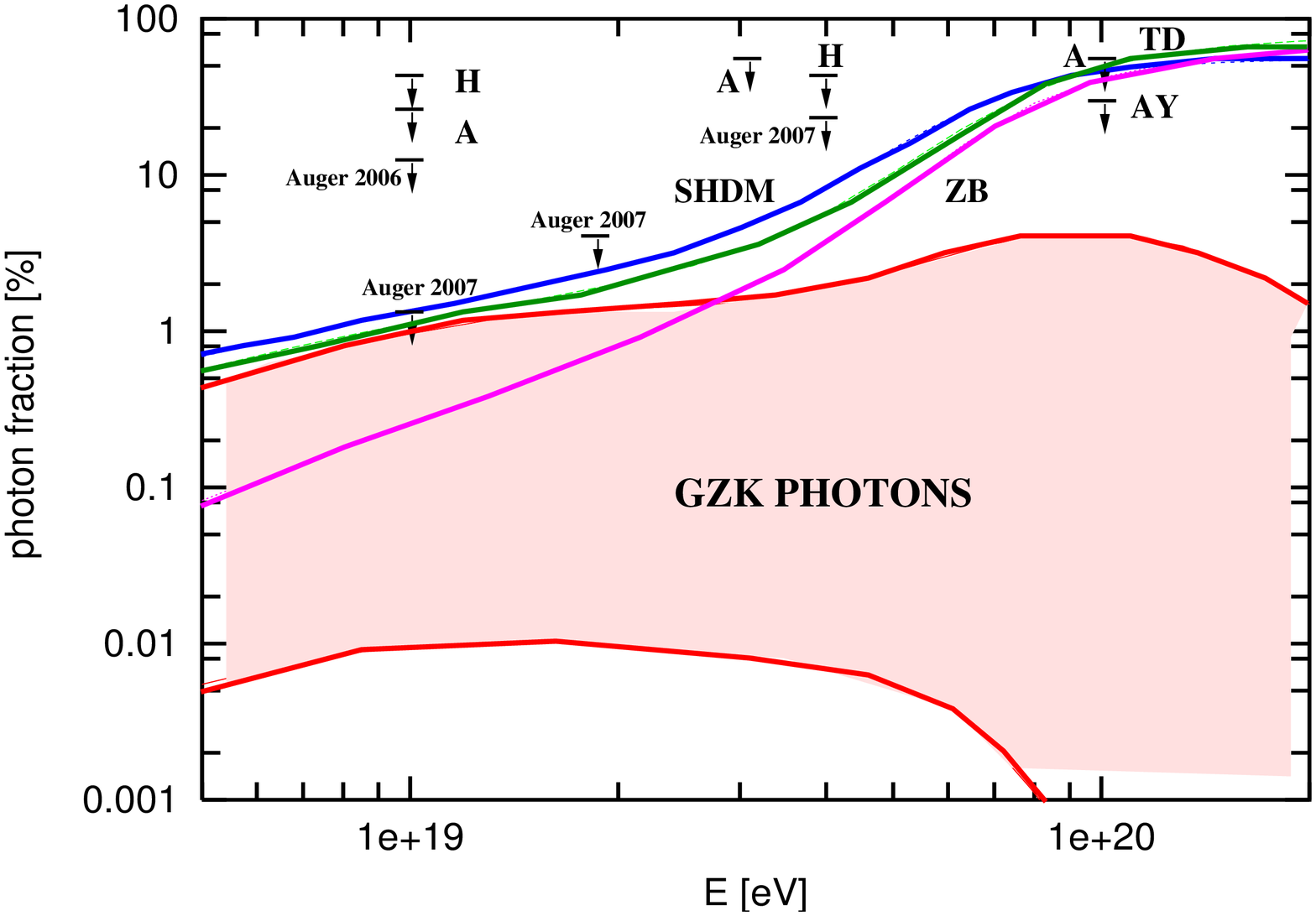}
\caption[...]{Photon fraction in percentage of the total 
 predicted integrated UHECR spectrum
above  the energy $E$ for (a) the AGASA spectrum (upper panel) and (b)
the HiRes spectrum (lower panel). The pink regions show the range of GZK photon
fractions expected if only nucleons are produced at 
the sources (see Sect. III). The curves labeled
 ZB (Z-bursts), TD (topological defects) and
SHDM  (Super Heavy Dark Matter model) show examples of
 minimum photon fractions predicted by these models (see Sect.IV).
Upper limits: {\bf A} from  AGASA, Ref.~\cite{agasa_composition_2} 
at $1-3\times 10^{19}$~eV, 
Ref~\cite{agasa_photon} and obtained with AGASA data at $10^{20}$ eV);
 {\bf AY} from the Yakutsk collaboration combining data from Yakutsk and AGASA, above  $1 \times 10^{20}$~eV~\cite{AgasaYakutskLimit};
 {\bf H} from Haverah Park~\cite{haverah}.
The 2006~\cite{Auger-photon-06} and 2007~\cite{Auger-photon-07} Auger upper bounds on the photon fraction are also shown.
}
\label{F14}
\end{figure}

\begin{figure}[ht]
\includegraphics[width=0.48\textwidth,clip=true,angle=0]{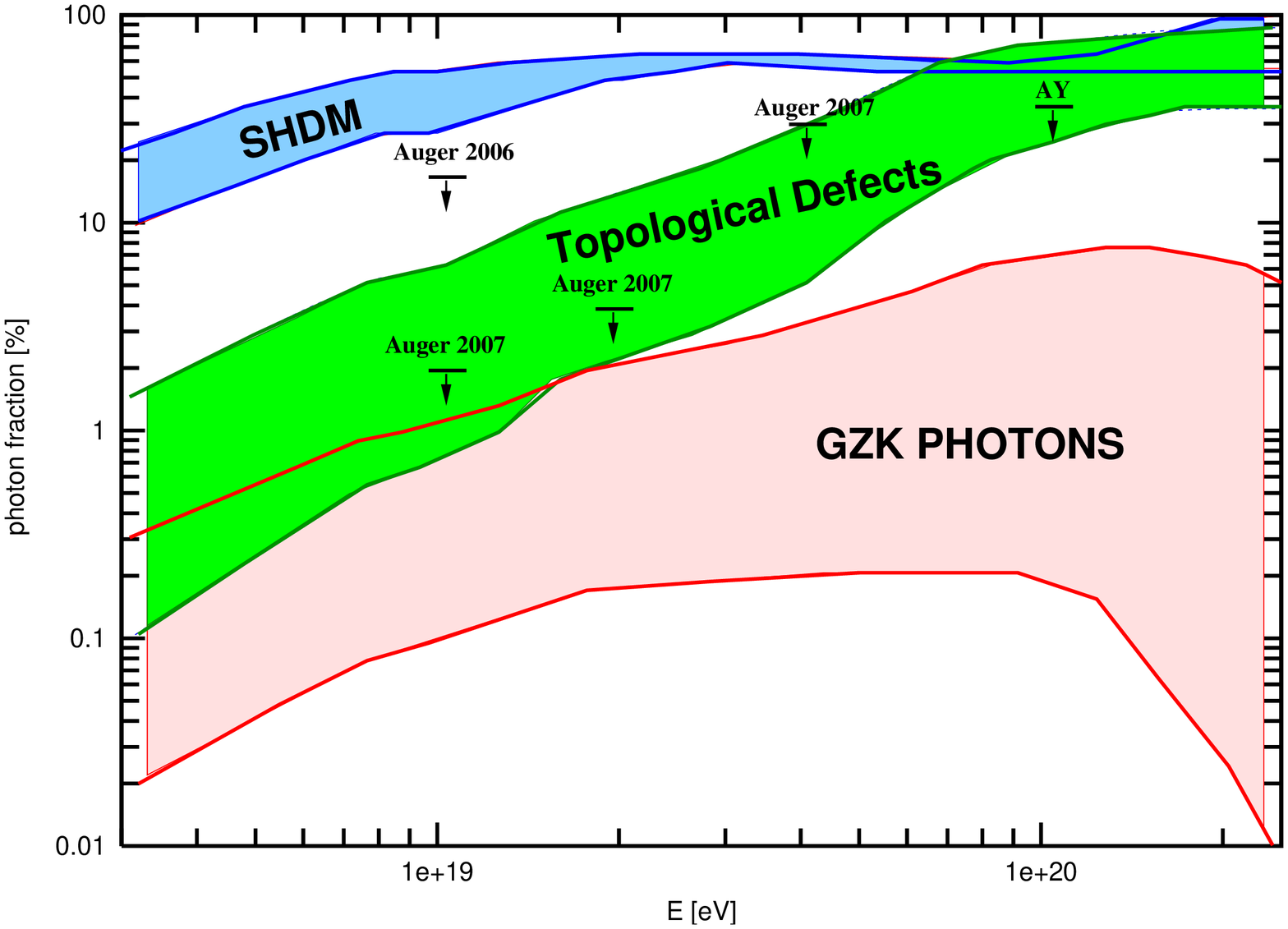}
\includegraphics[width=0.5\textwidth,clip=true,angle=0]{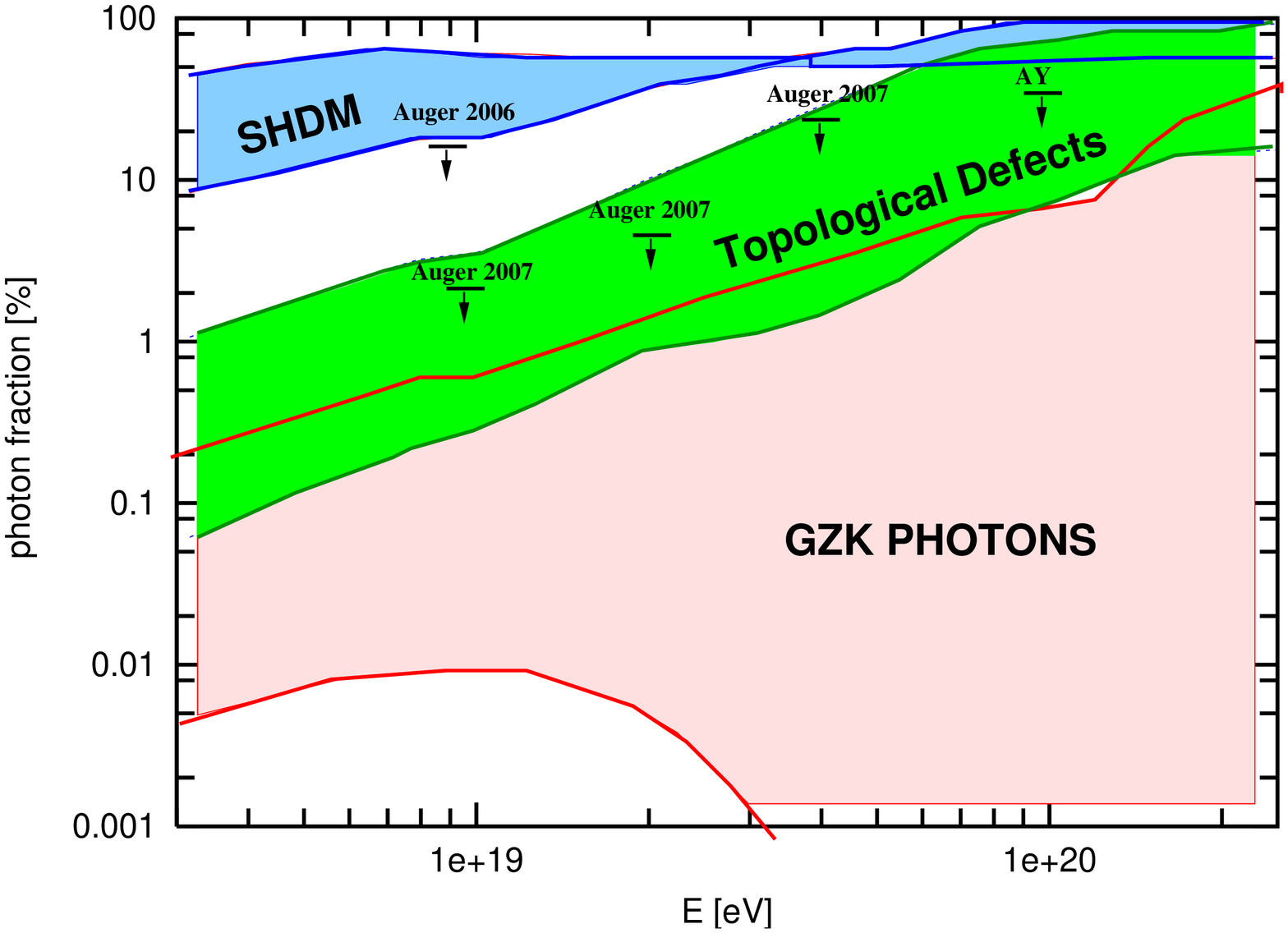}
\caption[...]{
Photon fraction in percentage of the total 
 predicted integrated UHECR spectrum
above  the energy $E$ for (a) the AGASA spectrum (upper panel) and (b)
the HiRes spectrum (lower panel).  Shown in pink  is the wider range of GZK photon
fractions expected if only nucleons are produced at 
the sources derived in Ref.~\cite{Gelmini:2007jy} (see Fig.~7 therein). Shown in green and blue
 are respectively the ranges of photon fractions
 in Fig.~\ref{F12-BIS} (for TD models) and in Fig.~\ref{F13-BIS} (for SHDM models) also obtained with the method of  Ref.~\cite{Gelmini:2007jy} (see the last paragraphs of IV.B and IV.C).   
The 2006~\cite{Auger-photon-06} and 2007~\cite{Auger-photon-07} Auger upper bounds on the photon fractionas well as the  upper bound by  the Yakutsk collaboration combining data from Yakutsk and AGASA, above  $1 \times 10^{20}$~eV~\cite{AgasaYakutskLimit} ({\bf AY})
 are also shown.
}
\label{F15}
\end{figure}
 
In Fig.~\ref{F14} we compare the range of GZK photon 
fractions we obtained in section III with
the minimal photon fractions predicted by the Top-Down 
 models shown in Figs.~10 to 13 and 15
 and  existing experimental upper bounds. Fig.~\ref{F14} shows the fraction of
 photons as percentage of   the total predicted 
integrated UHECR flux above the energy $E$ in every model. 

In Fig.~\ref{F14}a  and b the AGASA spectrum and the 
HiRes spectrum are assumed, respectively.
The ZB, TD and SHDM curves in Fig.~\ref{F14}
 correspond to the Z-burst, topological defects and
super heavy dark matter models  in Figs.~10 to 13 and 15. The pink bands show 
the range of GZK photons between the maximum and minimum fluxes obtained
in Sect. III. The upper and lower boundaries of the 
pink band in Fig.~\ref{F14}a are the photon 
curve in Fig.~\ref{F6}b and photon curve in Fig.~\ref{F6BIS}b,
 respectively. The upper and lower boundaries of the pink band  in 
Fig.~\ref{F14}b are the highest photon curve in Fig.~\ref{F7}b 
and the lowest photon curve 
of Fig.~\ref{F8}b, 
respectively. Notice how the GZK photon band depends on 
the assumed spectrum: the band for AGASA is above the 
band for HiRes, entirely separated from it.

 In Fig.~\ref{F15}  we compare  the range of GZK photon
fractions derived in Ref.~\cite{Gelmini:2007jy} with nucleons injected  by
the sources,  
with the maximum and minimum photon fractions in topological defects (necklaces) and superheavy dark matter models shown
 in Figs.~\ref{F12-BIS} and \ref{F13-BIS}. These were obtained with the same method of  Ref.~\cite{Gelmini:2007jy} and the heavy particle decay model described in the last paragraphs of the subsections IV.B and IV.C.  
 
 From  Fig.~\ref{F14} and Fig.~\ref{F15}  we conclude that at 
energies above $3 \times 10^{19}$~eV the minimum photon fraction predicted
 by Top-Down models is either larger or at most comparable to the maximum expected GZK photon ratio
 and the 2007 Auger~\cite{Auger-photon-07} and the Agasa-Yakutsk~\cite{AgasaYakutskLimit} upper bounds on the photon fraction strongly constrain Top-Down models, in particular SHDM models.

The differences between Figs. 17 and 18 are due to the different methods and models with which the photons fractions  were derived. The GZK photon fractions for the AGASA spectrum are lower in Fig.~\ref{F15} than in Fig.~\ref{F14} because  of the different fitting procedure and the different choice of $E_{\rm max}$ which can be only as high as 10$^{21}$~eV in Ref.~\cite{Gelmini:2007jy}, a more conservative value,  instead of 10$^{22}$~eV, the preferred value  for the AGASA spectrum in Section III.  

 The SHDM photon fractions are much higher in Fig. 18 than in Fig.17. The superheavy particle fragmentation functions used  to produce both figures are similar and the differences in the minimum photon fraction expected are due to the range of energies at which the SHDM is assumed to provide the bulk of UHECR: in Fig. 18 it is above 4 $\times 10^{19}$~eV and in Fig. 17 it is instead starting at energies closer to $10^{20}$ eV. However, in both cases the SHDM models studied either saturate or exceed the 2007 Auger bounds, in particular that at $1 \times 10^{19}$ eV, and the Agasa-Yakutsk bound at $1 \times 10^{20}$ eV. Thus, the Auger bounds by themselves already exclude as the dominant mechanism to produce UHECR the SHDM models considered here except at energies very close to $10^{20}$ eV\cite{Semikoz:2007wj}. Also the photon fractions given in Fig.2 of Ref.~\cite{Aloisio:2006yi}  are rejected by the 2007 Auger bound at  $1 \times 10^{19}$ eV.
There is another type  of SHDM models~\cite{Ellis:2005jc} in which the photon fraction can be smaller. Those with  the smallest photon fractions among tend to correspond to superheavy particles with larger mass and the constraint on the events predicted above experimental end point is important.  Some of these models are still allowed but very close to the existing photon limits, within a factor of two or so~\cite{Ellis-private}.

The topological defects models used  in Figs. 17 and 18 are different, that of Fig. 18 being in line with the more recent estimates of fragmentation functions in which the photon fraction is smaller than in older models. This is the main reason for the minimal photon ratios expected in these models to be smaller in Fig 18 than in Fig 17. These models are not ruled out by present photon fraction bounds however the photon fractions they predict are above 10\% at 1$\times 10^{20}$ eV. The present Agasa-Yakutsk limit
upper  limit  of $N_\gamma/N_{\rm tot}<36$\% strongly limits these models. So, either UHECR photons at  energies close to 10$^{20}$~eV will be detected,
or  better experimental  limits will be obtained in the future by
Auger. An upper limit  close to 10\% 
at those energies,  would  reject all Top-Down 
models as the origin of UHECR.

 Thus, the photon fraction
at energies above   10$^{19}$~eV, is a crucial test 
for Top-Down models. 
The only caveat to this
conclusion resides in considering that the evaluation~\cite{PB}
of the extragalactic radio background could be
wrong by several orders of magnitude, so that this 
background could be  larger than those of Ref.~\cite{PB} 
 by a large factor of 30 to 100 as suggested
in Ref.~\cite{Subir}, although there are
no specific arguments at present to justify these large factors.

We have shown in this paper that  either the detection of  UHECR photons
 or an improvement of the existing upper 
limits on the photon flux, is very important,
 both for Top-Down as well as  for Bottom-Up mechanisms to explain the UHECR.
SHDM and Z-burst models seem to be strongly disfavored by the present  experimental upper bounds on photon fraction. 
With astrophysical sources,
  the GZK photon flux is  important 
  to understand the initial proton or neutron 
spectrum emitted at the UHECR sources and the
 distribution of sources. UHECR photons may help 
us  to understand  the intervening 
  extragalactic  magnetic fields and radio background.
We have presented fits to both the AGASA and the HiRes UHECR spectra
with extragalactic nucleons, the GZK photons they produce and, when needed,
 an additional low
 energy component at energies below 10$^{19}$~eV (see section III). 
The band of expected GZK photon flux depends clearly on the UHECR spectrum
and also on the assumptions and procedure used (see Figs.~17 and 18). Once the
particular UHECR spectrum is fixed, the uncertainties in this flux due to the 
extragalactic nucleon model  and 
due to our ignorance of the intervening background are comparable
(see subsection II.E).
 Thus, extracting information on the extragalactic nucleon
 flux from the GZK photons
would require to have independent information on 
the extragalactic magnetic fields and
  radio background, and vice versa.

The detection of UHECR photons would open a new window for ultra-high energy
 astronomy and   help  
establish the UHECR sources.

\vspace{0.3cm}
{\bf Acknowledgments}

We thank I.~Tkachev for fruitful discussions and suggestions at early stages 
of this work. We also  thank S.~Troitsky for careful reading of the manuscript
 and for several important suggestions and corrections.  This work
 was supported in part by NASA grants NAG5-13399 and ATP03-0000-0057.
G.G was supported in part by the US DOE grant DE-FG03-91ER40662 
Task C.

\end{document}